\documentclass[lineno]{JFM-FLM_Au}

\usepackage[permil]{overpic}
\usepackage{url}
\usepackage{float}
\usepackage{booktabs}
\usepackage{placeins}
\usepackage{multirow}
\usepackage{xfrac} 
\usepackage{xcolor}
\usepackage{soul}

\lefttitle{Souza et al.}
\righttitle{arXiv}

\title{Triadic energy cascades and linear amplification mechanisms in supersonic turbulent boundary layers over curved walls}

\author{Lucas F. Souza\aff{1}, Hugo F. S. Lui \aff{1}, William R. Wolf \aff{1} \and Pedro S. Volpiani \aff{2}}

\affiliation{\aff{1}Department of Energy, School of Mechanical Engineering, University of Campinas, Campinas, Brazil
\aff{2}ONERA, The French Aerospace Lab, Meudon, France}

\corresau{W. R. Wolf, wolf@fem.unicamp.br}

\makeatletter
\def\ps@titlepage{%
  \leftskip\z@\let\@mkboth\@gobbletwo\vfuzz=5\p@
  \def\@oddhead{\vbox{\vspace*{-4pt}\hbox to \textwidth{\@j@urnal \hfill}}}%
  \def\@evenhead{\vbox{\vspace*{-4pt}\hbox to \textwidth{\@j@urnal \hfill}}}%
  \def\@oddfoot{\oddabsfooterflag}%
  \def\@evenfoot{\evenabsfooterflag}%
  \def\sectionmark##1{}%
  \def\subsectionmark##1{}%
}
\makeatother

\makeatletter
\def\pagelimitfooter{\bgroup%
  {\hbox to \textwidth{{\cppagefont\ifx\@volume\undefined\else\textbf{\@volume}\fi\ \ifx\@issue\undefined\ifpaper A\else\ifrapid R\else\ifpersp P\else\iffof F\else X\fi\fi\fi\fi1-\else\ifpaper A\else\ifrapid R\else\ifpersp P\else\iffof F\else X\fi\fi\fi\fi\@issue-\fi\thepage}\hfill}}%
\egroup}%
\makeatother

\makeatletter
\def\oddabsfooterflag{}
\def\evenabsfooterflag{}
\makeatother

\begin{document}
\maketitle

\begin{abstract}
Direct numerical simulations are performed to investigate supersonic turbulent boundary layers subjected to continuous concave and convex surface curvatures followed by a straight recovery section. Scale-resolved spectral turbulent kinetic energy (TKE) budgets, triadic energy-transfer bispectra, spectral proper orthogonal decomposition (SPOD), and global resolvent analysis are employed to elucidate the effects of compressibility and pressure gradients, and to distinguish the linear and nonlinear mechanisms at play. Over concave walls, bulk compression and the adverse pressure gradient preserve the mass-flux deficit while flow deceleration raises the momentum thickness, reducing the shape factor. Destabilizing curvature enhances pressure-strain redistribution, shifting near-wall turbulence toward a two-component state, and induces a shear-driven outer-layer production peak at low spanwise wavenumbers. Bispectral and SPOD analyses reveal that this outer-layer energy feeds a structured, multi-step triadic transfer toward fundamental large-scale and spanwise-uniform modes, which physically manifests as streamwise vortex breakdown rather than a three-dimensional inverse cascade. Global resolvent analysis further demonstrates that while nonlinear triadic interactions select the dominant scale amplitudes, the spatial topology of these coherent structures is dictated by linear mean-flow receptivity and spatially developing convective instabilities. Conversely, over convex walls, bulk dilatation and the favorable pressure gradient produce a severe near-wall density drop that increases the displacement thickness and shape factor despite flow acceleration. Stabilizing curvature and dilatation attenuate pressure fluctuations and suppress pressure-strain redistribution, breaking the near-wall turbulent cycle and trapping energy within elongated, one-component streamwise streaks, while a localized resolvent wavemaker energizes the downstream recovery region.
\end{abstract}



\section{Introduction}
\label{sec:introduction}

Understanding the structure and evolution of supersonic turbulent boundary layers is critical for the aerodynamic and thermal design of high-speed aerospace and propulsion systems, including scramjet intakes, external control surfaces, and supersonic turbine cascades \citep{lui_2024, HAMADA2026110029}. While decades of theoretical, experimental, and numerical efforts have established robust scaling laws for canonical flat-plate configurations \citep{vandriest_1951, bradshaw1977compressible, GANAPATHISUBRAMANI_CLEMENS_DOLLING_2006, Pirozzoli_2019}, practical engineering surfaces are characterized by complex surface curvature and spatially varying pressure gradients. In high-speed turbomachinery and aerodynamic bodies, supersonic boundary layers concurrently encounter intense streamline curvature alongside strong pressure gradients, where convex surfaces induce flow acceleration, favorable pressure gradients (FPGs), and bulk dilatation, while concave surfaces drive adverse pressure gradients (APGs) and bulk compression \citep{spina_1994, lui_2024, HAMADA2026110029}. 

\citet{Bradshaw_1974} demonstrated that bulk compression and dilatation act as extra rates of strain in addition to the mean shear, altering the Reynolds stresses by an order of magnitude larger than expected from the explicit terms in the Reynolds-stress transport equations. Streamline curvature further introduces centrifugal forces that can act independently of or alongside the pressure field. \citet{Thomann_1968} isolated curvature from pressure gradient effects to demonstrate that concave streamline curvature independently amplifies surface heat transfer by approximately 20\% relative to a flat plate subjected to an identical pressure distribution. These coupled geometric and thermodynamic effects introduce severe structural distortions, invalidating classical wall-scaling laws and necessitating an integrated investigation into the distinct mechanisms driven by curvature and pressure gradients \citep{Spalart_Watmuff_1993, spina_1994}.

The structural response of a supersonic turbulent boundary layer is dictated profoundly by the sign of the surface curvature. Concave surface curvature imposes an APG, streamline convergence, and bulk compression, generating an inherently destabilizing environment. The imbalance between the radial pressure gradient and centrifugal forces triggers centrifugal Görtler instabilities that manifest as counter-rotating streamwise vortices, causing profound turbulence amplification and structural reorganization \citep{Sun_Sandham_Hu_2019,Wang_Wang_Sun_Yang_Zhao_Hu_2019, Wu_Liang_Zhao, Zhang_Hao_Uy_2025}. Direct numerical simulations of supersonic boundary layers subjected to longitudinal concave walls by \citet{Wu_Liang_Zhao} demonstrated that flow compression and the accompanying density rise cause a noticeable boundary layer thinning, while simultaneously violating classical logarithmic and velocity-defect scaling laws. Reynolds stresses are amplified substantially throughout the layer, driven by a marked increase in turbulent kinetic energy production within the outer boundary layer, which energizes large-scale outer structures and intensifies their amplitude modulation footprint on near-wall streaks.

\citet{Wang_Wang_Sun_Yang_Zhao_Hu_2019} revealed that Görtler instabilities intensify streamwise extended vortices within hairpin packets, coalescing them into large-scale streamwise roll cells that promote sweep, ejection, and spanwise motions. This outer-layer vortex generation produces secondary bumps in streamwise and spanwise turbulence intensity profiles and displaces the peaks of wall-normal fluctuations and Reynolds shear stresses outward \citep{Wang_Wang_Sun_Yang_Zhao_Hu_2019}. Through resolvent analysis and large-eddy simulations, \citet{Zhang_Hao_Uy_2025} characterized the preferential spanwise wavelength of Görtler vortices at Mach 2.95, showing that the most amplified structures closely correspond to leading spectral proper orthogonal decomposition (SPOD) modes. 

When concave curvature is introduced into compression ramps, \citet{Kang_Lee_2026} demonstrated that varying the corner curvature modulates the interaction between counter-rotating structures and the detached shear layer: wall-normal-to-streamwise enstrophy tilting dominates the initial generation of streamwise enstrophy, while subsequent downstream stretching and vortex merging redistribute energy toward lower wavenumbers, governing low-frequency unsteadiness. These outer-layer structural modifications have parallel observations in subsonic and incompressible adverse pressure gradient boundary layers, where strong APGs promote a secondary outer production peak and intense energy diffusion toward the wall \citep{Skåre_Krogstad_1994}, induce deviations from the law of the wall and turbulence dissipation scaling \citep{Spalart_Watmuff_1993, leandro}, advect energetic small scales into the outer layer \citep{Tanarro_Vinuesa_Schlatter_2020}, and cause energetic structures to transition from wall-attached to wall-detached configurations with increasing velocity defect \citep{Gungor_Maciel_Gungor_2022}.

Over convex surfaces, the boundary layer experiences flow acceleration, FPG, and bulk dilatation. Under rapid expansion, the flow exhibits relaminarization tendencies, accompanied by a sharp reduction in turbulent kinetic energy (TKE) and Reynolds stress intensities \citep{Dussauge_Gaviglio_1987}. Employing rapid distortion approximations that neglect diffusive and dissipative terms while retaining production and pressure-strain interactions, \citet{Dussauge_Gaviglio_1987} and \citet{Smith1991} showed that bulk dilatation dominates the rapid evolution of Reynolds stresses, providing satisfactory agreement with experimental expansion data. 

Experimental investigations by \citet{Dawson_1994} demonstrated that surface pressure fluctuations across supersonic centered and gradual expansions undergo a significant reduction in overall root-mean-square levels and become strongly concentrated at low frequencies, revealing that small-scale turbulent fluctuations are selectively damped. Optical visualizations of hypersonic boundary layers over convex surfaces by \citet{Humble_2012} further revealed that while outer large-scale structures survive the expansion fan, they undergo differential acceleration that reorients them away from the wall, whereas small-scale structures are extinguished by bulk dilatation. 

Downstream of an expansion corner, the boundary layer enters a complex, non-equilibrium recovery process characterized by a distinct two-layer structure \citep{Sun_2017, Wang_2020}. Direct numerical simulations of supersonic boundary layers subjected to expansion corners \citep{Sun_2017} and external expansion waves \citep{Wang_2020} show that favorable pressure gradients widen the spanwise spacing of near-wall streaks and induce a characteristic two-layer turbulence structure. In this regime, the suppression of turbulence is far more pronounced in the outer layer than near the wall. Specifically, \citet{Sun_2017} found that inner-layer turbulence recovers toward local production-dissipation equilibrium more rapidly than the outer layer, where turbulence decays over an extended distance due to limited momentum and energy exchange across the layer. Furthermore, by isolating external expansion waves over a flat plate, \citet{Wang_2020} demonstrated that this two-layer behavior is primarily associated with bulk dilatation rather than wall curvature, with positive turbulent convection from the buffer layer to the outer region acting to compensate production and resist the decrease in turbulence. Under high-Mach-number conditions, this expansion-induced relaminarization suppresses high-frequency boundary-layer disturbances and alters downstream stability, rendering the recovering boundary layer exceptionally prone to premature separation when subsequently encountering a compression ramp \citep{Pandey_Casper_Beresh_2024}. 


To elucidate the physical mechanisms governing these contrasting amplification and attenuation phenomena, it is essential to examine the distribution and transport of TKE across discrete spatial scales. Classical TKE budgets provide valuable integral balances, but they obscure the multiscale pathways through which extra strain rates and pressure gradients alter scale-to-scale energy flows. In planar channel flows, spectral decomposition of the TKE budget by \citet{Mizuno_2016} revealed that spatial turbulent transport removes energy from large scales in the overlap layer and supplies small scales locally, while simultaneously driving downward energy fluxes that influence near-wall statistics. Extending these spectral budgets to high Reynolds numbers, \citet{Lee_Moser_2019} showed that turbulent production is heavily concentrated in very large-scale motions (VLSMs) within the streamwise velocity variance. This energy is then transferred nonlinearly across modes of varied orientations, redistributed to other velocity components via pressure–strain correlations, and ultimately dissipated nearly isotropically at the smallest scales. Furthermore, \citet{Cho_Hwang_Choi_2018} mapped the multi-scale balances between production and nonlinear transport, showing that while large motions transfer energy down to adjacent smaller scales, an inverse transfer from small to large scales concurrently energizes large-scale structures near the wall. 

In wall-bounded flows with pressure gradients and separation, \citet{schiavo} formulated an incompressible spectral TKE equation in Fourier space for flows with one homogeneous direction. These authors demonstrated that favorable and adverse pressure gradients profoundly restructure the distribution of turbulent processes along spanwise wavelengths, with APGs driving bi-directional energy transport toward both the wall and the channel core. In compressible inhomogeneous turbulence, the energy cascade is further complicated by variable density and dilatational fluctuations. Addressing this challenge, \citet{Arun_Sameen_Srinivasan_Girimaji_2021} formulated a scale-space energy transport equation for compressible turbulence, establishing that while turbulent production is driven by non-local interactions spanning the layer thickness, pressure-dilatation processes remain highly localized in scale space. In high-speed decelerating flows and shock-boundary layer interactions, spectral TKE transport analyses by \citet{Yu_Zhao_Tang_Yuan_Xu_2022} and \citet{Zhang_Han_Tang_Qu_Yan_2025} have shown that APGs selectively energize large-scale streaks, whose energy is transferred to cross-stream components via pressure-strain before cascading down to dissipative scales, underscoring the necessity of scale-resolved budgets in compressible non-equilibrium shear layers.

While one- and two-dimensional spectral budgets capture scale-dependent production, transport, and dissipation, they cannot uncover the phase-coupled nonlinear mechanisms that govern inter-scale energy exchange. The nonlinear advective term in the Navier-Stokes equations physically represents triadic interactions among wavenumber triplets. 
The theoretical framework of resonant triad interactions dates back to the foundational work of \citet{batchelor1953theory} in isotropic turbulence and \citet{Phillips_1960} on finite-amplitude surface waves, which demonstrated that third-order resonance conditions yield a sustained flux of energy between different scales \citep{Hammack_1993}. In turbulent flows, the bispectrum provides the rigorous mathematical foundation for quantifying these nonlinear couplings. Analytically, \citet{Kim_1980} established the direct relationship between the bispectral phase, three-wave coupling coefficients, and the direction and rate of power transfer between interacting modes. Pioneering experimental and theoretical investigations by \citet{Lii_Rosenblatt_Atta_1976}, \citet{HELLAND1979123}, and \citet{Herring_1980} linked measured bispectra directly to the one-dimensional turbulent energy transfer function, showing that wavenumber triplets contributing to vorticity production and spectral transfer are inherently non-local in wavenumber space. 

Closure models for the decay of homogeneous turbulence, including the direct-interaction approximation and test-field models examined by \citet{Herring_Orszag_Kraichnan_Fox_1974}, further demonstrated the critical role of triadic interactions in governing energy and enstrophy cascades. Exploring the elementary mechanics of individual triads, \citet{Waleffe} used a helical wave decomposition to classify triad interactions into forward and reverse classes, revealing that non-local triads produce a net reverse energy cascade from small to large scales alongside local catalytic forward transfers. In two-dimensional shear flows, \citet{Smyth_1992} formulated spectral transfer conservation laws that isolate triadic interactions into a source mode, a recipient mode, and a mediating advecting mode. Using this diagnosis, it was demonstrated that the transfer spectra are dominated by wave-mean flow interactions and downscale cascades driven by the deformation of small eddies during vortex mergers. Modern developments have successfully extended these concepts to complex inhomogeneous shear flows through modal bispectral techniques. 

\citet{Yeung_Chu_Schmidt_2026} introduced Triadic Orthogonal Decomposition (TOD), which identifies optimal coherent structures governing spectral momentum transfer and maps the localized regions of energy exchange across donor, recipient, and catalyst modes in an energy-budget bispectrum. In transitional shear layers, \citet{Nekkanti_Colonius_Schmidt_2025} combined spectral TKE budgets with SPOD to demonstrate that subharmonic vortex pairing is fueled both linearly by the mean flow and nonlinearly via triadic backscatter from the fundamental frequency. Furthermore, in high-speed transitional shock-boundary layer interactions (SBLIs), \citet{Ben_Hassan_Saidi_Wang_Fournier_Tenaud_Robinet_2025} applied bispectral mode decomposition (BMD) and SPOD to extract dominant triadic interactions within the separation bubble, identifying the precise triadic pathways that fuel low-frequency shock unsteadiness.

Despite significant advancements in understanding curvature effects, spectral energy budgets, and nonlinear triadic dynamics individually, critical gaps remain at their intersection. 
Detailed analyses of scale-resolved spectral TKE budgets and bispectral mode couplings have been largely confined to canonical incompressible channels, planar mixing layers, or idealized transitional flows. Consequently, the isolated influence of continuous concave and convex streamline curvature on scale-by-scale energy pathways in supersonic turbulent boundary layers remains unexplored. In particular, it remains fundamentally unknown how the competing mechanisms of centrifugal instability and bulk compression over concave surfaces, versus bulk dilatation and streamline stabilization over convex surfaces, systematically modify: (i) the spectral distribution of TKE production, spatial transport, and viscous dissipation across wall-normal heights and spanwise wavelengths; (ii) compressible pressure-dilatation and inter-component redistribution in scale space; and (iii) the underlying nonlinear triadic energy transfer pathways, specifically discriminating forward cascades from inverse backscatter and identifying active donor, recipient, and catalyst interactions.

To address these open questions, the present study conducts a scale-resolved investigation of supersonic turbulent boundary layers ($M_\infty = 2.0$, $Re_{\delta_i} = 7894$) developing over continuous concave and convex surfaces, followed by a straight recovery section, using direct numerical simulations (DNS). By combining mean-flow and Reynolds-stress diagnostics with scale-resolved spectral TKE budgets, triadic energy-transfer bispectra, SPOD, and global resolvent analysis, we isolate the effects of compressibility and pressure gradients to distinguish the linear and nonlinear mechanisms governing the concave and convex turbulent boundary layers. The remainder of this paper is organized as follows. In \S~\ref{sec:methodology}, we detail the numerical methodology and theoretical formulations, including the curvilinear DNS framework (\S~\ref{sec:DNS}), the scale-resolved spectral TKE budget (\S~\ref{sec:budget}), the triadic energy-transfer decomposition (\S~\ref{sec:triadic-analysis}), and the modal formulations for SPOD (\S~\ref{sec:SPOD}) and global resolvent analysis (\S~\ref{sec:linear}). The results are presented in \S~\ref{sec:results}, beginning with an overview of the instantaneous flow features (\S~\ref{sec:instantaneous_flow}), the mean flow properties and integral boundary-layer parameters (\S~\ref{sec:mean_flow}), and the Reynolds-stress anisotropy (\S~\ref{sec:turbulence_statistics}). Subsequently, the scale-resolved TKE budgets and triadic energy-transfer bispectra are examined in \S~\ref{sec:spectral_transport}, followed by the three-dimensional characterization of coherent structures via SPOD in \S~\ref{sec:coherent_structures} and the investigation of linear amplification mechanisms via global resolvent analysis in \S~\ref{sec:resolvent_analysis}. Finally, the main conclusions of the study are summarized in \S~\ref{sec:conclusions}.

\section{Theoretical Formulation and Numerical Methodology}
\label{sec:methodology}

\subsection{Direct numerical simulations}
\label{sec:DNS}

The numerical framework solves the three-dimensional compressible Navier-Stokes equations formulated in generalized curvilinear coordinates. The governing equations are integrated in time using the conservative variables $\mathbf{Q} = [\rho, \rho u, \rho v, \rho w, E_t]^{\mathrm{T}}$, which comprise density, momentum, and total energy per unit of volume. By mapping the physical domain in Cartesian coordinates $(x,y,z)$ onto a computational domain defined by the curvilinear coordinates $(\xi, \eta, \zeta)$, the governing equations take the form:

\begin{equation}
\frac{\partial \mathbf{Q}}{\partial t} = \frac{1}{\mathcal{J}} \left( - \frac{\partial \hat{\mathbf{F}}}{\partial \xi} - \frac{\partial \hat{\mathbf{G}}}{\partial \eta} - \frac{\partial \hat{\mathbf{H}}}{\partial \zeta} + \frac{\partial \hat{\mathbf{F}}_v}{\partial \xi} + \frac{\partial \hat{\mathbf{G}}_v}{\partial \eta} + \frac{\partial \hat{\mathbf{H}}_v}{\partial \zeta} \right) \mbox{ ,}
\end{equation}
\noindent
where $\mathcal{J} = \det(\partial(x,y,z)/\partial(\xi,\eta,\zeta))$ is the determinant of the Jacobian matrix. The terms $\hat{\mathbf{F}}$, $\hat{\mathbf{G}}$, and $\hat{\mathbf{H}}$ denote the transformed convective fluxes, while $\hat{\mathbf{F}}_v$, $\hat{\mathbf{G}}_v$, and $\hat{\mathbf{H}}_v$ represent the transformed viscous fluxes along the $\xi$, $\eta$, and $\zeta$ directions, respectively. The computational grids are constructed using a hyperbolic generator that enforces near-wall orthogonality. Hence, the $\xi$ and $\eta$ directions align tangentially and normally to the curved surfaces.


The governing equations are solved in non-dimensional form, with the length scale, velocity components, density, dynamic viscosity, pressure, and temperature normalized by the inflow boundary-layer thickness $\delta_i$, freestream speed of sound $c_\infty$, freestream density $\rho_\infty$, freestream viscosity $\mu_\infty$, $\rho_\infty c_\infty^2$, and $(\gamma-1)T_\infty$, respectively. More details about the governing equations can be found in the appendix \ref{app:governing-equations}. Here, $\gamma$ denotes the ratio of specific heats. With this choice of reference quantities, the non-dimensional freestream velocity is equal to the freestream Mach number. The flow conditions considered in this study correspond to a freestream Mach number of $M_\infty=2.0$ and a Reynolds number based on the freestream velocity and inflow boundary-layer thickness, $Re_{\delta_i}=7894$. The thermodynamic properties are modeled assuming a calorically perfect gas with $\gamma=1.4$ and a Prandtl number of $Pr=0.72$. The temperature dependence of the dynamic viscosity is determined using Sutherland's law, with a non-dimensional Sutherland constant of $0.368$. 

To accurately solve the governing equations, spatial discretization is performed using an optimized 11-point central finite-difference scheme \citep{Volpiani2024}, which achieves formal sixth-order accuracy over a broad wavenumber range. To suppress spurious grid-to-grid oscillations without compromising DNS fidelity, the non-dissipative central scheme is stabilized via an optimized 11-point, sixth-order, spatial filter \citep{Volpiani2024}. The physical accuracy of this approach is demonstrated in the results through comparisons with similar DNS data available in the literature, besides a rigorous resolution of the spectral turbulent kinetic energy budgets. Time marching is performed by an explicit, fifth-order, low-storage Runge-Kutta scheme with optimized coefficients to accurately resolve the temporal dynamics of turbulent structures while minimizing numerical dissipation, dispersion, and memory footprint \citep{Volpiani2024}. 
This framework is implemented in BASIC (version 2026.10), a modular high-fidelity Python-based flow solver~\citep{Volpiani2026}, recently extended to curvilinear coordinates and fully GPU-accelerated. 

\begin{figure}[htbp]
\centering
\begin{overpic}[trim = 0cm 0cm 0cm 0cm, clip,width=\textwidth]{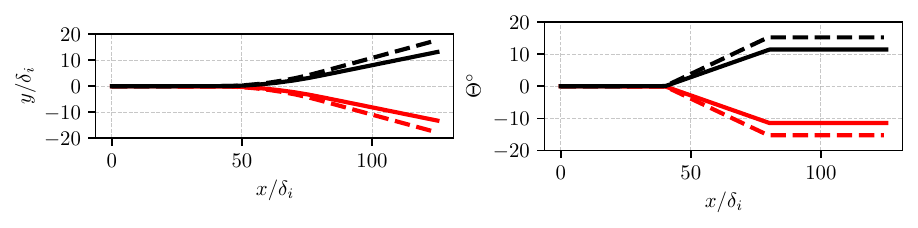}
\put(0,10){a)}
\put(500,10){b)}
\end{overpic}
\begin{overpic}[trim = 0cm 0cm 0cm 0cm, clip,width=\textwidth]{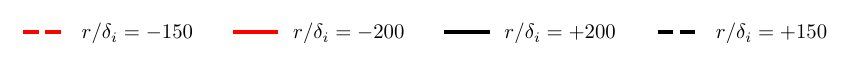}
\end{overpic}
\caption{Surface curvatures a) and wall turning angles b) along the streamwise direction for the investigated configurations.}
\label{fig:flow_config}
\end{figure}

Using the present numerical framework, direct numerical simulations (DNS) are performed to systematically investigate the effects of surface curvature on turbulent boundary layers.  Four curved configurations are considered, with radii of curvature $r/\delta_i=\pm150$ and $\pm200$. These configurations provide different curvature magnitudes and pressure-gradient regimes, encompassing adverse pressure gradients for concave surfaces (positive $r$) and favorable pressure gradients for convex surfaces (negative $r$).  The geometries of the different surfaces are compared in figure~\ref{fig:flow_config}(a), together with the corresponding streamwise evolution of the wall turning angle in figure~\ref{fig:flow_config}(b). The higher-curvature cases, $r/\delta_i=\pm150$, produce a total wall turning angle of approximately $15.28^\circ$, whereas the lower-curvature cases, $r/\delta_i=\pm200$, result in a total deflection of approximately $11.38^\circ$. 

\begin{figure}[htbp]
\centering
\begin{overpic}[trim = 0cm 0cm 0cm 0cm, clip,width=0.59\textwidth]{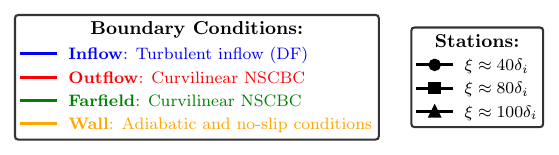}
\end{overpic}
\begin{overpic}[trim = 0cm 0cm 0cm 1.0cm, clip,width=0.99\textwidth]{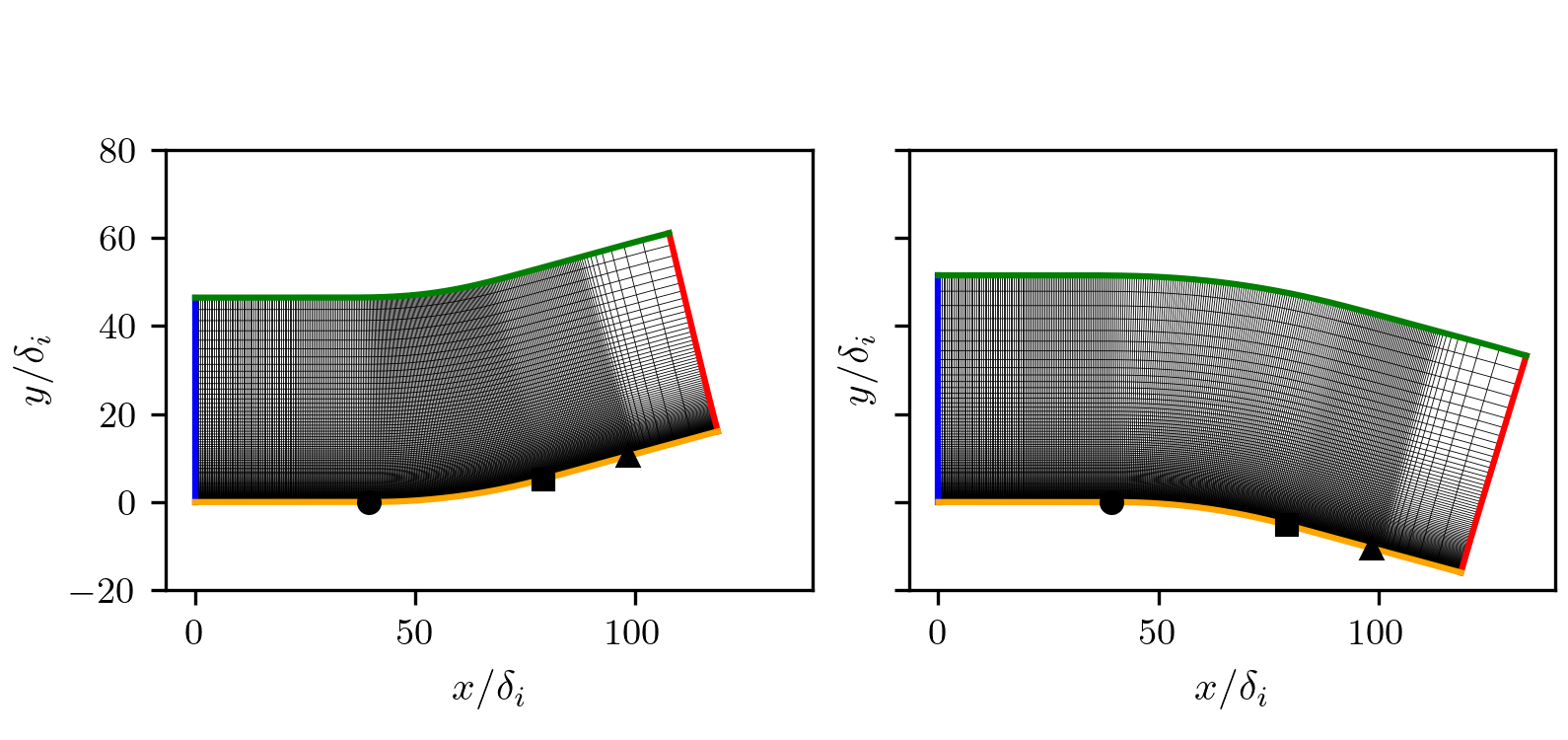}
\end{overpic}
\caption{Computational domain and boundary conditions for representative concave and convex curvature cases. For visual clarity, the depicted grids display every 10th point in the streamwise direction and every 4th point in the wall-normal direction. Grid resolution details are provided in viscous length scales in table \ref{tab:resolution}. 
}
\label{fig:bcs}
\end{figure}

The computational domain comprises several streamwise sections, as illustrated in figure~\ref{fig:bcs}. An inflow development region of length $40\delta_i$ is first employed to establish a fully developed turbulent boundary layer upstream of the curved section. This is followed by a constant-curvature region extending over an additional $40\delta_i$, within which the effects of curvature and the associated pressure gradient on the turbulent flow are examined. A flat-plate configuration ($r/\delta_i=\infty$) is also considered as a reference case.

Downstream of the curved section, a $20\delta_i$ recovery region, where the plate remains inclined but flat, is retained at full DNS resolution to examine the subsequent relaxation of the boundary layer. Finally, streamwise grid stretching is introduced near the outflow boundary to damp turbulent fluctuations and minimize spurious reflections at the domain outflow boundary. For all curved cases, the mesh is composed of $2688 \times 504 \times 324$ points in the streamwise, wall-normal, and spanwise directions, respectively, with domain lengths of $120\delta_i$, $48\delta_i$, and $10\delta_i$. Following the grid resolution of the DNS by \citet{Volpiani2018}, the flat-plate baseline configuration employs $2668\times384\times324$ grid points across domain lengths of $100\delta_i$, $18\delta_i$, and $10\delta_i$. Evaluated at a reference station immediately preceding the curvature, the spatial resolution in viscous wall units is $\Delta \xi^+ = 5.94$, $\Delta z^+ = 4.76$, and $\Delta \eta^+ = 0.46$ at the wall. More information about the grid viscous length scale over the curved sections and at the recovery part of the boundary layer is provided in table \ref{tab:resolution}.

\begin{table}
\centering
\caption{Grid resolution in wall units. $\Delta \eta^+_w$ is the first wall-normal
grid point ($\eta_1^+$) and $\Delta \eta^+_e$ is the wall-normal spacing at the
boundary-layer edge ($\eta=\delta_e$). $\Delta z^+$ is uniform across the boundary
layer, and $\Delta \xi^+$ varies by less than $2\%$ between the wall and the edge,
so only wall values are listed for these two quantities.}
\label{tab:resolution}
\begin{tabular}{llcccc}
\toprule
Curvature ($r/\delta_i$) & Station & $\Delta \xi^+_w$ & $\Delta z^+_w$ & $\Delta \eta^+_w$ & $\Delta \eta^+_e$ \\
\midrule
\multirow{2}{*}{$+150$} & $\xi=60$ & 7.33 & 5.79 & 0.54 & 4.43 \\
 & $\xi=95$ & 9.42 & 7.44 & 0.70 & 8.21 \\
\midrule
\multirow{2}{*}{$+200$} & $\xi=60$ & 7.01 & 5.54 & 0.52 & 4.23 \\
 & $\xi=95$ & 8.52 & 6.73 & 0.63 & 7.00 \\
\midrule
\multirow{2}{*}{$-200$} & $\xi=60$ & 4.74 & 3.75 & 0.35 & 3.23 \\
 & $\xi=95$ & 3.28 & 2.59 & 0.24 & 2.92 \\
\midrule
\multirow{2}{*}{$-150$} & $\xi=60$ & 4.39 & 3.47 & 0.33 & 3.08 \\
 & $\xi=95$ & 2.35 & 1.86 & 0.17 & 2.27 \\
\bottomrule
\end{tabular}
\end{table}

At the upstream boundary, a synthetic turbulent inflow is imposed using the digital filter technique to generate the incoming turbulence. This method synthesizes turbulence from random white noise tailored to satisfy predetermined length scales and reference Reynolds stresses \citep{Xie2008, Lund}. Specifically, the spatial integral length scales for the streamwise, wall-normal, and spanwise disturbances are prescribed as $0.6\delta_i$, $0.3\delta_i$, and $0.3\delta_i$, respectively. The inflow conditions are based on the DNS database from \cite{Volpiani2018} at $Re_\tau=140$. At the outflow and top boundaries, the curvilinear version of the Navier-Stokes characteristic boundary condition (NSCBC) \citep{curv_NSCBC} is applied. This implementation generalizes the formulation from \cite{Poinsot1992} to accommodate the curved and inclined boundaries present in this configuration. A no-slip, adiabatic wall condition is enforced along the bottom surface, and periodic boundary conditions are applied in the spanwise direction.

\begin{figure}[htbp]
\centering
\begin{overpic}[trim = 0cm 0cm 0cm 0cm, clip,width=0.56\textwidth]{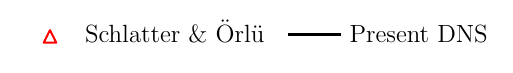}
\end{overpic}
\begin{overpic}[trim = 0cm 0cm 0cm 0cm, clip,width=0.9\textwidth]{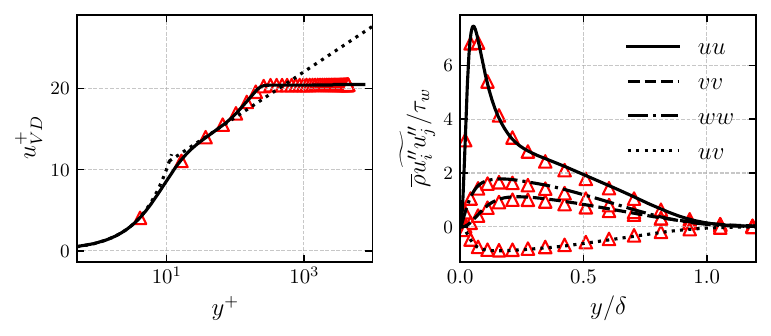}
\end{overpic}
\caption{Validation of the present results at $\xi/ \delta_i = 31$. Left: Inner-scaled Van Driest transformed mean velocity profile ($u^+_{\text{VD}}$). Right: Favre-averaged Reynolds stress components scaled as $\overline{\rho} \widetilde{u''_i u''_j} / \tau_w$. Both profiles are compared against the incompressible DNS database of \cite{SCHLATTER_2010} at a matched momentum thickness Reynolds number of $Re_{\theta,i}=Re_{\delta_2} = 670$.}
\label{fig:validation}
\end{figure}

To verify that the synthetic inflow successfully transitions into a physical turbulent boundary layer, statistical profiles are extracted at $\xi/ \delta_i = 31$, well upstream of the curvature onset. Figure~\ref{fig:validation} compares the present computational results with the incompressible DNS database of \cite{SCHLATTER_2010} at a matched momentum thickness Reynolds number of $Re_{\delta_2} = 670$. The left panel shows the inner-scaled Van Driest transformed mean velocity profile ($u^+_{\text{VD}}$), which accurately captures the viscous sublayer and logarithmic region. The right panel presents the Favre-averaged Reynolds stress components, appropriately scaled by the wall shear stress as $\overline{\rho} \widetilde{u''_i u''_j} / \tau_w$. The good agreement between the present results and the reference data for both the mean velocity and Reynolds stress profiles confirms that the turbulent fluctuations have reached a structural equilibrium state before entering the curved section. Further details regarding the state of the incoming boundary layer and the local spatial resolution are provided in table~\ref{tab:reynolds} for two reference stations: $\xi/\delta_i = 31.2$, which corresponds to the validation station matched at $Re_{\delta_2} = 670$, and $\xi/\delta_i = 40.0$, located immediately at the onset of the curved section. At these stations, the table reports the momentum-thickness Reynolds numbers based on wall viscosity ($Re_{\delta_2} = \rho_e U_e \theta/\mu_w$) and edge viscosity ($Re_\theta = \rho_e U_e \theta/\mu_e$), the friction Reynolds number ($Re_\tau = \delta_{99}/\ell_v$), the compressible shape factor ($H = \delta^*/\theta$), and the skin-friction coefficient ($C_f$). In addition, the local grid spacings in inner units are listed for the streamwise ($\Delta \xi_w^+$), spanwise ($\Delta z_w^+$), and wall-normal directions at the wall ($\Delta \eta^+_w$) and at the boundary-layer edge ($\Delta \eta^+_e$), confirming that the mesh resolution strictly satisfies direct numerical simulation requirements prior to the curvature.

\begin{table}
\centering
\caption{Boundary-layer parameters of the flat-plate case at the reference
stations. $Re_{\delta_2}=\rho_e U_e \theta/\mu_w$,
$Re_\theta=\rho_e U_e \theta/\mu_e$, $Re_\tau=\delta_{99}/\ell_v$,
$H=\delta^*/\theta$ is the compressible shape factor and $C_f$ the
skin-friction coefficient. Grid spacings in wall units are defined as in
table~\ref{tab:resolution}.}
\label{tab:reynolds}
\setlength{\tabcolsep}{4pt}
\begin{tabular}{cccccccccc}
\toprule
$\xi$ & $Re_{\delta_2}$ & $Re_\theta$ & $Re_\tau$ & $H$ & $C_f\times 10^{3}$ & $\Delta \xi_w^+$ & $\Delta z_w^+$ & $\Delta \eta^+_w$ & $\Delta \eta^+_e$ \\
\midrule
$31.2$ & $670$ & $1113$ & $230$ & $3.55$ & $3.21$ & $5.99$ & $4.74$ & $0.44$ & $3.72$ \\
$40.0$ & $747$ & $1239$ & $250$ & $3.51$ & $3.09$ & $5.94$ & $4.69$ & $0.44$ & $4.09$ \\
\bottomrule
\end{tabular}
\end{table}

\subsection{Spectral turbulent kinetic energy budgets}
\label{sec:budget}

To investigate the scale-by-scale distribution and spatial transport of turbulent kinetic energy, the TKE budget is formulated in the spectral domain by exploiting the flow spanwise homogeneity and periodicity of the computational domain. A Fourier transform is applied in the homogeneous direction, with $\widehat{(\cdot)}$ and $\widehat{(\cdot)}^*$ denoting the Fourier coefficient and its complex conjugate, respectively. The present approach allows a spectral representation of the TKE budget as a function of the streamwise and wall-normal coordinates and the spanwise wavenumber. This formulation enables the contributions to the TKE budget to be examined over spanwise scales and along the curved boundary layers, providing insight into the mechanisms through which concave and convex surface curvatures modify the distribution and transport of TKE \citep{Yu_Zhao_Tang_Yuan_Xu_2022}. The instantaneous variables are decomposed using Favre ($f = \tilde{f} + f''$) and Reynolds averaging ($f = \bar{f} + f'$).

The budget equations are formulated in a Cartesian coordinate system $(x,y)$. However, due to the surface curvature, the spatial gradients required to evaluate the budget terms are computed on the curvilinear grid $(\xi, \eta)$ using the chain rule. For instance, the derivative of an arbitrary variable $f$ with respect to the Cartesian $x$-coordinate is evaluated as $\partial f/\partial x = (\partial f/\partial \xi)(\partial \xi/\partial x) + (\partial f/\partial \eta)(\partial \eta/\partial x)$.
For the individual terms composing the TKE budget, $\Re$ indicates the time-averaged real part of a complex quantity. All variables are non-dimensional by freestream quantities ($u_\infty, \rho_\infty$).

The spectral TKE budget is defined as 
\begin{equation}
    P + C + DT + T + DV + \varepsilon + \Pi + DP + B = 0 \mbox{ ,}
\label{eq:spectral_tke_budget}
\end{equation}
where the spectral Reynolds stress tensor is defined as
\begin{equation}
  \hat{Q}_{ij}(x,y,k_z) =  \frac{1}{2}\Re\left[ \widehat{\rho u_i''} \, \widehat{u_j''}^* + \widehat{u_i''} \, \widehat{\rho u_j''}^* \right] \mbox{ .}
\end{equation}
The physical mechanisms contributing to the spectral TKE budget are grouped according to their underlying physical processes. Spectral production ($P$) represents the transfer of energy from the mean flow to the turbulent fluctuations through interactions with the mean velocity gradients 
\begin{equation}
    P = - \hat{Q}_{ij}\frac{\partial\tilde{u_i}}{\partial x_j}
\mbox{ ,} 
\label{eq:production}
\end{equation}
whereas mean convection ($C$) describes the spatial transport of spectral TKE by the mean flow
\begin{equation}
    C= - \frac{1}{2} \frac{\partial( \tilde{u}_j  \hat{Q}_{ii})}{\partial x_j} \mbox{ .}
\label{eq:mean_convection}
\end{equation}

The nonlinear turbulent advection contains two physically distinct contributions: transport of TKE through physical space and redistribution of TKE among spanwise wavenumbers. To distinguish these processes explicitly, the nonlinear term is decomposed into a spatial flux divergence ($DT$) and a conservative inter-scale transfer term ($T$). The former is given by
\begin{equation}
 DT = - \frac{1}{4} \frac{\partial}{\partial x_j} \Re \left[ \widehat{(\rho u_i'' u_j'')} \widehat{u_i''}^* + \widehat{(u_i'' u_j'')} \widehat{\rho u_i''}^* \right], \quad j \in \{1,2\} \mbox{ ,}
 \label{eq:turb_difusion}
\end{equation}
and represents the divergence of the turbulent transport of spectral TKE in the streamwise and wall-normal directions. The restriction $j\in\{1,2\}$ reflects the fact that $DT$ accounts exclusively for turbulent transport in the non-homogeneous streamwise and wall-normal directions. 
The $j=3$ contribution is retained in the nonlinear term and, after Fourier transformation, contributes to the spectral inter-scale transfer rather than to the physical-space flux divergence.
The remaining contribution is defined as
\begin{equation}
T = - \frac{1}{2} \Re \left[ \widehat{\left( \frac{\partial (\rho u_i'' u_j'')}{\partial x_j} \right)} \widehat{u_i''}^* + \widehat{\left( u_j'' \frac{\partial u_i''}{\partial x_j} \right)} \widehat{\rho u_i''}^* \right] - DT \mbox{ .}
\label{eq:interscale_transfer}
\end{equation}
By construction, subtracting $DT$ from the nonlinear contribution removes the physical-space transport component, leaving $T$ as the term responsible for redistributing TKE across spanwise wavenumbers. Consequently, $T$ is conservative with respect to the total TKE when integrated over all spanwise wavenumbers, $(\sum_{k_z}T(k_z)=0)$, and therefore represents the genuine inter-scale transfer associated with nonlinear interactions.

Although the flow is homogeneous in the spanwise direction, the derivative appearing in equation~\ref{eq:interscale_transfer} includes the spanwise derivative of the Fourier-transformed fields. Since the transformed quantities are complex-valued, differentiation in the homogeneous direction is performed spectrally according to $\partial_z\hat{f} = \mathrm{i} k_z \hat{f}$,
where $k_z$ denotes the spanwise wavenumber. In contrast, derivatives with respect to the $x$ and $y$ coordinates are evaluated directly in physical space with the chain rule on the curvilinear mesh. This treatment allows the nonlinear term to retain the contribution of spanwise gradients while separating physical-space transport from inter-scale redistribution.

Viscous and pressure effects contribute to the spatial transport, redistribution, and dissipation of TKE. The viscous terms comprise viscous diffusion ($DV$), which represents the spatial transport of spectral TKE by the fluctuating viscous stresses, and viscous dissipation ($\varepsilon$), which accounts for the irreversible transfer of TKE into internal energy. These contributions are given by

\begin{equation}
    DV = \frac{1}{2 Re} \Re \left[ \frac{\partial \widehat{\tau_{ij}'}}{\partial x_j} \widehat{u_i''}^* + \widehat{\tau_{ij}'} \left(\frac{\partial \widehat{u_i''}}{\partial x_j}\right)^* + \widehat{\left( \frac{1}{\rho} \frac{\partial \tau_{ij}'}{\partial x_j} \right)} \widehat{\rho u_i''}^* + \widehat{\left( \frac{\tau_{ij}'}{\rho} \right)} \widehat{\left(\rho\frac{\partial  u_i''}{\partial x_j}\right)}^* \right] \mbox{ ,}
\label{eq:viscous_diffusion}
\end{equation}
and
\begin{equation}
    \varepsilon = - \frac{1}{2 Re} \Re \left[ \widehat{\tau_{ij}'} \left(\frac{\partial \widehat{u_i''}}{\partial x_j}\right)^* + \widehat{\left(\frac{\tau_{ij}'}{\rho}\right)} \widehat{\left( \rho\frac{\partial u_i''}{\partial x_j}\right)}^* \right] \mbox{ .}
\label{eq:dissipation}
\end{equation}

The pressure contributions are similarly decomposed into pressure diffusion ($DP$) and pressure-strain ($\Pi$). Pressure diffusion represents the spatial transport of spectral TKE by fluctuating pressure, whereas pressure-strain describes the exchange between turbulent kinetic and internal energy associated with pressure-velocity interactions. They are defined as
\begin{equation}
    DP = - \frac{1}{2} \Re \left[ \frac{\partial \widehat{p'}}{\partial x_i} \widehat{u_i''}^* + \widehat{p'} \left(\frac{\partial \widehat{u_i''}}{\partial x_i}\right)^* + \widehat{\left( \frac{1}{\rho} \frac{\partial p'}{\partial x_i} \right)} \widehat{\rho u_i''}^* + \widehat{\left(\frac{p'}{\rho}\right)} \widehat{\left(\rho \frac{\partial u_i''}{\partial x_i}\right)}^* \right] \mbox{ ,}
\label{eq:pressure_diffusion}
\end{equation}
and
\begin{equation}
    \Pi = \frac{1}{2} \Re \left[ \widehat{p'} \left(\frac{\partial \widehat{u_i'}}{\partial x_i}\right)^* + \widehat{\left(\frac{p'}{\rho}\right)} \widehat{\left( \rho\frac{\partial u_i'}{\partial x_i}\right)}^* \right] \mbox{ .}
\label{eq:pressure_strain}
\end{equation}
Following the decomposition of the nonlinear contributions discussed above, $DV$ and $DP$ represent spatial transport of spectral TKE, whereas $\varepsilon$ and $\Pi$ represent local transfers involving turbulent kinetic energy. In particular, $\varepsilon$ acts as a sink of TKE, while $\Pi$ couples the turbulent kinetic-energy and internal-energy budgets in the compressible flow \citep{Yu_Zhao_Tang_Yuan_Xu_2022}.

Finally, the mass-diffusion term ($B$) accounts for contributions arising from density fluctuations and, in particular, from the distinction between Reynolds and Favre averaging \citep{Yu_Zhao_Tang_Yuan_Xu_2022}. This term therefore vanishes in the incompressible, constant-density limit. We note that the expression adopted here incorporates a necessary correction to an algebraic oversight in the original derivation. A detailed derivation and discussion of this correction are provided in Appendix~\ref{app:mass_diffusion}.
\begin{equation}
B = \frac{1}{2}\Re \left[ \widehat{u''_i}  - \frac{\widehat{\rho} \widehat{u''_i}^*}{\bar{\rho}} + \widehat{\left(\frac{1}{\rho}\right)} \widehat{\rho u''_i}^* \right] \left( - \frac{\partial \bar{p}}{\partial x_i} + \frac{\partial \bar{\tau}_{ik}}{\partial x_k} \right) + \frac{1}{2}\Re \left[ \frac{\widehat{\rho}^* \widehat{u''_i}}{\bar{\rho}} \right] \frac{\partial  \bar{R}_{ik}}{\partial x_k} \mbox{ .}
\label{eq:mass_diffusion}
\end{equation}

\subsection{Triadic interactions and spectral energy transfer}
\label{sec:triadic-analysis}

To elucidate the mechanisms underlying the turbulent cascade, the net nonlinear transfer is further decomposed to resolve energy exchanges between individual scales. Because the nonlinear terms involve products of fluctuating fields, their Fourier transforms are expressed as discrete convolutions in wavenumber space \citep{Nekkanti_Colonius_Schmidt_2025, Cho_Hwang_Choi_2018}. For two arbitrary fluctuating fields $a$ and $b$, the Fourier transform of their product is
%
\begin{equation}
    \widehat{(ab)}(\mathrm{r}) = \sum_{\mathrm{d}} \widehat{a}(\mathrm{d}) \widehat{b}(\mathrm{c}) = \sum_{\mathrm{d}} \widehat{a}(\mathrm{d}) \widehat{b}(\mathrm{r-d}) \mbox{ .}
\end{equation}
Here, each pair of interacting modes satisfies the triadic resonance condition ($\mathrm{d} + \mathrm{c} = \mathrm{r}$, or $\mathrm{c}=\mathrm{r}-\mathrm{d}$), 
where ($\mathrm{d}$), ($\mathrm{c}$), and ($\mathrm{r}$) denote the donor, catalyst, and recipient modes, respectively. Applying this decomposition to the nonlinear advective terms allows the contribution of each triad to the spectral energy transfer to be identified. 

We apply the donor-recipient framework of \citet{Yeung_Chu_Schmidt_2026, Smyth_1992} to compressible flows, accounting for the density-weighted velocity correlations that arise from the use of Favre fluctuations. Within this framework, ($\mathrm{d}$) denotes the donor mode from which energy is transferred, ($\mathrm{r}$) denotes the recipient mode, and ($\mathrm{c}$) mediates the nonlinear interaction between them. The contribution of an individual triad to the nonlinear energy transfer is then written as
%
\begin{equation}
    \mathcal{T}(\mathrm{d}, \mathrm{c}, \mathrm{r}) = -\frac{1}{2} \Re \left[ \frac{\partial}{\partial x_j} \left( \widehat{\rho u_i''}(\mathrm{d}) \widehat{u_j''}(\mathrm{c}) \right) \widehat{u_i''}^*(\mathrm{r}) + \left( \widehat{u_j''}(\mathrm{c}) \frac{\partial \widehat{u_i''}(\mathrm{d})}{\partial x_j} \right) \widehat{\rho u_i''}^*(\mathrm{r}) \right] 
    \mbox{ .}
\end{equation}
The triadic interaction above contains both physical-space transport and inter-scale spectral transfer. These contributions are separated by applying the product rule to the spatial derivatives in the non-homogeneous directions. The resulting physical-space turbulent transport associated with a given triad is
\begin{equation}
    DT(\mathrm{d}, \mathrm{c}, \mathrm{r}) = - \frac{1}{4} \Re \left[ \frac{\partial}{\partial x_j}  \left( \widehat{\rho u_i''}(\mathrm{d}) \widehat{u_j''}(\mathrm{c}) \widehat{u_i''}^*(\mathrm{r}) + \widehat{u_i''}(\mathrm{d}) \widehat{u_j''}(\mathrm{c}) \widehat{\rho u_i''}^*(\mathrm{r}) \right) \right], \quad j \in \{1,2\} \mbox{ .}
\end{equation}

Subtracting the physical-space transport contribution from the total triadic interaction isolates the spectral component associated with energy exchange between the donor and recipient modes as
\begin{equation}
    T_{\text{spectral}}(\mathrm{d}\to \mathrm{r}) = \mathcal{T}(\mathrm{d}, \mathrm{c}, \mathrm{r}) - \ DT(\mathrm{d}, \mathrm{c}, \mathrm{r}) \mbox{ .}
\end{equation}
The resulting quantity represents the net energy transfer associated with the specific donor-recipient pair mediated by the catalyst mode. Summation over all donor and catalyst modes recovers the net inter-scale transfer at the recipient wavenumber. Because the nonlinear transfer is conservative, the resulting spectral transfer redistributes turbulent kinetic energy among spanwise scales without changing the total spectral energy.
%


\subsection{Spectral proper orthogonal decomposition} 
\label{sec:SPOD}

To identify the energetically dominant space-time coherent structures associated with specific spanwise wavenumbers and temporal frequencies, we employ Spectral Proper Orthogonal Decomposition (SPOD) \citep{Towne_Schmidt_Colonius_2018}. Let $\mathbf{q}(\mathbf{x},t)$ denote the multivariate state vector of physical fluctuations, comprising the density, velocity components, and temperature,
\begin{equation}
    \mathbf{q}(x, y, z, t) = \Big[ \rho'(\mathbf{x},t), \, u'(\mathbf{x},t), \, v'(\mathbf{x},t), \, w'(\mathbf{x},t), \, T'(\mathbf{x},t) \Big]^{\mathrm{T}} \mbox{ ,}
\end{equation}
where $\mathbf{x}=(x,y,z)$ denotes the spatial coordinates. Exploiting the spanwise homogeneity of the flow, a discrete spatial Fourier transform is first applied in the homogeneous $z$ direction,
\begin{equation}
    \check{\mathbf{q}}(x, y, k_z, t) = \sum_{m=0}^{N_z-1} \mathbf{q}(x, y, z, t) \, \exp\left( -i k_z z_m \right) \mbox{ ,}
\end{equation}
where $k_z$ denotes the spanwise wavenumber. The resulting Fourier coefficients are complex-valued and describe the spatial structure of the fluctuations at a prescribed spanwise wavenumber.

For each target $k_z$, the resulting time series is partitioned into $N_b$ overlapping blocks according to Welch's method. Each block $n\in{1,2,\ldots,N_b}$ contains $N_{\mathrm{fft}}$ snapshots sampled at a uniform time interval $\Delta t$. A Hamming window $w(t)$ is applied to each block to reduce spectral leakage, after which a temporal discrete Fourier transform (DFT) is performed:
\begin{equation}
    \widehat{\mathbf{q}}^{(n)}(x, y, k_z, f) = \sum_{j=0}^{N_{\text{fft}}-1} w(j\Delta t) \, \check{\mathbf{q}}^{(n)}(x, y, k_z, j\Delta t) \, \exp\left( -i 2\pi f j \Delta t \right) \mbox{ .}
\end{equation}

Because the spanwise Fourier coefficients $\check{\mathbf{q}}$ are generally complex-valued, the temporal Fourier coefficients at a fixed $k_z$ do not possess the usual conjugate symmetry of a real-valued time signal. Consequently, the positive- and negative-frequency components are treated as distinct realizations in the SPOD analysis. Assuming reflectional symmetry of the spanwise statistics, the positive-frequency realizations and the complex conjugates of the corresponding negative-frequency realizations can be combined to increase the statistical convergence of the SPOD estimate, following \cite{SPOD_2}. The resulting augmented snapshot ensemble is defined as
%
%
\begin{equation}
    \mathbf{\mathrm{Q}}_f = \frac{1}{\sqrt{2}}\Big[ \mathbf{\mathrm{Q}_{\text{pos}}}(f), \, \mathbf{\mathrm{Q}}_{\text{neg}}^*(f) \Big] \mbox{ ,}
\end{equation}
where $\mathbf{\mathrm{Q}_{\text{pos}}}$ and $\mathbf{\mathrm{Q}}_{\text{neg}}$ contain the Fourier realizations from the positive- and negative-frequency components, respectively. The matrix $\mathbf{\mathrm{Q}}_f$ has dimensions $M \times 2N_b$, where $M=N_{\text{vars}} N_x N_y$ is the number of spatial and multivariate degrees of freedom in the $(x,y)$ plane for the prescribed spanwise wavenumber $k_z$.
%

To provide a physically meaningful measure of the energy content of the compressible fluctuations, the SPOD modes are computed using a generalized inner product over the spatial domain $\Omega_{xy}$,
\begin{equation}
    \langle \widehat{\mathbf{q}}_1, \, \widehat{\mathbf{q}}_2 \rangle_{\mathbf{W}} = \int_{\Omega_{xy}} \widehat{\mathbf{q}}_2^{\mathrm{H}} \, \mathbf{W} \, \widehat{\mathbf{q}}_1 \, \mathrm{d}\Omega \approx \widehat{\mathbf{q}}_2^{\mathrm{H}} \, \mathbf{W}_{\text{global}} \, \widehat{\mathbf{q}}_1 \mbox{ ,}
\end{equation}
where $\mathrm{H}$ denotes the conjugate transpose. The global weight matrix is written as $\mathbf{W}_{\text{global}} = \mathbf{W}_{\text{p-vars}} \otimes \mathbf{W}_{\text{grid}}$ and accounts for both the spatial integration $\mathbf{W}_{\text{grid}} = \text{diag}(\mathrm{d}A)$ and the relative scaling of the thermodynamic and kinematic variables $\mathbf{W}_{\text{p-vars}}$. The former matrix represents the quadrature weights associated with the discrete mesh element $\mathrm{d}A$, while the variable-weight matrix is based on the Chu compressible-energy norm \citep{Chu1965}
\begin{equation}
    \mathbf{W}_{\text{p-vars}} = \text{diag}\left( 
        \frac{\bar{T}}{\bar{\rho} \gamma M_\infty^2}, \, 
        \bar{\rho}, \, 
        \bar{\rho}, \, 
        \bar{\rho}, \, 
        \frac{\bar{\rho}}{\bar{T} \gamma (\gamma - 1) M_\infty^2} 
    \right) \mbox{ .}
\end{equation}

For each $(k_z,f)$ pair, SPOD identifies the orthogonal set of spatial structures that optimally represents the cross-spectral density under this energy-weighted inner product. The modes are obtained efficiently using the method of snapshots. Specifically, the reduced cross-spectral kernel matrix is defined as
\begin{equation}
    \mathbf{K}_f = \mathbf{\mathrm{Q}}_f^\mathbf{{\mathrm{H}}} \, \mathbf{W}_{\text{global}} \, \mathbf{\mathrm{Q}}_f \mbox{ ,}
\end{equation}
which has dimensions $2N_b\times2N_b$. The SPOD eigenvalues and expansion coefficients are obtained from
%
\begin{equation}
    \mathbf{K}_f \, \mathbf{v}_l = \lambda_l \, \mathbf{v}_l \mbox{ ,}
\end{equation}
where $\lambda_l$ represents the energy associated with the $l$-th SPOD mode at the prescribed $(k_z,f)$ pair. The corresponding spatial modes are reconstructed as
\begin{equation}
    \boldsymbol{\Phi}_l(k_z, f) = \frac{1}{\sqrt{\lambda_l}} \, \mathbf{\mathrm{Q}}_f \, \mathbf{v}_l \mbox{ .}
\end{equation}
The normalization ensures orthonormality with respect to the chosen energy inner product.
The leading mode ($l=1$) therefore represents the most energetic coherent structure at the prescribed spanwise wavenumber and temporal frequency. By examining the spatial distribution and relative energy of the leading SPOD modes across selected $(k_z,f)$ pairs, we identify the dominant coherent structures associated with the spectral dynamics of the turbulent boundary layer. Frequencies are reported in non-dimensional form, as the Strouhal number, which is defined as $St = f \delta_{\mathrm{i}} / U_\infty$.

\subsection{Resolvent analysis}
\label{sec:linear}

To perform global resolvent analysis within the framework previously described, the instantaneous state is decomposed into a time-averaged base flow $\bar{\mathbf{q}}(x, y)$ obtained from the DNS and an infinitesimal disturbance $\mathbf{q}'(x, y, z, t)$, such that $\mathbf{q} = \bar{\mathbf{q}} + \mathbf{q}'$ with $\|\mathbf{q}'\| \ll \|\bar{\mathbf{q}}\|$. Assuming statistical homogeneity along the spanwise direction, the perturbations are expanded into cross-sectional Fourier modes:

\begin{equation}
\mathbf{q}'(x, y, z, t) =  \hat{\mathbf{q}}(x, y) \, e^{i (\beta z - \omega t)} ,
\label{eq:fourier_ansatz}
\end{equation}
where $\beta$ is the non-dimensional spanwise wavenumber, $\omega = 2\pi \text{St}$ is the circular frequency ($\text{St} = f \delta_{\mathrm{i}} / U_\infty$ being the Strouhal number), and $\hat{\mathbf{q}}(x, y) \in \mathbb{C}^{5 N_x N_y}$ is the complex disturbance 
vector. 

Linearizing the Navier-Stokes residual operator $\mathbf{R}(\mathbf{q})$ about the base flow $\bar{\mathbf{q}}$ yields the forced disturbance evolution system
\begin{equation}
\frac{\partial \mathbf{q}'}{\partial t} = \mathbf{A}(\bar{\mathbf{q}}, \beta) \mathbf{q}' + \mathbf{f}',
\label{eq:linearized_evolution}
\end{equation}
where $\mathbf{f}'$ denotes an external forcing and $\mathbf{A} \in \mathbb{C}$ is the discrete global Jacobian matrix:
\begin{equation}
\mathbf{A} = \left. \frac{\partial \mathbf{R}}{\partial \mathbf{q}} \right|_{\bar{\mathbf{q}}, \beta}.
\end{equation}
In the present framework, $\mathbf{A}$ is constructed directly by differentiating the discrete spatial residual operator of the DNS solver via forward-mode automatic differentiation in JAX paired with graph coloring. Because $\mathbf{A}$ is obtained from the exact discrete residual routines, it inherits all numerical boundary conditions from the forward DNS solver without requiring separate discretization or ad-hoc linearization. However, instead of the digital filter employed for turbulence generation at the inflow during the forward simulation, Dirichlet boundary conditions are prescribed at the inlet when evaluating the Jacobian matrix, which corresponds to $\mathbf{q}' = 0$ at the inflow plane. Applying the temporal Fourier transform ($\partial / \partial t \to -i\omega$) to equation~\eqref{eq:linearized_evolution} yields the frequency-domain dynamical system
\begin{equation}
(i\omega \mathbf{I} - \mathbf{A}) \hat{\mathbf{q}} = \hat{\mathbf{f}}.
\label{eq:frequency_domain}
\end{equation}

To restrict actuation and observation to physically relevant subdomains and target state components, an input operator $\mathbf{B}$ and an output restriction operator $\mathbf{C}$ are introduced such that the system response is observed as $\hat{\mathbf{y}} = \mathbf{C} \hat{\mathbf{q}}$ under a localized input forcing $\mathbf{B}\hat{\mathbf{f}}$. The generalized input-output resolvent operator is then expressed as
\begin{equation}
\mathcal{R}(\omega) = \mathbf{C} (i\omega \mathbf{I} - \mathbf{A})^{-1} \mathbf{B}.
\label{eq:resolvent_operator}
\end{equation}
The optimal linear amplification is quantified as the maximum gain between the output response and input forcing
\begin{equation}
\sigma^2(\omega) = \max_{\hat{\mathbf{f}} \neq \mathbf{0}} \frac{\|\hat{\mathbf{y}}\|_{\mathbf{W}_C}^2}{\|\hat{\mathbf{f}}\|_{\mathbf{W}_B}^2} = \max_{\hat{\mathbf{f}} \neq \mathbf{0}} \frac{\hat{\mathbf{y}}^H \mathbf{W}_C \hat{\mathbf{y}}}{\hat{\mathbf{f}}^H \mathbf{W}_B \hat{\mathbf{f}}},
\label{eq:gain_opt}
\end{equation}
where $\mathbf{W}_B$ and $\mathbf{W}_C$ are Hermitian, positive-definite weight matrices defining the energy inner products on the curvilinear mesh.

To maintain consistency with the SPOD modal decomposition, the inner product weight matrices $\mathbf{W}_B$ and $\mathbf{W}_C$ are evaluated under the compressible Chu energy norm introduced in the previous section. In terms of primitive fluctuations $\mathbf{s}' = [\rho', u', v', w', T']^T$, the point-wise energy density is given by $e_{\text{Chu}} = \frac{1}{2} \mathbf{s}'^H \mathbf{W}_{\text{p-vars}} \mathbf{s}'$. However, because the discrete Jacobian $\mathbf{A}$ and state perturbations are formulated in conservative variables $\mathbf{q}' = [\rho', (\rho u)', (\rho v)', (\rho w)', E_t']^T$, we introduce a point-wise linear transformation matrix $\mathbf{T}_m$ such that $\mathbf{s}' = \mathbf{T}_m \mathbf{q}'$. Defining the auxiliary thermodynamic coefficient $\alpha = \gamma (\gamma - 1) M_\infty^2 / \bar{\rho}$ and specific kinetic energy $\bar{k} = \frac{1}{2}(\bar{u}^2 + \bar{v}^2 + \bar{w}^2)$, the transformation operator takes the explicit form:

\begin{equation}
\mathbf{T}_m = \begin{bmatrix}
1 & 0 & 0 & 0 & 0 \\[6pt]
-\dfrac{\bar{u}}{\bar{\rho}} & \dfrac{1}{\bar{\rho}} & 0 & 0 & 0 \\[6pt]
-\dfrac{\bar{v}}{\bar{\rho}} & 0 & \dfrac{1}{\bar{\rho}} & 0 & 0 \\[6pt]
-\dfrac{\bar{w}}{\bar{\rho}} & 0 & 0 & \dfrac{1}{\bar{\rho}} & 0 \\[6pt]
\alpha \bar{k} - \dfrac{\bar{T}}{\bar{\rho}} & -\alpha \bar{u} & -\alpha \bar{v} & -\alpha \bar{w} & \alpha
\end{bmatrix}.
\label{eq:Tm_matrix}
\end{equation}
Accounting for the discrete cell area (or volume) $\Delta A$, the local weight matrix in conservative variables is evaluated via the congruence transformation
\begin{equation}
    \mathbf{W}_{\text{c-vars}} = \frac{1}{2} \Delta A \, \mathbf{T}_{m}^T \mathbf{W}_{\text{p-vars}} \mathbf{T}_{m}.
\label{eq:W_congruence}
\end{equation}
As a consequence of its lower-triangular structure, the determinant of $\mathbf{T}_m$ is the product of its diagonal entries, $\det(\mathbf{T}_m) = \alpha / \bar{\rho}^3 = \gamma (\gamma - 1) M_\infty^2 / \bar{\rho}^4$. Because $\bar{\rho} > 0$, $\gamma > 1$, and $M_\infty > 0$, $\det(\mathbf{T}_m)$ remains strictly positive across the entire domain, ensuring that $\mathbf{T}_m$ is unconditionally invertible and non-singular. Substituting $\mathbf{s}' = \mathbf{T}_{m} \mathbf{q}'$ into the local Chu energy density and integrating over all grid cells with volumes $\Delta V = J^{-1}$, the global Chu inner product in conservative variables becomes 
\begin{equation}
E_{\text{Chu}} = \mathbf{q}'^H \mathbf{W_{\text{c-vars}}} \mathbf{q}' .
\end{equation}

Because $\mathbf{W}_{\text{c-vars}}$ is symmetric positive-definite, the weight matrices can be factored as $\mathbf{W}_B = \mathbf{F}_B^T \mathbf{F}_B$ and $\mathbf{W}_C = \mathbf{F}_C^T \mathbf{F}_C$ via Cholesky decomposition. The optimization problem~\eqref{eq:gain_opt} then reduces to the singular value decomposition of the weighted resolvent operator
\begin{equation}
\mathcal{R}_w(\omega) = \mathbf{F}_C \mathbf{C} (i\omega \mathbf{I} - \mathbf{A})^{-1} \mathbf{B} \mathbf{F}_B^{-1} = \sum_{j=1}^{K} \sigma_j(\omega) \, \mathbf{u}_{w,j} \mathbf{v}_{w,j}^H,
\label{eq:weighted_resolvent}
\end{equation}
where $\sigma_j$ represents the energy gain of the $j$-th mode at frequency $\omega$. The corresponding unit-energy physical forcing and response modes in conservative variables are recovered via:
\begin{equation}
\hat{\mathbf{f}}_j = \mathbf{B} \mathbf{F}_B^{-1} \mathbf{v}_{w,j}, \qquad 
\hat{\mathbf{q}}_j = \mathbf{C}^T \mathbf{F}_C^{-1} \mathbf{u}_{w,j}.
\label{eq:phys_modes}
\end{equation}

\section{Results}
\label{sec:results}

This section presents results for turbulent boundary layers developing over concave and convex surfaces with different radii of curvature. For all investigated cases, the flow is characterized by a freestream Mach number of $M_\infty = 2.0$ and an inflow boundary-layer thickness Reynolds number of $Re_{\delta_i} = 7894$. The working fluid is modeled as a calorically perfect gas with $\gamma = 1.4$ and $Pr = 0.72$.

\subsection{Instantaneous flow features}
\label{sec:instantaneous_flow}

\begin{figure}
\centering
\begin{overpic}[trim = 0cm 0cm 0cm 0cm, clip,width=0.49\textwidth]{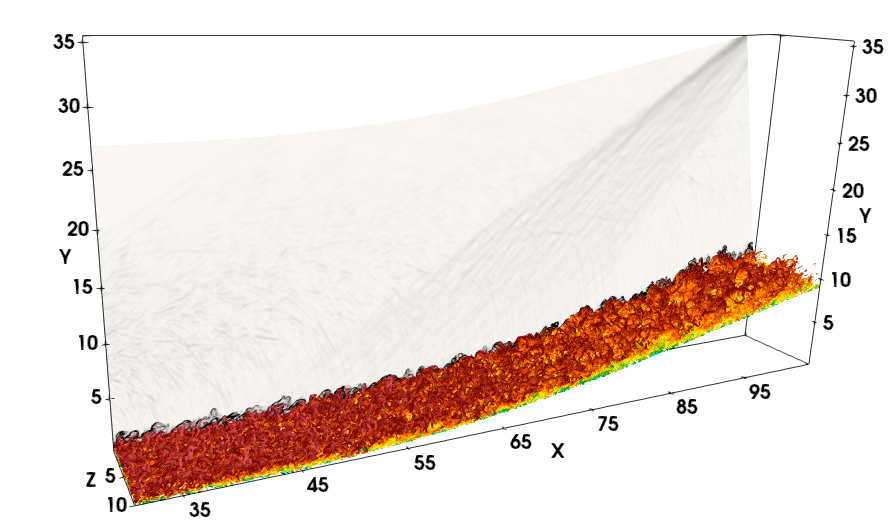}
\put(300,450){$r/\delta_i = +200$}
\put(150,450){$a)$}
\end{overpic}
\begin{overpic}[trim = 0cm 0cm 0cm 0cm, clip,width=0.49\textwidth]{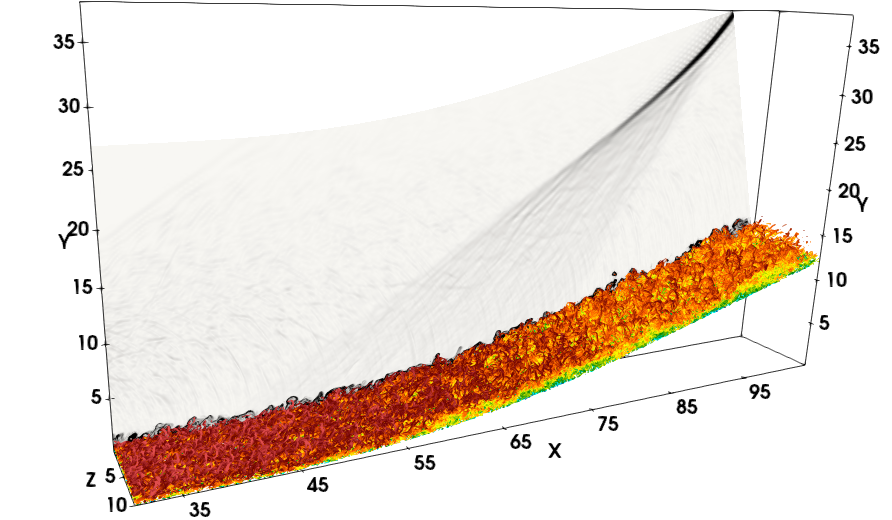}
\put(300,450){$r/\delta_i = +150$}
\put(150,450){$b)$}
\end{overpic}
\vspace{\baselineskip}
\vspace{\baselineskip}
\begin{overpic}[trim = 0cm 0cm 0cm 0cm, clip,width=0.49\textwidth]{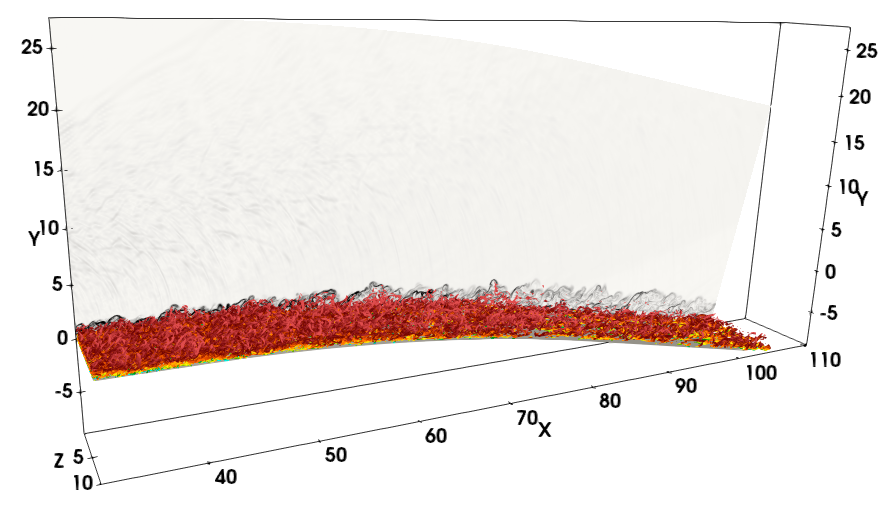}
\put(300,450){$r/\delta_i = -200$}
\put(150,450){$c)$}
\end{overpic}
\begin{overpic}[trim = 0cm 0cm 0cm 0cm, clip,width=0.49\textwidth]{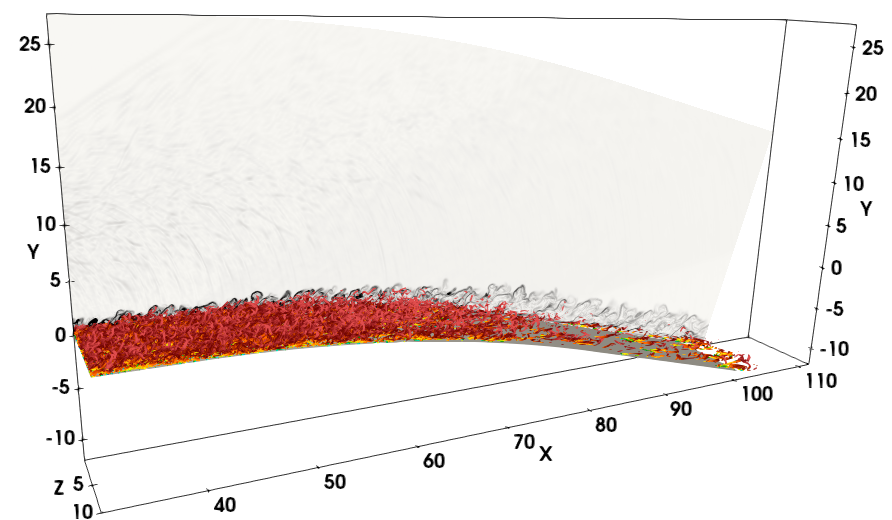}
\put(300,450){$r/\delta_i = -150$}
\put(150,450){$d)$}
\end{overpic}
\caption{Instantaneous flow fields in the curved section ($x \in [30, 100]$). Isosurfaces of $\mathcal{Q} = 1.0 \,c_\infty^2/\delta_i^2$ are colored by streamwise velocity $u$, with background contours of density gradient magnitude $|\nabla \rho|$: (a) concave surface with $r/\delta_i = +200$; (b) concave surface with $r/\delta_i = +150$; (c) convex surface with $r/\delta_i = -200$; (d) convex surface with $r/\delta_i = -150$.}
\label{fig:inst}
\end{figure}

The instantaneous flow fields within the curved region, extending from $x=30$ to $100$, are shown in figure~\ref{fig:inst} for the concave and convex configurations at $r/\delta_i=\pm200$ and $r/\delta_i=\pm150$. Turbulent structures are identified using isosurfaces of the $\mathcal{Q}$-criterion, with $\mathcal{Q}=1.0 \,c_\infty^2/\delta_i^2$, colored by the streamwise velocity component $u$. A background plane showing the magnitude of the density gradient, $|\nabla\rho|$, is also included to visualize the compression-wave system developing in the outer flow.

The top panels, figures~\ref{fig:inst}(a,b), show the flow over the concave surfaces. The turbulent structures undergo deformation as they evolve through the curved region. In the outer flow, the $|\nabla\rho|$ field reveals compression waves generated by the surface turning. For the milder curvature, $r/\delta_i=200$ (figure~\ref{fig:inst}a), these waves remain relatively weak and dispersed in the outer flow. With increasing curvature, $r/\delta_i=150$ (figure~\ref{fig:inst}b), the compression waves become more concentrated and progressively coalesce into an oblique shock wave outside the boundary layer.
The bottom panels, figures~\ref{fig:inst}(c,d), show the corresponding convex configurations. In these cases, the turbulent structures become progressively less populated towards the downstream end of the curved region. This attenuation is more pronounced for the stronger curvature, $r/\delta_i=-150$ (figure~\ref{fig:inst}d), where a substantial reduction in the visible turbulent structures is observed towards $x\approx100$.

\begin{figure}[htbp]
\centering
\begin{overpic}[trim = 0cm 4cm 0cm 3cm, clip,width=0.90\textwidth]{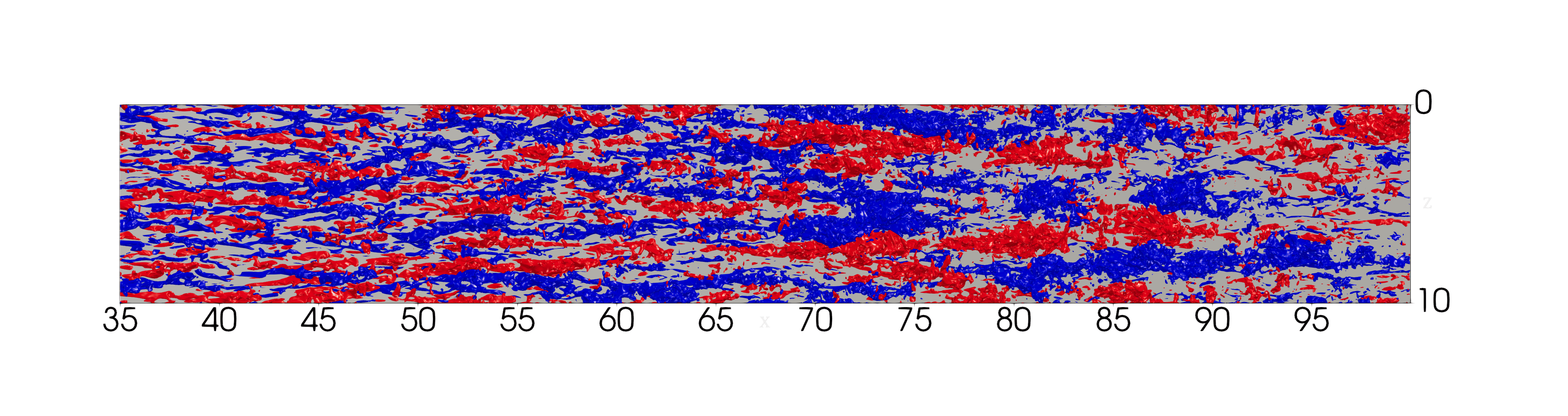}
\put(-20,40){\rotatebox{90}{$r/\delta_i = +150$}}
\put(930,105){\rotatebox{0}{$z/\delta_i$}}
\end{overpic}
\begin{overpic}[trim = 0cm 4cm 0cm 7cm, clip,width=0.90\textwidth]{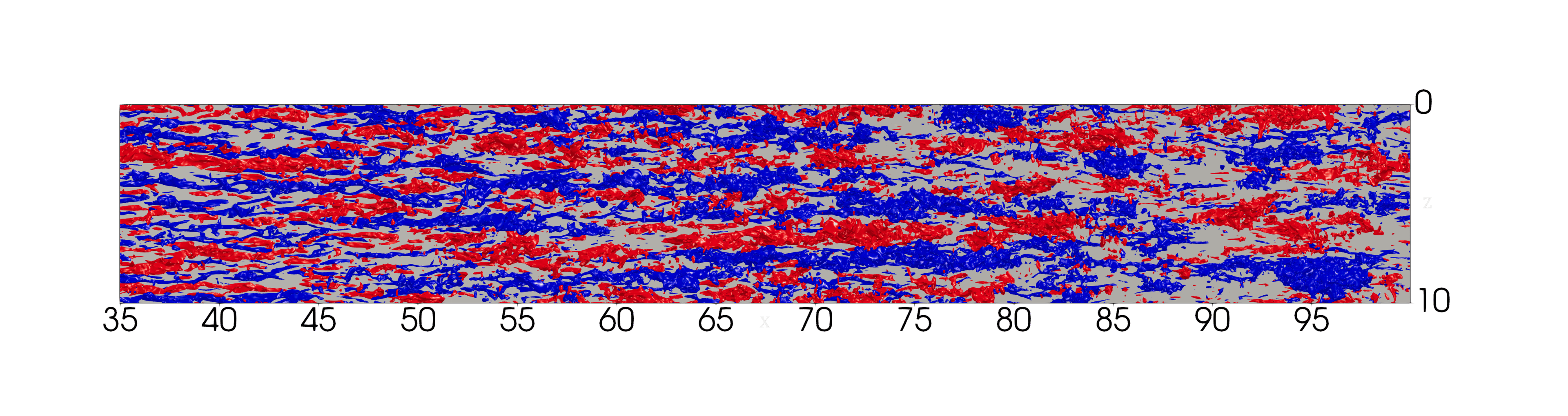}
\put(-20,40){\rotatebox{90}{$r/\delta_i = +200$}}
\put(930,110){\rotatebox{0}{$z/\delta_i$}}
\end{overpic}
\begin{overpic}[trim = 0cm 4cm 0cm 7cm, clip,width=0.90\textwidth]{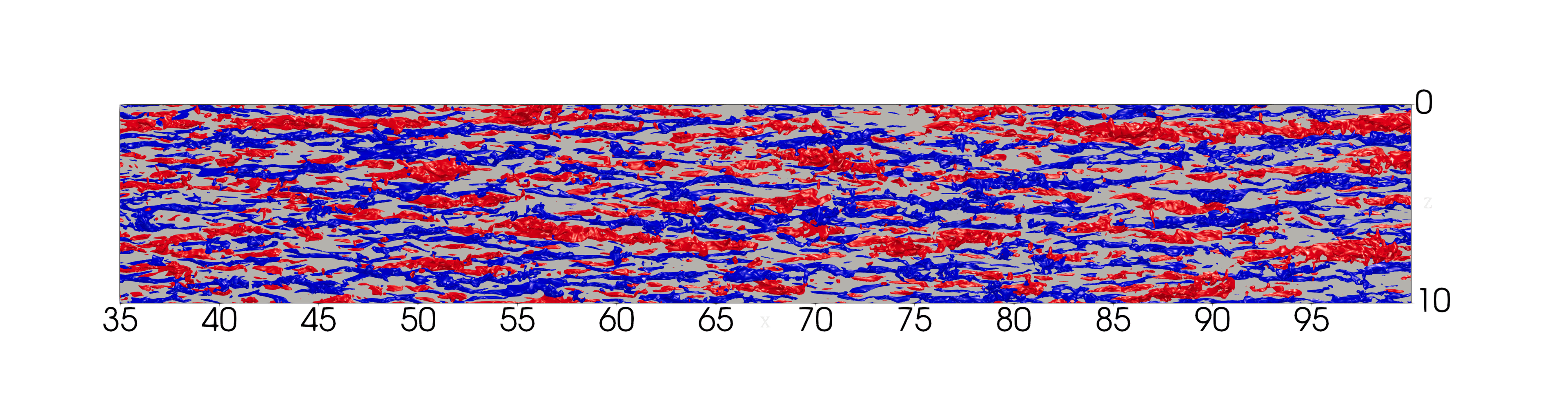}
\put(-20,50){\rotatebox{90}{$r/\delta_i = \infty$}}
\put(930,110){\rotatebox{0}{$z/\delta_i$}}
\end{overpic}
\begin{overpic}[trim = 0cm 4cm 0cm 7cm, clip,width=0.90\textwidth]{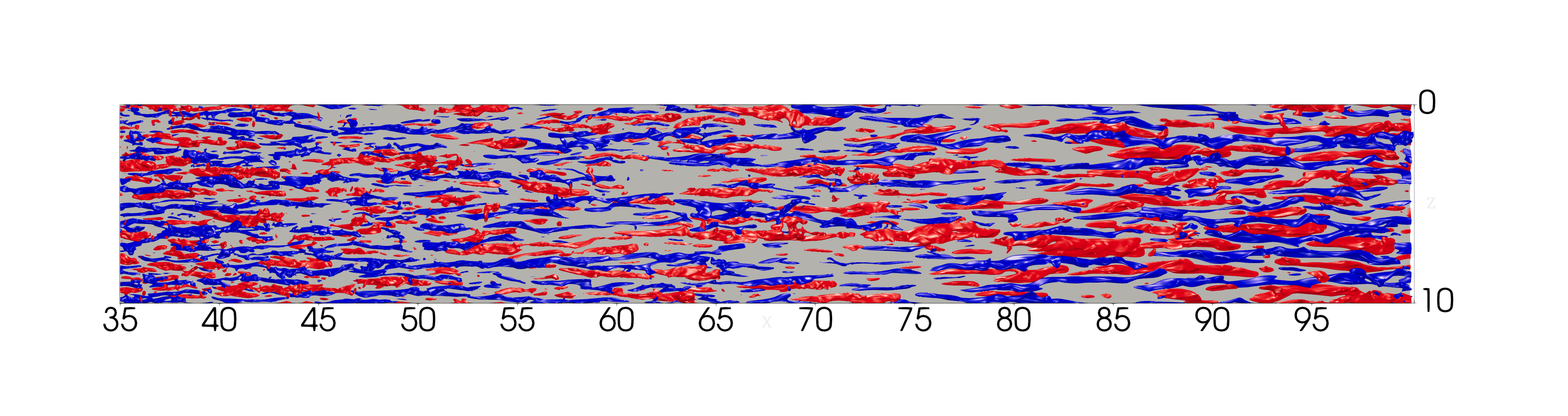}
\put(-20,40){\rotatebox{90}{$r/\delta_i = -200$}}
\put(930,110){\rotatebox{0}{$z/\delta_i$}}
\end{overpic}
\begin{overpic}[trim = 0cm 4cm 0cm 7cm, clip,width=0.90\textwidth]{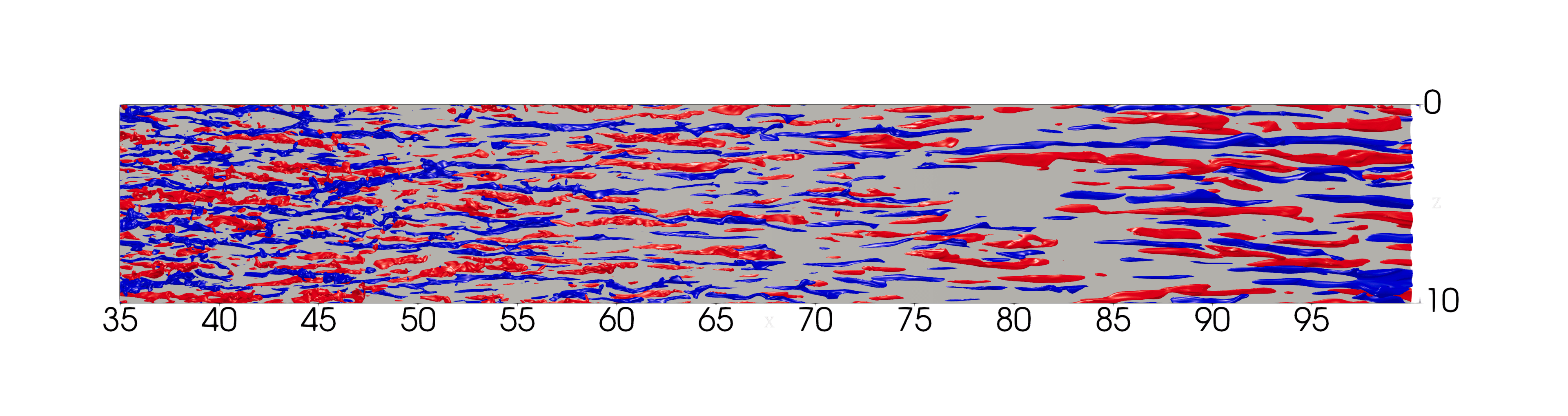}
\put(-20,40){\rotatebox{90}{$r/\delta_i = -150$}}
\put(930,110){\rotatebox{0}{$z/\delta_i$}}
\put(475,-10){\rotatebox{0}{$x/\delta_i$}}
\end{overpic}
\caption{Isosurfaces of streamwise velocity fluctuations at $u_{\xi}'/U_\infty = 0.15$, illustrating the spatial organization of near-wall streaky structures across different surface curvatures. Top to bottom: convex cases ($r/\delta_i = -150, -200$), flat-plate baseline ($r/\delta_i = \infty$), and concave cases ($r/\delta_i = 200, 150$).}
\label{fig:fluc}
\end{figure}

To examine how surface curvature modifies the near-wall flow structure, figure~\ref{fig:fluc} shows isosurfaces of streamwise velocity fluctuations at $u_{\xi}'/U_\infty = 0.15$. The panels are ordered from top to bottom by surface curvature, starting with the concave configurations ($r/\delta_i = +150$ and $+200$), passing through the flat-plate reference ($r/\delta_i = \infty$), and ending with the convex cases ($r/\delta_i = -200$ and $-150$).

In all cases, the characteristic streaky structures generated in the upstream flat-plate ($x/\delta_i<40$) boundary layer enter the curved section ($40<x/\delta_i<80$), where they undergo distinct spatial reorganizations. Over the concave configurations ($r/\delta_i > 0$), the flow exhibits larger structures with greater spanwise extent. As the flow progresses along the concave section, the streak-like structures become increasingly broadened in the spanwise direction. Near the beginning of the recovery region, these elongated structures give way to more compact coherent motions, marking a change in the spatial organization of the flow.

In contrast, over the convex surfaces ($r/\delta_i < 0$), the streak density along the spanwise direction decreases over the curved segment, resulting in fewer, more spaced streaks \citep{Wang_2020}. This attenuation is consistent with the visualizations in terms of the $Q$-criterion fields for the convex surfaces (figures~\ref{fig:inst}(c,d)). Further downstream, in the recovery region ($80<x/\delta_i<100$), where the plate remains inclined but is flat, the streaky structures undergo an apparent amplification. Although this downstream development contrasts with the reorganization observed over the curved section, the instantaneous flow fields alone do not establish whether it reflects renewed turbulent activity or the amplification of disturbances within a predominantly laminar flow. These features are examined further in the following sections.

\subsection{Mean flow properties}
\label{sec:mean_flow}

\begin{figure}[htbp]
\centering
\begin{overpic}[trim = 0cm 0cm 0cm 0cm, clip, width=0.99\textwidth]{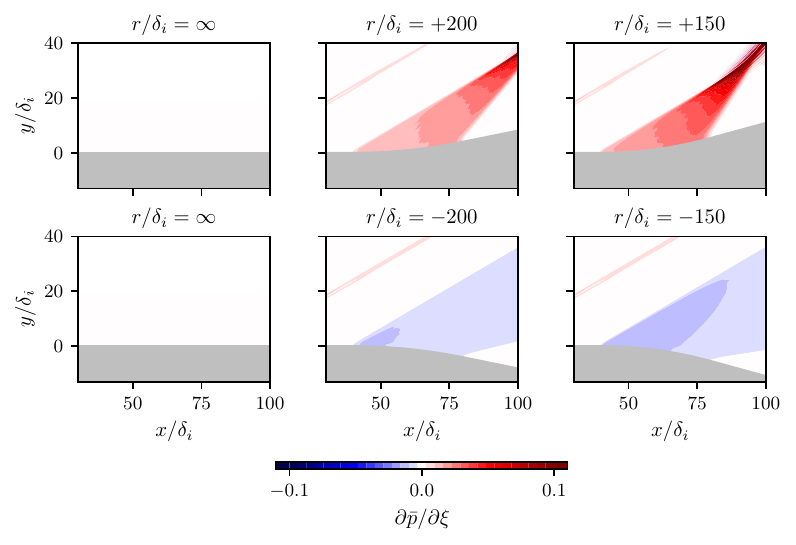}
\end{overpic}
\caption{Contours of the mean streamwise pressure gradient $\partial \overline{p}/\partial\xi$. Columns from left to right: flat plate baseline, $r/\delta_i = \pm200$, and $r/\delta_i = \pm150$. Top row: concave geometries; bottom row: convex geometries. }
\label{fig:mean_flow}
\end{figure}

The macroscopic effects of surface curvature on boundary layer development are first assessed through an examination of the time-averaged flow fields. Figure~\ref{fig:mean_flow} illustrates contours of the mean streamwise pressure gradient ($\partial\overline{p}/\partial \xi$) across all configurations investigated. The cases are arranged from left to right starting with the baseline flat plate, followed by progressively stronger surface curvatures of $r/\delta_i = \pm200$ and $r/\delta_i = \pm150$. The top row presents the concave configurations, while the bottom row displays the corresponding convex geometries.

As anticipated, geometric curvature induces distinct streamwise and wall-normal pressure gradients that dictate the flow behavior. The concave surfaces impose an adverse pressure gradient and compression waves, leading to flow deceleration. Although the compressive turning over concave walls generates an oblique shock wave in the outer stream, the computational domain and curvature onset are designed such that this shock remains far from the wall and does not impinge upon the boundary layer. Thus, the near-wall modifications are governed strictly by streamline curvature and pressure gradients rather than shock-boundary layer interactions. Over the convex surfaces, the flow experiences a favorable pressure gradient accompanied by expansion waves, driving a mean flow acceleration.

A direct consequence of the streamline curvature is the variation in external flow. Under these conditions, the velocity outside the boundary layer does not attain a uniform, asymptotic plateau, rendering standard flat-plate definitions of the boundary layer edge (such as $\overline{u} = 0.99 U_\infty$) ambiguous \citep{ Wu_Liang_Zhao, Wang_Wang_Sun_Yang_Zhao_Hu_2019,griffin-moin-bl,HAMADA2026110029}. To circumvent this limitation and obtain a consistent definition of the boundary layer thickness $\delta$ across all geometries, an equivalent inviscid tangential velocity profile $u_{\mathrm{inv}}$ is reconstructed along the boundary layer from the conservation of total enthalpy, following \cite{griffin-moin-bl}:
\begin{equation}
    h + \frac{1}{2}\left(\bar{u}_{\xi}^2 + \bar{u}_{\eta}^2\right) = h_\infty + \frac{1}{2}\left(u_{\xi,\infty}^2 + u_{\eta,\infty}^2\right) \mbox{ ,}
    \label{eq:total_enthalpy}
\end{equation}
where $u_{\xi}$ and $u_{\eta}$ are the tangential and wall-normal velocity components, respectively, and the subscript $\infty$ denotes freestream quantities. For a calorically perfect gas with specific heat ratio $\gamma$, the static enthalpy is given by $h = \frac{\gamma}{\gamma - 1}\frac{p}{\rho}$. Rearranging \eqref{eq:total_enthalpy} yields the reconstructed inviscid velocity
\begin{equation}
    u_{\mathrm{inv}} = \left[ \frac{2\gamma}{\gamma - 1} \left( \frac{p_\infty}{\rho_\infty} - \frac{\bar{p}}{\bar{\rho}} \right) + u_{\xi,\infty}^2 + u_{\eta,\infty}^2 - \bar{u}_{\eta}^2 \right]^{1/2},
    \label{eq:inviscid_velocity}
\end{equation}
where $\bar{p}$, $\bar{\rho}$, and $\bar{u}_\eta$ are the Reynolds-averaged pressure, density, and wall-normal velocity profiles, respectively.

While the inviscid reconstruction accurately captures the flow field in the outer region where viscous effects are negligible, it systematically departs from the true velocity within the boundary layer owing to the viscous effects. The boundary layer is therefore identified as the region where the actual mean tangential velocity, $\bar{u}_\xi$, deviates from this reconstructed profile, $u_{\mathrm{inv}}$. This deviation is quantified by the ratio $\mathcal{C} = \bar{u}_\xi^2 / u_{\mathrm{inv}}^2$. The boundary layer thickness, $\delta$, is subsequently defined as the wall-normal location where the viscous profile recovers $99\%$ of the local inviscid value, corresponding to the threshold $\mathcal{C} = 0.99^2$. This diagnostic criterion is adopted because it provides highly consistent estimates of $\delta$ across all investigated configurations without relying on additional empirical parameters.

To characterize the boundary layer state along the flat plate and curved configurations, the streamwise development of the primary integral quantities is examined. The compressible displacement thickness $\delta^*$ and momentum thickness $\theta$ are defined as

\begin{equation}
    \delta^*(\xi) = \int_{0}^{\delta} \left( 1 - \frac{\bar{\rho}\bar{u}_\xi}{\rho_e u_{\xi,e}} \right) \mathrm{d}\eta,
    \label{eq:delta_star}
\end{equation}

\begin{equation}
    \theta(\xi) = \int_{0}^{\delta} \frac{\bar{\rho}\bar{u}_\xi}{\rho_e u_{\xi,e}} \left( 1 - \frac{\bar{u}_\xi}{u_{\xi,e}} \right) \mathrm{d}\eta,
    \label{eq:theta}
\end{equation}
where 
$\rho_e$ and $u_{\xi,e}$ represent the density and tangential velocity values evaluated at the boundary layer edge $\eta = \delta$. The kinematic shape factor is subsequently defined as $ H(\xi) = \delta^*(\xi)/\theta(\xi)$.

\begin{figure}[htbp]
\centering
\begin{overpic}[trim = 0cm 0cm 0cm 0cm, clip, width=0.99\textwidth]{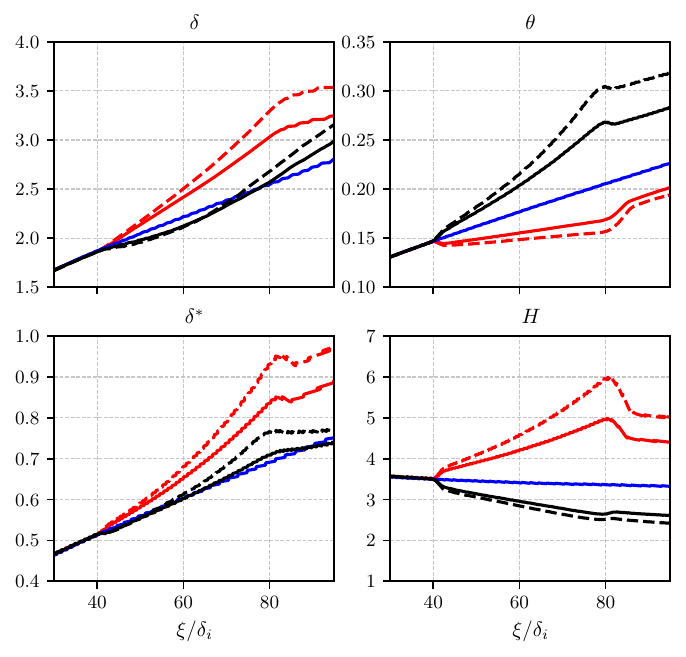}
\end{overpic}
\begin{overpic}[trim = 0cm 0cm 0cm 0cm, clip, width=0.99\textwidth]{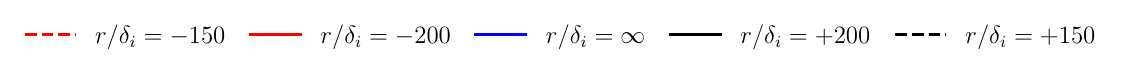}
\end{overpic}
\caption{Streamwise evolution of boundary layer properties: boundary layer thickness $\delta$ (top left), momentum thickness $\theta$ (top right), displacement thickness $\delta^*$ (bottom left), and shape factor $H$ (bottom right). Blue solid line: flat plate; red lines: convex/FPG cases; black lines: concave/APG cases. Solid lines denote $r/\delta_i = \pm200$, while dashed lines denote $r/\delta_i = \pm 150$.}

\label{fig:bl_properties}
\end{figure}

The streamwise evolution  of the integral boundary layer parameters is compared in figure~\ref{fig:bl_properties}. Prior to the curvature onset, all trajectories collapse onto the baseline flat-plate profile, verifying the inflow consistency and statistical convergence over all numerical simulations. Downstream of the curvature initiation, distinct dynamical trends emerge depending on the geometric configuration. 

The evolution of the boundary layer thickness, $\delta$, exhibits a pronounced sensitivity to the applied curvature. Over convex surfaces, the combined effects of the favorable pressure gradient and bulk dilatation drive a substantial thickening of the boundary layer \citep{spina_1994, lui_2024, HAMADA2026110029}. This growth scales directly with the curvature magnitude, resulting in a maximum thickness for the strongest curvature case ($r/\delta_i = -150$, red dashed lines). Conversely, for the concave surfaces (black lines), the adverse pressure gradient and bulk compression initially compress the boundary layer, rendering it slightly thinner than the flat-plate baseline (blue line). As the flow progresses downstream, however, the boundary layer gradually thickens, with $\delta$ eventually surpassing the flat-plate thickness near $\xi \approx 70\delta_i$. This structural evolution is in agreement with the physical mechanisms described by \citet{spina_1994}, in which bulk dilatation and bulk compression inherently govern the respective thickening and thinning of compressible boundary layers.

\begin{figure}[htbp]
\centering
\begin{overpic}[trim = 0cm 0cm 0cm 0cm, clip,width=0.99\textwidth]{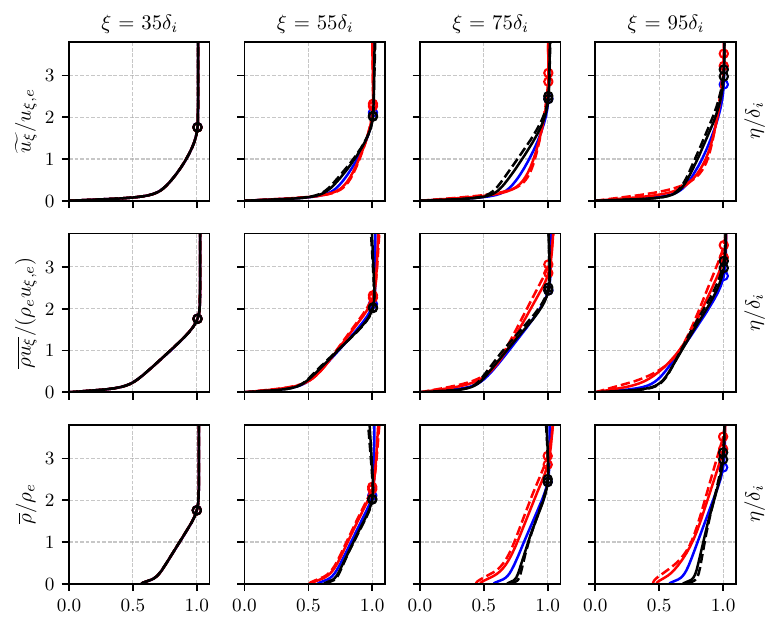}
\end{overpic}
\begin{overpic}[trim = 0cm 0cm 0cm 0cm, clip,width=0.99\textwidth]{figure_boundary_layer_legend.pdf}
\end{overpic}
\caption{Wall-normal profiles of the Favre-averaged streamwise velocity (top), mean streamwise momentum (center), and mean density (bottom) at selected streamwise stations. 
All quantities are normalized by their respective local boundary-layer-edge properties to illustrate the spatial evolution of the momentum and velocity deficits.}
\label{fig:velocity_profiles}
\end{figure}

The integral boundary-layer properties in figure~\ref{fig:bl_properties} reveal opposite responses to concave and convex surface curvature. Compared to the flat plate, the momentum thickness, $\theta$, increases over the concave surfaces subjected to APGs, whereas it decreases over the convex surfaces subjected to FPGs. The displacement thickness, $\delta^*$, exhibits a different trend, increasing for all curved configurations relative to the flat-plate baseline, with the largest increase occurring over the convex surfaces. These contrasting responses lead to markedly different evolutions of the shape factor, $H=\delta^*/\theta$: $H$ increases substantially in the FPG cases but decreases in the APG cases. To elucidate the physical mechanisms underlying these trends, we examine the wall-normal profiles of the mean flow in figure~\ref{fig:velocity_profiles}, considering the Favre-averaged streamwise velocity, the mean streamwise momentum, and the mean density, arranged from top to bottom, at selected streamwise stations ($x/\delta_{i} = 35, 55, 75,$ and $95$). The profiles follow the same color and line-style convention of figure~\ref{fig:bl_properties}, with all quantities normalized by their respective local boundary-layer-edge values. The boundary-layer edge is identified by a color marker in each profile. This normalization facilitates comparisons of the velocity and mass-flux deficits for the different curvature configurations, thus providing a physical interpretation of the observed variations in the integral quantities.

Over the convex surfaces (red lines), the FPG accelerates the flow, producing a fuller streamwise velocity profile than in the flat-plate (blue line) and concave cases (black lines). Nevertheless, closer inspection of the profiles at $x/\delta_{i} = 75$ and $95$ shows that the near-wall velocity gradient decreases compared to the flat-plate and concave surfaces, indicating a reduction in the wall shear stress. 
The associated thermodynamic response is characterized by a pronounced reduction in density, which persists when the profiles are normalized by their respective boundary-layer-edge values. This density variation is central to understanding the increase in displacement thickness. Indeed, in the present supersonic boundary layer, $\delta^*$ is governed by the integrated mass-flux deficit, rather than by the velocity deficit alone. The reduction in the normalized density therefore counteracts the increase in velocity and produces a larger mass-flux deficit, resulting in a substantial increase in $\delta^*$ despite the flow acceleration. This behavior is consistent with the compressibility effects discussed by different authors \citep{spina_1994,Sun_2017,HAMADA2026110029}. In contrast, the momentum thickness decreases because the reduction in the velocity deficit outweighs the contribution of the changing density-weighted mass flux. The simultaneous increase in $\delta^*$ and decrease in $\theta$ accounts for the pronounced rise in $H$ over the convex surfaces \citep{Sun_2017}.

The response over the concave surfaces is considerably different. The APG decelerates the flow and increases the velocity deficit relative to the local boundary-layer-edge velocity. At the same time, the associated compression increases the density relative to the corresponding FPG cases. These competing effects leave the normalized mean streamwise momentum profiles relatively close to the flat-plate baseline, consistent with the comparatively small changes in $\delta^*$. However, the momentum thickness increases substantially, reflecting the larger velocity deficit in the decelerated flow. Because this increase in $\theta$ is not accompanied by a comparable increase in $\delta^*$, the shape factor decreases over the concave surfaces. The opposite trends in $H$ over the convex and concave surfaces thus arise from the distinct responses of the velocity and density fields to the imposed pressure gradients. These results highlight the importance of compressibility in determining the integral properties of supersonic turbulent boundary layers over curved surfaces and demonstrate that the resulting shape-factor trends differ from those commonly associated with incompressible boundary layers subjected to favorable and adverse pressure gradients \citep{spina_1994, HAMADA2026110029}.

\begin{figure}[htbp]
\centering
\begin{overpic}[trim = 0cm 0cm 0cm 0cm, clip,width=0.99\textwidth]{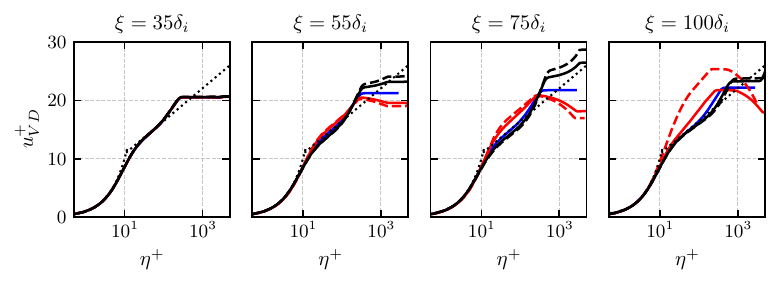}
\end{overpic}
\begin{overpic}[trim = 0cm 0cm 0cm 0cm, clip,width=0.99\textwidth]{figure_boundary_layer_legend.pdf}
\end{overpic}
\caption{Van Driest transformed velocity profiles at selected streamwise stations.}
\label{fig:van-driest-velocity}
\end{figure}

To further investigate how the different surface curvatures and their induced pressure gradients disrupt the internal equilibrium of the boundary layers, the Van Driest transformed \citep{vandriest_1951} velocity profiles at representative streamwise stations are presented in figure~\ref{fig:van-driest-velocity}. While the outer-scaled distributions in figure~\ref{fig:velocity_profiles} demonstrated the macroscopic impact of dilatation and compression, inner scaling elucidates how these non-equilibrium effects penetrate across the viscous sublayer, logarithmic, and wake regions. Across all stations, the viscous sublayer remains resilient and well-collapsed onto the linear law ($u^+ = y^+$) \citep{Wang_Wang_Sun_Yang_Zhao_Hu_2019, leandro, pirozzoli_2026}. However, further away from the wall, distinct departures from canonical equilibrium emerge. For the APG cases over concave curvatures, the wake component is amplified, while sustained downstream deceleration introduces a downward deflection of the logarithmic layer, qualitatively consistent with previous observations \citep{Spalart_Watmuff_1993, Wu_Liang_Zhao, leandro}. On the other hand, the FPG cases over convex surfaces yield opposing behavior, characterized by an upward departure from the flat-plate baseline profile that distorts the logarithmic region \citep{Spalart_Watmuff_1993, Wang_2020}, followed by a weakened wake component. Moreover, increasing convex curvature leads to progressively larger departures from the canonical inner-scaled velocity profiles, potentially indicating incipient flow relaminarization along the downstream flat recovery section.  

\begin{figure}[htbp]
\centering
\begin{overpic}[trim = 0cm 0cm 0cm 0cm, clip,width=0.99\textwidth]{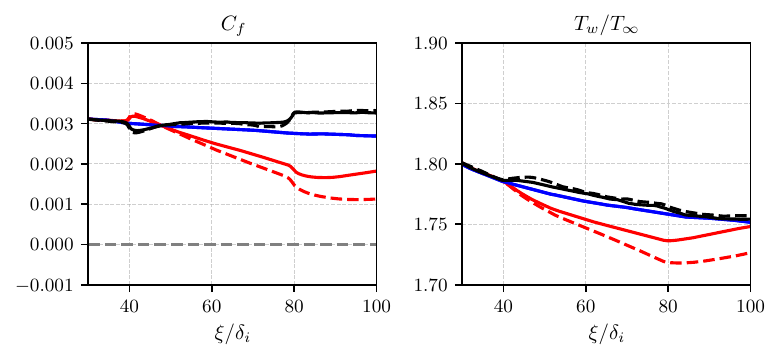}
\end{overpic}
\begin{overpic}[trim = 0cm 0cm 0cm 0cm, clip,width=0.99\textwidth]{figure_boundary_layer_legend.pdf}
\end{overpic}
\caption{Streamwise evolution of skin friction coefficient and wall temperature.}
\label{fig:cf_cp}
\end{figure}

The changes in the shape factor and mean velocity profiles induced by surface curvature, together with the associated pressure gradients and compressibility effects, are reflected in the streamwise evolution of the skin-friction coefficient ($C_f$) and wall temperature ($T_w$), as shown in figure~\ref{fig:cf_cp}. 
In the FPG cases over convex surfaces, flow expansion is accompanied by a reduction in wall temperature and, consequently, in the near-wall dynamic viscosity. This reduction in viscosity, together with the lower near-wall velocity gradient observed in figure~\ref{fig:velocity_profiles}, leads to a decrease in wall shear stress and a pronounced reduction in $C_f$ along the convex surfaces \citep{spina_1994}. Conversely, in the APG cases over concave surfaces, the wall temperature remains close to the flat-plate baseline. The variations in $C_f$ are therefore primarily associated with changes in the near-wall velocity gradients induced by the adverse pressure gradient. Downstream of the curved section, within the flat recovery region, $C_f$ exhibits a discernible increase in the concave cases. This recovery coincides with the progressive relaxation of the Van Driest-transformed velocity profiles towards the canonical logarithmic law, as shown in figure~\ref{fig:van-driest-velocity}.


\subsection{Turbulence anisotropy and Reynolds stresses}
\label{sec:turbulence_statistics}

To characterize the anisotropy state of the present supersonic turbulent boundary layers, componentality contours are computed using the invariants of the Reynolds stress anisotropy tensor \citep{Emory2014VisualizingTA}. The Reynolds stress tensor is constructed from Favre-averaged fluctuations, $\widetilde{u_i'' u_j''} = \overline{\rho u_i'' u_j''} / \bar{\rho}$, which are non-dimensionalized by the square of the inflow freestream velocity, $U_\infty^2$. The turbulent kinetic energy is defined as half of its trace, $k = \frac{1}{2} \widetilde{u_i'' u_i''}$. The non-dimensional Reynolds stress anisotropy tensor is then given by $a_{ij} = \widetilde{u_i'' u_j''}/(2k) - \delta_{ij}/3$, where $\delta_{ij}$ is the Kronecker delta.

At each spatial location, the ordered eigenvalues of $a_{ij}$ ($\lambda_1 \ge \lambda_2 \ge \lambda_3$) define the barycentric coordinates representing the three limiting states of turbulence: the one-component state ($C_1 = \lambda_1 - \lambda_2$), the two-component state ($C_2 = 2(\lambda_2 - \lambda_3)$), and the three-component isotropic state ($C_3 = 3\lambda_3 + 1$). To enhance visual contrast across distinct structural regimes, the empirical contrast transformation proposed by \citet{Emory2014VisualizingTA} is applied, $C_{ic}^* = (C_i + 0.65)^{5.00}$. The results are mapped onto an RGB color space
\begin{equation}
\begin{bmatrix}
R \\
G \\
B
\end{bmatrix}
=
C_{1c}^*
\begin{bmatrix}
1 \\
0 \\
0
\end{bmatrix}
+
C_{2c}^*
\begin{bmatrix}
0 \\
1 \\
0
\end{bmatrix}
+
C_{3c}^*
\begin{bmatrix}
0 \\
0 \\
1
\end{bmatrix},
\end{equation}
where red, green, and blue correspond to the one-component, two-component, and isotropic states, respectively. 

\begin{figure}[htbp]
\centering
\includegraphics[width=0.99\textwidth]{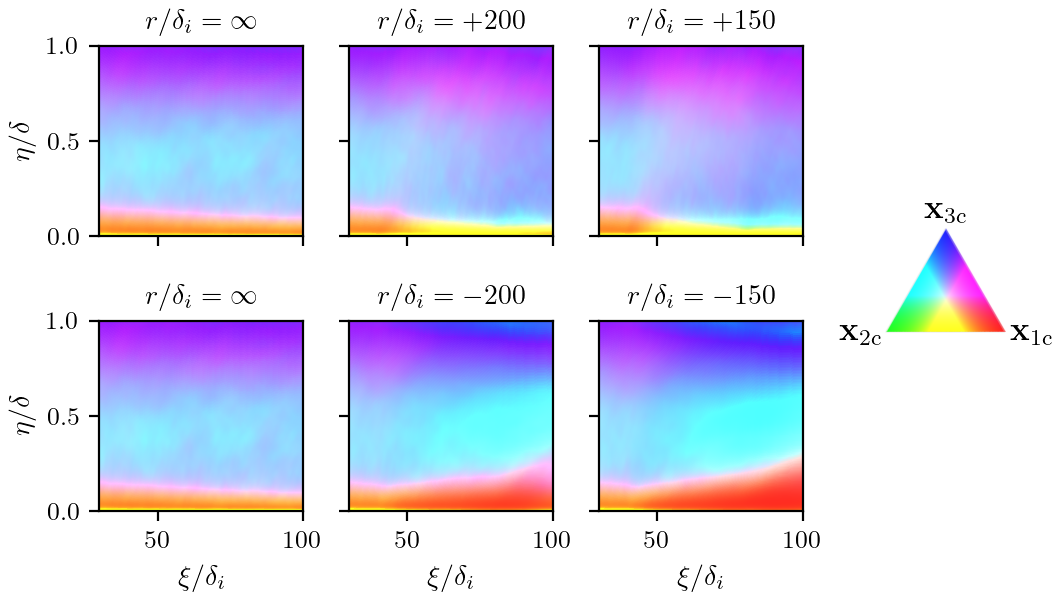}
\caption{Componentality contours based on the barycentric map of the Reynolds stress anisotropy tensor for $x \in [35, 100]$ against the normalized wall-normal coordinate $\eta/\delta$. Top: flat plate, concave $r/\delta_i = +200$, and concave $r/\delta_i = +150$. Bottom: flat plate, convex $r/\delta_i = -200$, and convex $r/\delta_i = -150$. The color legend on the right indicates the limiting turbulence states: red (one-component), green (two-component), and blue (three-component, isotropic).}
\label{fig:anisotropy-state}
\end{figure}

\begin{figure}[htbp]
\centering
\includegraphics[width=0.99\textwidth]{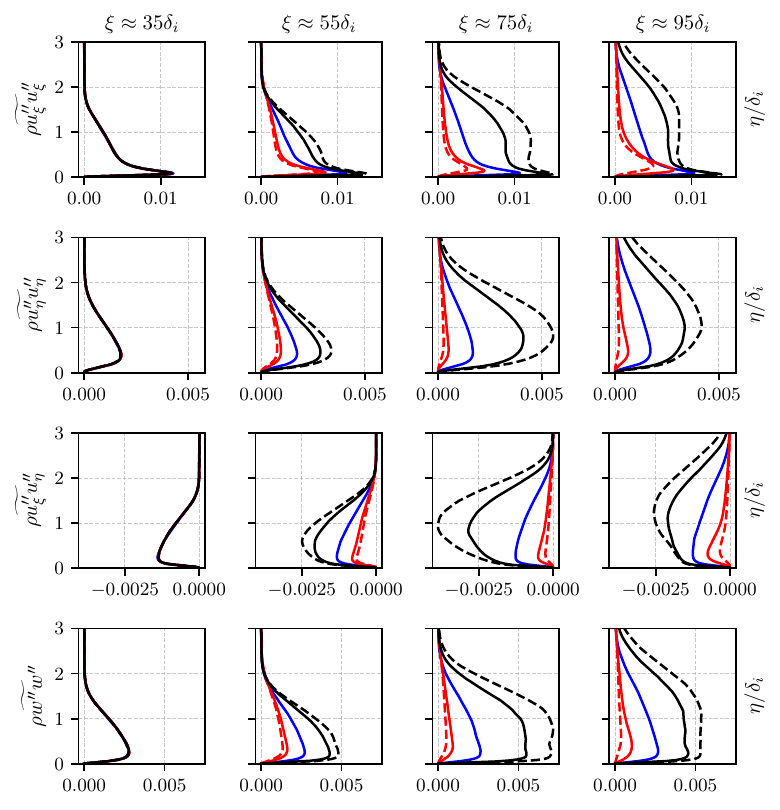}
\includegraphics[width=0.99\textwidth]{figure_boundary_layer_legend.pdf}
\caption{Streamwise evolution of density-scaled  Favre-averaged Reynolds stress profiles, $\bar{\rho}\widetilde{u_i'' u_j''}/\rho_{\infty}U^2_{\infty}$ 
at streamwise stations $\xi/\delta_i = 35, 55, 75, 95$. 
}
\label{fig:reynolds_stress_density}
\end{figure}

\begin{figure}[htbp]
\centering
\includegraphics[width=0.99\textwidth]{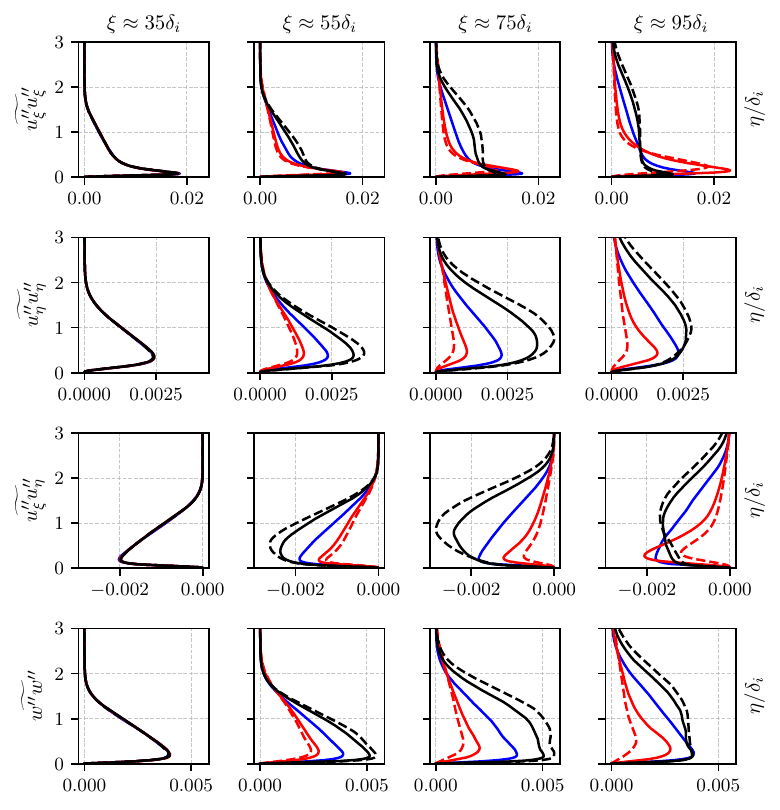}
\includegraphics[width=0.99\textwidth]{figure_boundary_layer_legend.pdf}
\caption{Streamwise evolution of kinematic Favre-averaged Reynolds stress profiles, $\widetilde{u_i'' u_j''}/U^2_{\infty}$ 
at streamwise stations $\xi/\delta_i = 35, 55, 75, 95$.}
\label{fig:reynolds_stress_favre}
\end{figure}

Figure~\ref{fig:anisotropy-state} presents the spatial distribution of these componentality contours across the curved domain ($\xi/\delta_i \in [35, 100]$), plotted against the normalized wall-normal coordinate $\eta/\delta$. To support the physical interpretation of these structural maps, figures~\ref{fig:reynolds_stress_density} and~\ref{fig:reynolds_stress_favre} show the corresponding profiles of the tangential ($\widetilde{u_{\xi}'' u_{\xi}''}$), wall-normal ($\widetilde{u_{\eta}'' u_{\eta}''}$), spanwise ($\widetilde{u_z'' u_z''}$), and shear ($\widetilde{u_{\xi}'' u_{\eta}''}$) Reynolds stress components at four streamwise stations ($\xi/\delta_i = 35, 55, 75$, and $95$), comparing the density-scaled stresses ($\bar{\rho} \widetilde{u_i'' u_j''}$) against the kinematic Favre stresses ($\widetilde{u_i'' u_j''}$).

Upstream of the curved surfaces ($\xi/\delta_i \le 35$), all cases exhibit an identical canonical structure: the near-wall region is dominated by the one-component state (red), reflecting the presence of the elongated streamwise velocity streaks identified in \S~\ref{sec:instantaneous_flow}. Away from the wall, the turbulence progressively transitions toward a more isotropic state in the outer region of the boundary layers.

Over the concave surfaces (top row of figure~\ref{fig:anisotropy-state}), the near-wall turbulence reorganizes rapidly upon entering the curved section, shifting from a one-component state toward an intermediate one-two-component state (yellow) near the wall, and then a more isotropic state in the outer layer. The Reynolds stress profiles demonstrate that this reorganization is driven by the amplification of the spanwise velocity fluctuations ($\widetilde{u_z'' u_z''}$) near the wall, which is associated with the emergence of wider coherent structures, as shown in figure~\ref{fig:fluc}. Concurrently, both the tangential ($\widetilde{u_{\xi}'' u_{\xi}''}$) and wall-normal ($\widetilde{u_{\eta}'' u_{\eta}''}$) components are amplified under the adverse pressure gradient, which in turn leads to an increase in the turbulent shear stress ($\widetilde{u_{\xi}'' u_{\eta}''}$) across the curved region.

Over the convex surfaces (bottom row of figure~\ref{fig:anisotropy-state}), the one-component region broadens significantly, occupying a larger fraction of the boundary layer thickness. This structural change is explained by the Reynolds stress profiles in figures~\ref{fig:reynolds_stress_density} and~\ref{fig:reynolds_stress_favre}. The convex curvatures induce a damping of both the wall-normal ($\widetilde{u_{\eta}'' u_{\eta}''}$) and spanwise ($\widetilde{u_z'' u_z''}$) fluctuations. Because these cross-stream components are strongly suppressed, the turbulent kinetic energy becomes heavily concentrated in the streamwise component, driving the state toward a one-component limiting behavior. 

Comparing the density-scaled stresses (figure~\ref{fig:reynolds_stress_density}) with the kinematic stresses (figure~\ref{fig:reynolds_stress_favre}) clarifies the underlying scaling. In the density-scaled profiles, all stress components appear to be reduced over the convex wall. However, this apparent decrease in the streamwise stress is primarily a thermodynamic consequence of flow expansion and the accompanying decrease in local density, as shown in figure~\ref{fig:velocity_profiles}. In the kinematic Favre-averaged profiles (figure~\ref{fig:reynolds_stress_favre}), the outer-layer streamwise fluctuations are indeed damped by flow expansion compared to the baseline and concave cases \citep{Smith1991}, while the near-wall peak of $\widetilde{u_{\xi}'' u_{\xi}''}$ remains of the same order of magnitude as in the flat-plate and concave configurations. Further downstream in the recovery region ($\xi/\delta_i = 95$), the streamwise stress undergoes strong amplification, with the milder convex curvature ($r/\delta_i = -200$ solid red line) exhibiting the highest near-wall $\widetilde{u_{\xi}'' u_{\xi}''}$ peak.

\subsection{Production, transport, and inter-scale transfer of turbulent kinetic energy}
\label{sec:spectral_transport}

To elucidate the physical mechanisms governing the generation and redistribution of turbulent kinetic energy across scales, we examine the scale-resolved TKE budgets for surfaces with different curvatures. Since the flow is statistically homogeneous in the spanwise direction ($z$), the governing equations are Fourier-transformed in this direction, yielding the spectral representation of the TKE budget. Throughout this section, the spectral TKE budget terms are non-dimensionalized using the inflow freestream velocity $U_\infty$ as the reference velocity scale.

\begin{figure}[htbp]
\centering
\begin{overpic}[trim = 0cm 0cm 0cm 0cm, clip,width=0.99\textwidth]{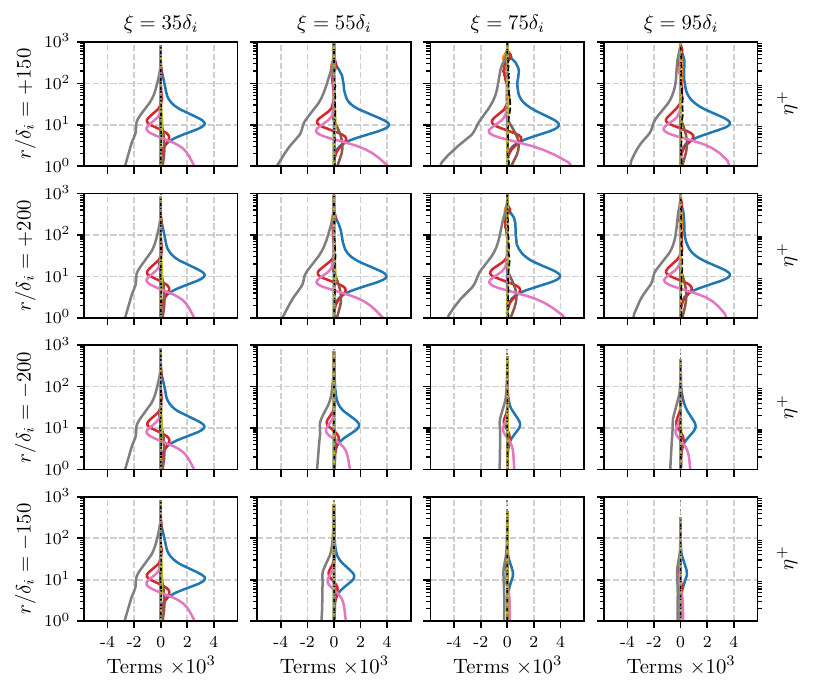}
\end{overpic}
\begin{overpic}[trim = 0cm 0cm 0cm 0cm, clip,width=0.99\textwidth]{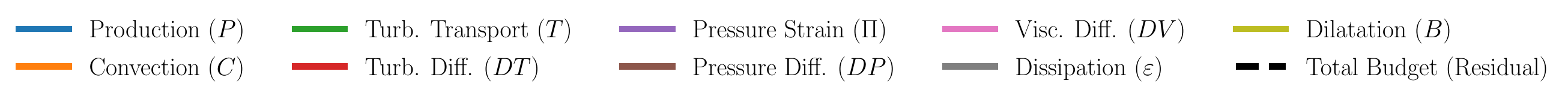}
\end{overpic}
\caption{Turbulent kinetic energy (TKE) budgets integrated over all spanwise wavenumbers. The budget terms are non-dimensionalized using the inflow freestream velocity $U_\infty$. Rows (top to bottom) present cases with varying curvature radii, from $r/\delta_i = +150$ to $r/\delta_i = -150$. Columns (left to right) display the streamwise development at stations $\xi/ \delta_i = 35, 55, 75$, and $95$, where station $\xi/ \delta_i = 95$ corresponds to the straight recovery region downstream of the curved section.}
\label{fig:tke-budgets}
\end{figure}

Before examining the scale-by-scale spectral distributions, we evaluate the total TKE budget obtained by integrating the spectral transport terms over all spanwise wavenumbers. Figure~\ref{fig:tke-budgets} displays the streamwise evolution of these total TKE budgets from $\xi = 35\delta_i$ to $95 \delta_i$, comparing the curved geometries. For all stations, the sum of the budget terms yields a negligible residual, validating both the statistical convergence of the dataset and the numerical accuracy of the DNS methodology. At the upstream station ($\xi= 35\delta_i $), prior to the onset of curvature, all cases exhibit the canonical flat-plate distribution, showing good agreement with existing literature, where turbulent production peaks sharply in the buffer layer around $\eta^+ \approx 10$, where it is primarily balanced by viscous dissipation and near-wall diffusion.

As the boundary layer develops along the curved section ($\xi/\delta_i = 55\text{--}75$), the TKE budgets diverge substantially depending on the sign of the surface curvature. Over the concave geometries, the adverse pressure gradient and destabilizing effects of curvature cause an amplification of turbulent production, accompanied by proportional increases in dissipation and viscous diffusion. Within the near-wall region, the pressure diffusion term assumes an increasingly prominent role, exhibiting a gradual rise in magnitude along the curved surface. In addition to the primary near-wall peak, a secondary production peak emerges in the outer region of the boundary layer, as typically observed for adverse pressure gradient turbulent boundary layers \citep{Skåre_Krogstad_1994, Wu_Liang_Zhao, Wang_Wang_Sun_Yang_Zhao_Hu_2019,Gungor_Maciel_Gungor_2022, leandro}. This augmented outer-layer production is balanced by corresponding increases in dissipation, mean convection and turbulent diffusion. 

Conversely, the convex surfaces exhibit an apparent reduction for all budget terms, along with a modest outward shift of the near-wall production peak toward $10 < \eta^+ < 20$. When interpreting these trends, it is essential to consider the thermodynamic consequences of bulk dilatation. Because the budget equation governs the density-weighted turbulent kinetic energy ($\rho k$ with $k = 0.5u''_iu''_i$), the terms inherently scale with the local mean density $\bar{\rho}$. As identified in the Reynolds stress analysis of \S~\ref{sec:turbulence_statistics}, the flow expansion over the convex wall causes a marked decrease in $\bar{\rho}$. This thermodynamic effect decreases the overall magnitude of the budget terms, masking the fact that the kinematic velocity fluctuations ($\widetilde{u_{\xi}'' u_{\xi}''}$) sustain energy near the surface of the convex case, as shown in figure~\ref{fig:reynolds_stress_favre}. Similarly, bulk compression over the concave wall raises the local density, contributing 
to the overall elevation of the budget terms.

\begin{figure}[htbp]
\centering
\begin{overpic}[trim = 0cm 0cm 0cm 0cm, clip,width=0.99\textwidth]{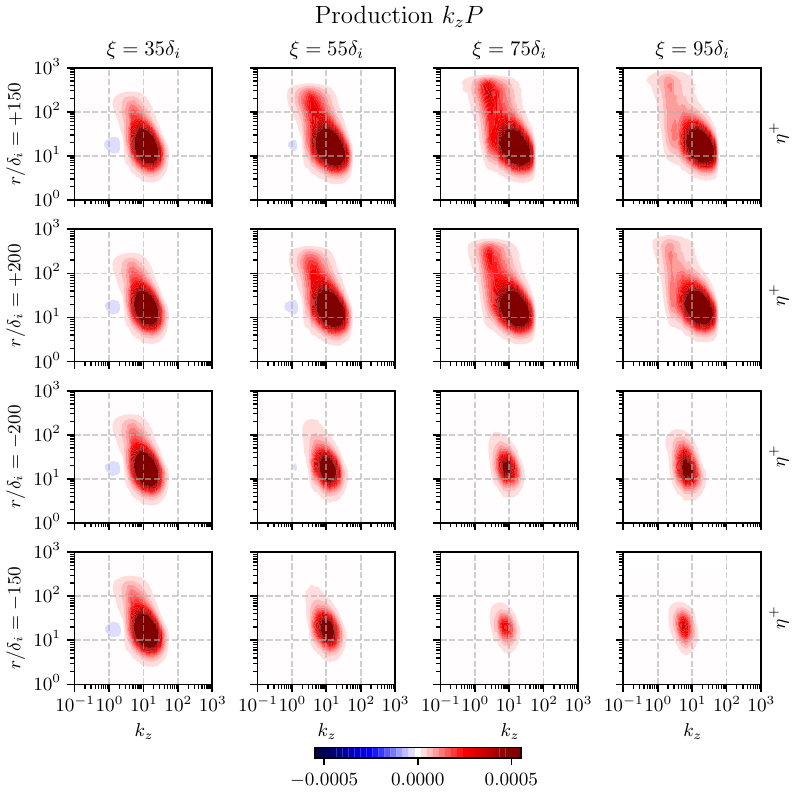}
\end{overpic}
\caption{Premultiplied spectral production of TKE, $k_z P$, as a function of spanwise wavenumber $k_z$ and wall-normal coordinate $\eta^+$. Rows (top to bottom): from $r/\delta_i = +150$ to $r/\delta_i = -150$. Columns (left to right): streamwise stations $\xi/ \delta_i = 35, 55, 75$, and $95$, where $\xi/ \delta_i = 95$ lies in the downstream recovery region.}
\label{fig:tke-prod}
\end{figure}

To examine how TKE generation is partitioned across scales in the spanwise direction, figure~\ref{fig:tke-prod} displays contours of premultiplied spectral production, $k_z P(k_z)$, as a function of the spanwise wavenumber $k_z$ and the inner-scaled wall-normal coordinate $\eta^+$. The production term represents the scale-specific rate at which turbulent fluctuations extract kinetic energy from the mean flow. At the upstream station ($\xi = 35\delta_i$), all cases exhibit a single production peak centered in the buffer layer at $\eta^+ \approx 10\text{-}12$ across intermediate wavenumbers, characteristic of the canonical near wall behavior. Over the concave surfaces, however, the emergence of the secondary production peak in the outer region of the boundary layer is distinctly concentrated at lower wavenumbers (longer spanwise wavelengths) \citep{Gungor_Maciel_Gungor_2022}. This demonstrates that the outer energization and amplification of Reynolds stresses identified earlier are driven by large-scale coherent motions, consistent with the widened, spanwise-coherent structures observed in the instantaneous flow visualizations of figure~\ref{fig:fluc} in \S~\ref{sec:instantaneous_flow}.

This scale-specific energization is found to be predominantly driven by the mean shear component, defined as $P_{\xi \eta} = -\hat{Q}_{\xi \eta} \frac{\partial \widetilde{u}_\xi}{\partial \eta}$, where $\hat{Q}_{\xi \eta}$ is the spectral shear Reynolds stress and $\partial \widetilde{u}_\xi / \partial \eta$ is the mean wall-normal velocity gradient (see appendix \ref{app:tensor-rotation}).
Figure~\ref{fig:prod-dec} shows the streamwise evolution of the total spectral production $P$ across the boundary layer and  highlights the non-equilibrium recovery dynamics downstream of the curvature ($\xi/ \delta_i \ge 80$). In this straight recovery section, the turbulence does not immediately relax to canonical flat-plate values, displaying pronounced upstream history effects. For the convex configurations, production remains severely suppressed throughout the recovery domain. Conversely, for the concave cases, although the outer secondary production peak progressively decays upon exiting the curved wall, the near-wall buffer-layer peak remains elevated above the upstream baseline, sustaining a state of enhanced turbulence production well into the recovery region.

\begin{figure}[htbp]
\centering
\begin{overpic}[trim = 0cm 0cm 0cm 0cm, clip,width=0.99\textwidth]{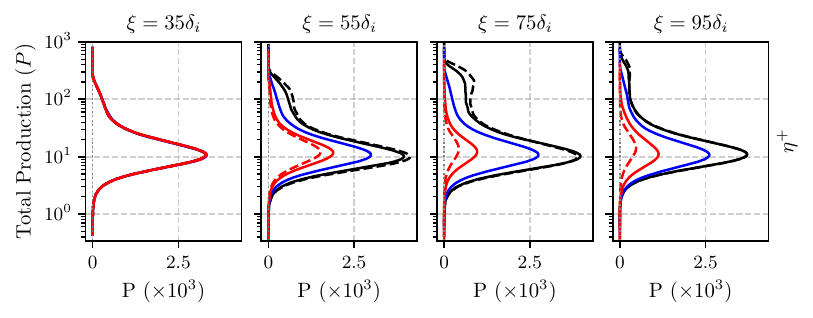}
\end{overpic}
\begin{overpic}[trim = 0cm 0cm 0cm 0cm, clip,width=0.99\textwidth]{figure_boundary_layer_legend.pdf}
\end{overpic}
\caption{Streamwise evolution of integrated TKE production profiles $P$ across stations $\xi/ \delta_i = 35, 55, 75$, and $95$.}
\label{fig:prod-dec}
\end{figure}

The spatial modulation of energy production and its persistent non-equilibrium memory naturally raise the question of how turbulent kinetic energy is redistributed among scales by non-linear triadic interactions. In the spectral budget of turbulent kinetic energy, this inter-scale transport is governed by the turbulent transport term $T$. By definition, $T$ is purely conservative, accounting for the spectral redistribution of energy across spanwise wavelengths ($\lambda_z$) without net generation or dissipation, such that its integral across all scales vanishes identically ($\sum_{k_z}T(k_z)=0$, as shown in figure~\ref{fig:tke-budgets}).

\begin{figure}[htbp]
\centering
\begin{overpic}[trim = 0cm 0cm 0cm 0cm, clip, width=0.99\textwidth]{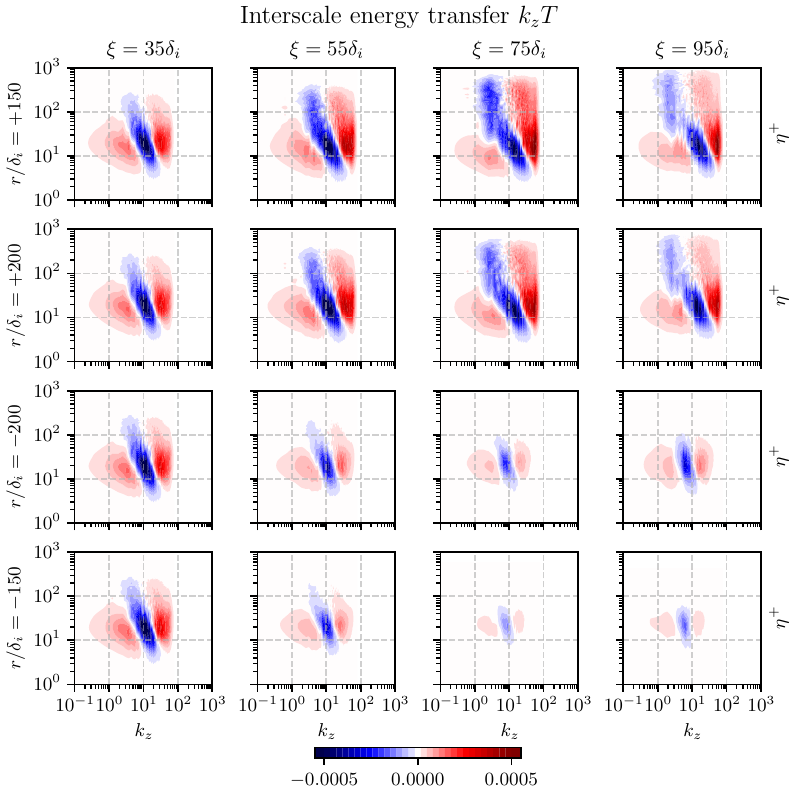}
\end{overpic}
\caption{Premultiplied spectral inter-scale energy transfer rate, $k_z T$, as a function of spanwise wavenumber $k_z$ and wall-normal coordinate $\eta^+$. Rows (top to bottom) correspond to cases with varying curvature radii: from $r/\delta_i = +150$ to $r/\delta_i = -150$. Columns (left to right) display the streamwise development at stations $\xi/ \delta_i = 35, 55, 75$, and $95$, where $\xi/ \delta_i = 95$ lies in the downstream recovery region.}
\label{fig:tke-T}
\end{figure}

The spatial distribution of $T$ is shown in figure~\ref{fig:tke-T} across the investigated curvatures. At the upstream flat-plate station, prior to the curvatures ($\xi/ \delta_i = 35$), an excellent collapse is observed for all configurations, confirming numerical consistency and recovering canonical boundary-layer features \citep{Cho_Hwang_Choi_2018, Lee_Moser_2019,Yu_Zhao_Tang_Yuan_Xu_2022}. The spectrum exhibits a dual-direction transfer, wherein energy is extracted from intermediate scales and redistributed simultaneously toward both smaller (forward transfer) and larger spanwise scales (backward transfer). This backward energy transfer is widely observed in the 1D spanwise spectra of TKE for various flow configurations \citep{Cho_Hwang_Choi_2018,Lee_Moser_2019, schiavo, Yu_Zhao_Tang_Yuan_Xu_2022}. Notably, \cite{Yu_Zhao_Tang_Yuan_Xu_2022} observed the scale-transfer term to collapse into a strictly unidirectional large-to-small cascade upon approaching a shock wave-boundary layer interaction (SBLI). In the present study, the qualitative bidirectional mechanism is preserved across all curvature cases, despite the concave configurations imposing an APG. This contrast highlights an important physical distinction, suggesting that the suppression of inverse energy transfer observed in SBLI is fundamentally attributable to the presence of the shock wave itself and its associated abrupt jump conditions, rather than the accompanying APG. 

Nevertheless, surface curvature significantly modifies the wall-normal distribution of this spectral energy transfer. Over the concave walls, the emergence of the secondary production peak extends non-linear activity deep into the boundary layer wake region. Within this outer layer, the large spanwise scales exhibit a prominent forward transfer of energy toward smaller scales. While some energy continues to be inversely transferred toward larger spanwise structures, this mechanism is more confined to the near-wall region. Over the convex walls, non-linear inter-scale transfer is heavily suppressed and remains strictly confined to the buffer layer, consistent with the localized persistence of turbulent production in this stabilizing regime.

\begin{figure}[htbp]
\centering
\begin{overpic}[trim = 0cm 0cm 0cm 0cm, clip, width=0.99\textwidth]{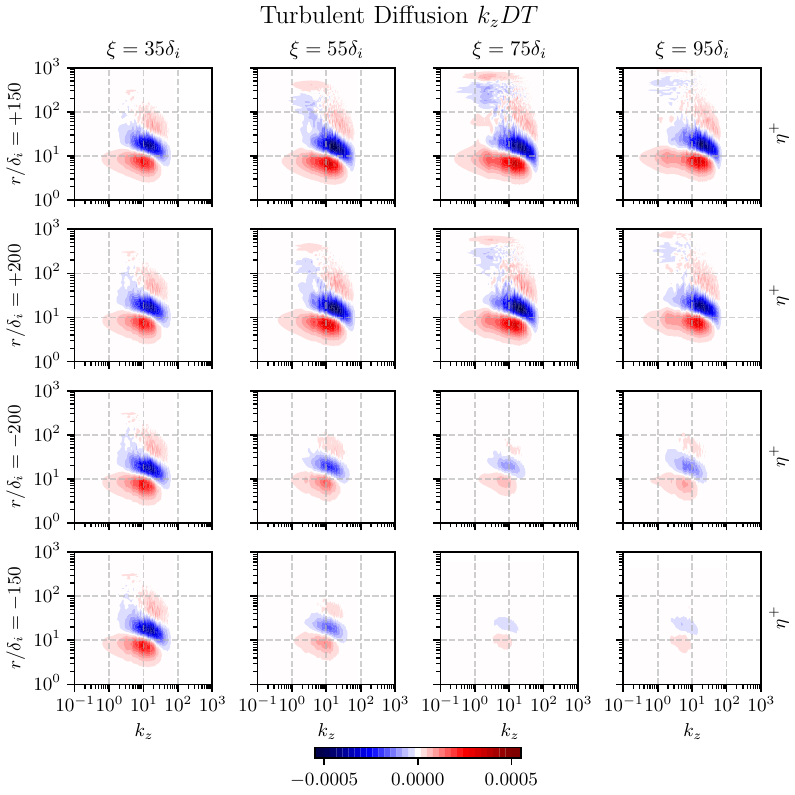}
\end{overpic}
\caption{Premultiplied spectral turbulent diffusion, $k_z DT$, as a function of spanwise wavenumber $k_z$ and wall-normal coordinate $\eta^+$. Rows (top to bottom) correspond to cases with varying curvature radii: from $r/\delta_i = +150$ to $r/\delta_i = -150$. Columns (left to right) display the streamwise development at stations $\xi/ \delta_i = 35, 55, 75$, and $95$, where $\xi/ \delta_i = 95$ lies in the downstream recovery region.}
\label{fig:tke-DT}
\end{figure}

To elucidate the spatial redistribution of turbulent kinetic energy along the boundary layers, figure~\ref{fig:tke-DT} presents contours of the spectral turbulent spatial transport (or turbulent diffusion), $DT$. This term dictates how turbulent energy, once produced, is transported by turbulent fluctuations along the boundary layer where, for each wavenumber, the spatial integral vanishes. In the upstream flat-plate baseline ($\xi/\delta_i = 35$), the canonical spatial transport mechanism is clearly observed with the turbulent diffusion extracting energy from the highly energetic buffer layer (the primary site of turbulent production) and redistributing it simultaneously inward toward the viscous sublayer and outward toward the logarithmic region.

Along the concave curvature, the spatial redistribution of TKE becomes more complex. Specifically, energy associated with large spanwise structures is increasingly extracted from the logarithmic layer and transported outward into the wake region. The intensity of this outward transport is amplified with the curvature radius, as the magnitude of the APG increases. For the concave surfaces, two regions of turbulent transport can be observed in the plots, one similar to the flat-plate baseline at higher spanwise wavenumbers, and another in the wake region, at lower spanwise wavenumbers.
Conversely, over the convex surfaces, the FPG systematically suppresses the outward transport of energy into the wake region relative to the upstream baseline. By station $\xi/\delta_{i} = 75$ for the strongest convex curvature ($r/\delta_{i} = -150$), the outward diffusion is negligible. Instead, turbulent energy is predominantly transported inward from the buffer layer toward the viscous sublayer.

\begin{figure}[htbp]
\centering
\begin{overpic}[trim = 0cm 0cm 0cm 0cm, clip,width=0.99\textwidth]{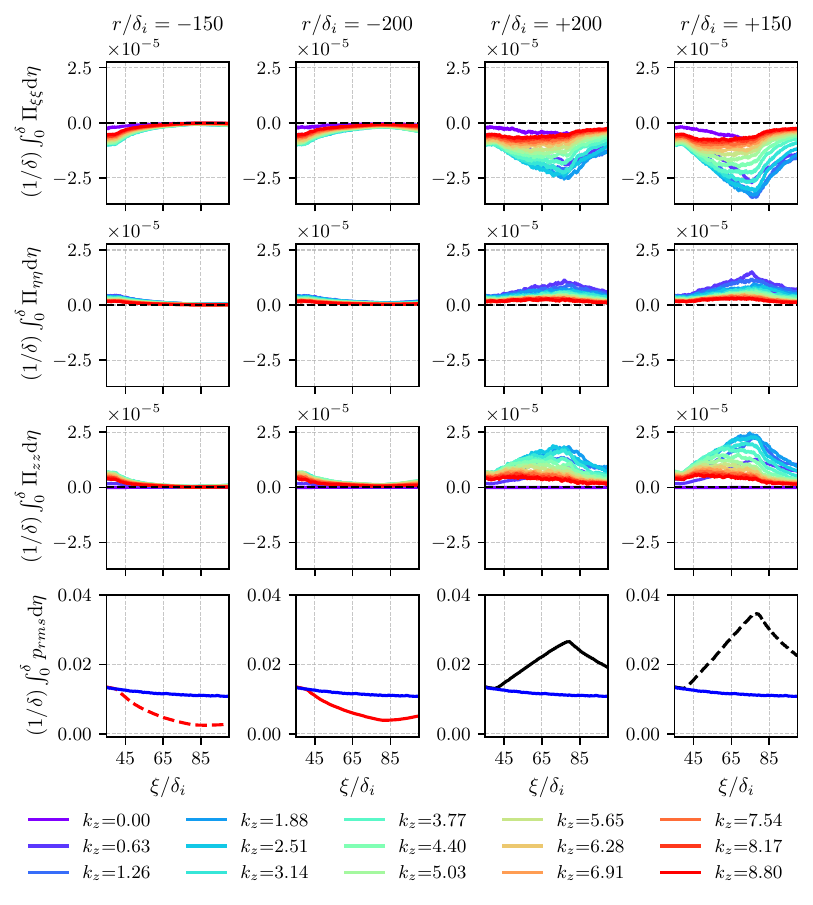}
\end{overpic}
\caption{Streamwise evolution of the boundary-layer-integrated pressure-strain redistribution terms and pressure fluctuations for selected low spanwise wavenumbers. Rows from top to bottom: streamwise component $(1/\delta)\int \Pi_{\xi \xi} \, \mathrm{d}\eta$, wall-normal component $(1/\delta)\int \Pi_{\eta \eta} \, \mathrm{d}\eta$, spanwise component $(1/\delta)\int \Pi_{zz} \, \mathrm{d}\eta$, and the spanwise-averaged root-mean-square pressure $(1/\delta)\int p_{\text{rms}} \, \mathrm{d}\eta$. Columns from left to right: convex configurations ($r/\delta_{i} = -150$, $-200$) and concave configurations ($r/\delta_{i} = +200$, $+150$). In the last row, the line shapes and colors follow the styles of previous figures for the various curvatures.}
\label{fig:tke-Pi}
\end{figure}

While the turbulent transport $(T)$ is responsible for energy redistribution across wavenumbers, pressure-strain $(\Pi)$ acts locally in spectral space to redistribute energy among velocity components. In wall-bounded flows, this mechanism governs the redistribution of energy from the highly energized streamwise velocity component into the cross-stream fluctuations, a process that is fundamentally responsible for the generation of streamwise vortices \citep{Cho_Hwang_Choi_2018, Lee_Moser_2019, Gungor_Maciel_Gungor_2022, Zhang_Han_Tang_Qu_Yan_2025}. Because the local Mach number inside the boundary layer is low, the fluctuating velocity field behaves in a quasi-incompressible manner ($\partial u_i''/\partial x_i \approx 0$), meaning that the sum of these terms across all velocity components (the pressure dilatation) and wavenumbers is small. Nevertheless, the individual diagonal components are substantial in magnitude and drive a considerable energy transfer between the velocity directions. To provide a compact representation that clearly captures the main streamwise trends, these individual terms are analyzed integrated over the boundary layer thickness. Figure~\ref{fig:tke-Pi} illustrates the streamwise evolution of these integrated terms across the selected low-wavenumber range ($k_z \le 8.80$) for both concave and convex geometries.

Under the stabilizing favorable pressure gradients over the convex surfaces, the inter-component energy redistribution is almost completely suppressed, coinciding with a pronounced attenuation of the root-mean-square pressure fluctuations, $p_{\mathrm{rms}}$ \citep{Dawson_1994}. This behavior can be rationalized through the Poisson equation governing pressure fluctuations in compressible boundary layers \citep{Smith1991}
\begin{equation}
    -\nabla^2 p' = 2\bar{\rho} \frac{\partial u_i'}{\partial x_j} \frac{\partial \tilde{u}_j}{\partial x_i} + \frac{1}{\widetilde{T}} \frac{\partial T'}{\partial x_i} \frac{\partial \bar{p}}{\partial x_i} \mbox{ .}
\label{eq:poisson_prms}
\end{equation}
In this formulation, the first source term captures the interaction between fluctuating velocity gradients and the mean shear, scaled directly by the mean density, $\bar{\rho}$, while the second term accounts for the coupling between turbulent temperature fluctuations and the mean pressure gradient. Over the convex wall, the magnitude of these source terms is substantially diminished through two coupled mechanisms. First, convex streamline curvature exerts a stabilizing centrifugal effect that dampens cross-stream velocity fluctuations, thereby diminishing the fluctuating velocity gradient tensor $\partial u_i'/\partial x_j$. Second, the rapid bulk dilatation decreases the local mean density, $\bar{\rho}$, directly scaling down the source strength of the primary velocity-gradient term. Furthermore, the overall suppression of turbulent mixing damps the temperature fluctuations, $T'$, thereby attenuating the intensity of the second term. The simultaneous reduction of turbulent fluctuations and the thermodynamic decrease in mean density severely weaken the source of $-\nabla^2 p'$, directly accounting for the collapse of $p_{\mathrm{rms}}$ along the convex surface.


The severe dampening of the pressure-strain term corroborates the prominent shift toward a one-component turbulence state (see figure \ref{fig:anisotropy-state}) in these regions. Crucially, this suppression also explains the anomalous accumulation of energy and the elevated magnitudes of the streamwise Reynolds stress ($\widetilde{u_{\xi}'' u_{\xi}''}$) observed downstream in the flat-plate recovery section ($x/\delta_{\text{in}} > 80$, as seen in figure~\ref{fig:reynolds_stress_favre}). Because the attenuated pressure-strain term is unable to effectively transfer the turbulent kinetic energy, which is predominantly produced in the streamwise component, into the cross-stream directions, the energy remains within the streamwise fluctuations \citep{Wang_2020}. Consequently, this disruption suppresses the transverse mechanisms and effectively breaks the autonomous near-wall turbulent cycle described by \citet{Lee_Moser_2019}.

Over the concave walls, where the turbulent boundary layers develop under APGs, the magnitude of the pressure-strain redistribution is significantly enhanced. This intensified inter-component transfer directly fuels the near-wall increase in the transverse Reynolds stresses, driving the shift toward the two-component turbulence state documented previously in figure \ref{fig:anisotropy-state}. As evidenced by the bottom panel of figure~\ref{fig:tke-Pi}, this enhanced redistribution is tightly coupled to a substantial amplification of the pressure fluctuations ($p_{\text{rms}}$) relative to the flat-plate baseline, which facilitates the energy transfer as can be seen from equation~\ref{eq:pressure_strain}.

To characterize the physical onset of the large-scale structures, we shift the focus towards the scale-by-scale generation and subsequent spectral redistribution of TKE. The evolution of the spectral state is primarily governed by two mechanisms: the spectral production term $P(k_z)$, which dictates the injection of energy by mean flow distortions, and the nonlinear energy transfer term $T(k_z)$, which accounts for the triadic interactions coupling different spanwise scales. Because the nonlinear transfer arises from the convective terms in the Navier-Stokes equations, it acts as a spectral convolution that redistributes energy across wavenumbers without altering the total global kinetic energy. To evaluate the net effect of these specific mechanisms on the boundary layer as a whole, $P(k_z)$ and $T(k_z)$ are integrated along the wall-normal direction as in \cite{Nekkanti_Colonius_Schmidt_2025}. The spatial turbulent diffusion term is purposefully excluded from this balance; since it merely redistributes kinetic energy across the wall-normal coordinate, its integral across the boundary layer thickness identically vanishes, allowing us to strictly isolate the dynamics of energy injection and its spectral redistribution.
\begin{figure}[htbp]
\centering
\begin{overpic}[trim = 0cm 0cm 0cm 0cm, clip,width=0.99\textwidth]{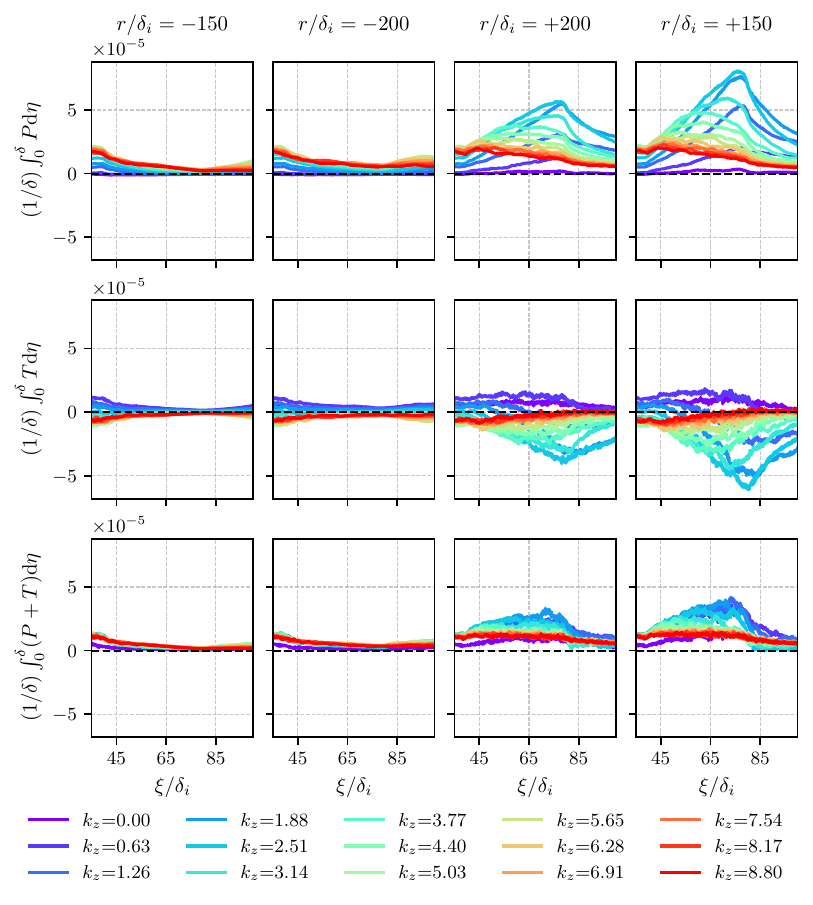}
\end{overpic}
\caption{Streamwise evolution of boundary-layer-integrated TKE budget terms for low spanwise wavenumbers. Rows from top to bottom: spectral production $(1/\delta)\int P \, \mathrm{d}\eta$, nonlinear transfer $(1/\delta)\int T \, \mathrm{d}\eta$, and their sum. Columns from left to right: convex configurations ($r/\delta_{i} = -150$, $-200$) and concave configurations ($r/\delta_{i} = +200$, $+150$).}
\label{fig:tke-bl-integration}
\end{figure}

To focus on the most energetic turbulent features, we restrict our analysis to the low-wavenumber range $k_z \in [0, 8.80]$. Figure~\ref{fig:tke-bl-integration} presents the boundary-layer-integrated production, nonlinear transfer (turbulent transport), and their sum. Looking first at the production term (top row), upon entering the curved section ($40 < x/\delta_{i} < 50$), concave geometries exhibit an initial amplification of wavenumbers around $k_z \approx 5.6$, which corresponds to the modulation of the incoming boundary layer streaks identified in figure~\ref{fig:fluc}. However, as the flow progresses further into the concave section ($50 < \xi/\delta_i < 80$), the destabilizing curvature and adverse pressure gradient strongly excite even lower spanwise wavenumbers, with production peaking at $k_z = 2.51$ and $1.88$. This low-wavenumber amplification corresponds directly to the outer secondary production peak discussed previously, and to the scales involved in the inter-component energy transfer shown in figure~\ref{fig:tke-Pi}. Over the convex surfaces, this amplification of large-scale structures is absent and the production peak remains anchored at $k_z \approx 5.65$, undergoing only a general attenuation in magnitude. Downstream in the recovery region ($80 < \xi/\delta_i < 100$), the production of low-wavenumber structures drops rapidly in the concave cases, while the convex cases experience a small amplification of the $k_z = 5.65$ streaky structures, corroborating the instantaneous fluctuation fields in figure~\ref{fig:fluc}.

Because the flow is highly nonlinear, analyzing production in isolation provides an incomplete picture of the resulting structural state. The nonlinear inter-scale transfer through the turbulent transport term $T(k_z)$ (middle row of figure~\ref{fig:tke-bl-integration}) must be considered to understand how the injected TKE is dynamically redistributed. Over concave walls, the nonlinear transfer acts as a net sink, extracting energy from the heavily produced intermediate scales ($k_z = 1.88$ and $2.51$). This extracted energy is subsequently channeled into a net gain for even larger scales, specifically targeting $k_z = 0.63$ and the spanwise-uniform two-dimensional mode ($k_z = 0$). For the convex cases, the magnitude of the nonlinear transfer diminishes proportionately with the pressure gradient and the overall suppression of TKE. Consequently, the sum of production and nonlinear transfer (bottom row) dictates the net growth of these modes, confirming that energy over the concave surface ultimately pools within the large-scale range of $k_z = 0.63\text{-}1.88$.
\begin{figure}[htbp]
\centering
\begin{overpic}[trim = 0cm 0cm 0cm 0cm, clip,width=0.99\textwidth]{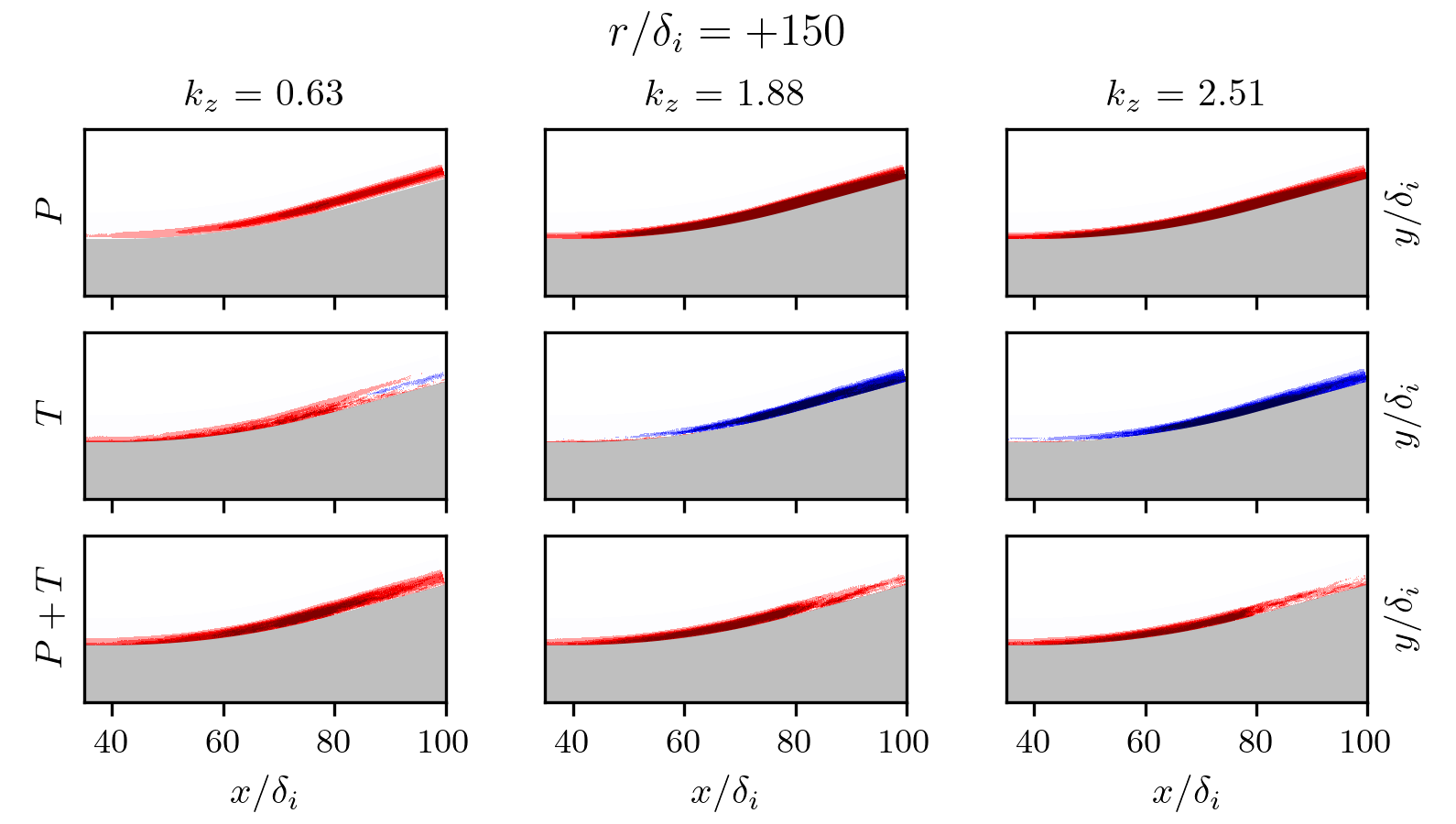}
\end{overpic}
\caption{Spatial distributions of spectral production $P(k_z)$, nonlinear transfer $T(k_z)$, and their sum for selected spanwise wavenumbers in the concave case $r/\delta_i = +150$, illustrating localized energy extraction and injection.}
\label{fig:tke-P-mode}
\end{figure}

The integrated balances in figure~\ref{fig:tke-bl-integration} suggest a directed spectral cascade that transfers energy from $k_z > 0.63$ down to the fundamental large-scale mode $k_z = 0.63$ and spanwise-uniform oscillations. Because $T(k_z)$ represents the net budget from interactions with all other scales, we inspect the spatial distributions of these specific modes in figure~\ref{fig:tke-P-mode} to confirm their spatial correlation. Results are presented for the concave surface with $r/\delta_i = +150$. The contours reveal that the physical regions where energy is removed via nonlinear interactions at $k_z = 1.88$ and $2.51$ directly overlap with the regions of nonlinear energy gain at $k_z = 0.63$. This occurs more prominently in the range $60<x<80$ where these larger spanwise structures are observed in figure \ref{fig:fluc}.

\begin{figure}[htbp]
\centering
\begin{overpic}[trim = 0cm 0cm 0cm 0cm, clip,width=\textwidth]{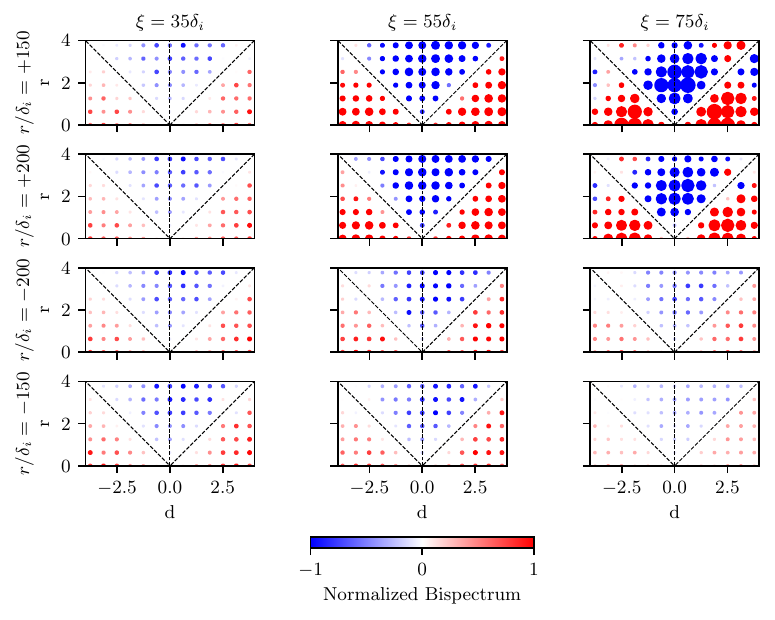}
\end{overpic}
\caption{Bispectrum of the turbulent kinetic energy at different $r/\delta_{i}$. Each marker represents a donor (d) or recipient (r) spanwise wavenumber $k_z$ with the distances between markers being $\Delta k_z \approx 0.63$.}
\label{fig:tke-bispectrum}
\end{figure}

To explicitly trace the inter-scale energy pathways, figure~\ref{fig:tke-bispectrum} presents the TKE bispectrum, which expands the convolution sum to isolate discrete triadic interactions (as formulated in \S~\ref{sec:triadic-analysis}). The quantities plotted represent a spatial average of the spectral energy transfer, evaluated at each reported streamwise station $\xi$ as
\begin{equation}
    \langle T(\mathrm{d} \to \mathrm{r}, \xi, \eta) \rangle = \frac{1}{2\delta_{i}\delta} \int_{\xi-\delta_{i}}^{\xi+\delta_{i}} \int_{0}^{\delta} T(\mathrm{d} \to \mathrm{r}; \xi', \eta') \, \mathrm{d}\eta' \, \mathrm{d}\xi',
    \label{eq:bispectrum_average}
\end{equation}
where $\delta$ is the local boundary layer thickness. For comparison purposes, these area-averaged contours are then normalized by the maximum absolute energy exchange recorded at the upstream flat-plate station ($x/\delta_{i} = 35$), where this third-order statistic demonstrates excellent consistency across all computational domains. The computed bispectra strictly satisfy the triadic conservation properties outlined by \citet{Yeung_Chu_Schmidt_2026}. Specifically, the energy transferred from wavenumber $k_d$ to $k_r$ is equal and opposite to the transfer from $k_r$ to $k_d$, representing a conservative exchange. Consequently, this conservation pair renders the bispectral transfer map perfectly antisymmetric with respect to the diagonal lines $k_d = k_r$ and $k_d = -k_r$. Additionally, the bispectrum exhibits exact symmetry with respect to the $k_r = 0$ axis. Besides, since the lower half-plane ($k_r < 0$) contains only redundant information, it is omitted from the visualized contours. Furthermore, it is crucial to highlight a fundamental distinction between the present analysis and the bispectral interpretation described by \citet{Yeung_Chu_Schmidt_2026}. Because the time- and spanwise-averaged flow is explicitly subtracted from the velocity field prior to our analysis, the $k_z = 0$ component exclusively represents unsteady, spanwise-uniform fluctuations rather than a steady mean flow. As a result, the diagonal $k_d = k_r$ does not represent mean advection, nor does the $k_d = 0$ line represent turbulent production (i.e., energy injection from the mean flow into the coherent structures). Instead, interactions along these boundaries characterize nonlinear exchanges strictly within the fluctuating field involving these spanwise-uniform motions.

Examining the physical pathways, the bispectral contours reveal a prominent difference between the opposing geometries. Over the convex surfaces, the inter-scale energy transfer remains qualitatively very similar to the flat-plate baseline; the primary modification is a general attenuation of the triadic interaction strengths, sustained by the bulk dilatation and the consequent suppression of turbulence. In contrast, the concave curvatures fundamentally alter inter-scale energy exchange. The bispectral map for the concave region clearly captures a direct inverse transfer of energy from intermediate scales ($k_z \in [1.26, 3.14]$) down to $k_z = 0.63$. By normalizing these scales by the fundamental wavenumber $k_{z,0} = 0.63$, this transfer can be mapped as interactions between integer harmonic modes. Distinct triadic pathways emerge wherein energy is fed from modes 2, 3, 4, and 5 into mode 1, mediated by a range of catalytic scales. This establishes a structured, multi-step energy cascade involving prominent triads such as $5 \to 3 \to 1$, $5 \to 2 \to 1$, $4 \to 3 \to 1$, $4 \to 2 \to 1$, and $3 \to 2 \to 1$. Notably, this spanwise upscale reaches its maximum intensity at station $\xi/\delta_i = 75$ over the concave surface, where the transfer from mode 3 to mode 1 emerges as the most prominent triadic interaction. This inverse transfer mechanism aligns with the recent findings of \citet{Kang_Lee_2026} in a compression corner environment, where an analogous upscale redistribution of energy was directly attributed to the nonlinear pairing process of G{\"o}rtler instabilities.

Furthermore, the bispectrum isolates the energy transferred to the spanwise-uniform oscillations ($k_z = 0$). This specific transfer is driven by the self-interaction of the individual harmonic modes ($k_z/k_{z,0} \in [1, 6]$). In contrast, the inter-scale energy transfers directed toward larger, non-zero spanwise scales actively feed the development of progressively wider spanwise coherent motions. While earlier budget analyses demonstrated that turbulent production preferentially excites large-scale motions in the outer layer along the concave surfaces, the bispectrum reveals that these energetic structures also come to dominate the global energy transfer dynamics, collectively sustaining the curvature-modified boundary layer.

It is important to note, however, that the observed direction of energy transfer, while manifesting as an inverse cascade in the spanwise wavenumber domain, does not necessarily imply a global inverse turbulent energy cascade \citep{Cho_Hwang_Choi_2018, Lee_Moser_2019}. The turbulent structures within the boundary layer are inherently three-dimensional, whereas the present bispectral analysis is projected exclusively onto a single spectral dimension $k_z$. A transfer towards larger spanwise scales may reflect a geometric reconfiguration of the eddies rather than a true multi-dimensional upscale cascade. To definitively elucidate this mechanism, it is necessary to measure the spatial coherence of these structures in both the streamwise and wall-normal directions. Motivated by this requirement, the subsequent section employs SPOD to fully extract and characterize the three-dimensional coherent structures that dominate this energy exchange process. Because kinetic energy budgets and bispectral transfers exhibit strong qualitative similarity across both concave and convex geometries, being primarily modulated by the magnitude of the adverse and favorable pressure gradients, the ensuing SPOD analysis will focus exclusively on the strongest curvature cases ($r/\delta_i = \pm 150$).


\subsection{Characterization of coherent structures}
\label{sec:coherent_structures}

\begin{figure}[htbp]
\centering
\begin{overpic}[trim = 0cm 0cm 0cm 0cm, clip, width=\textwidth]{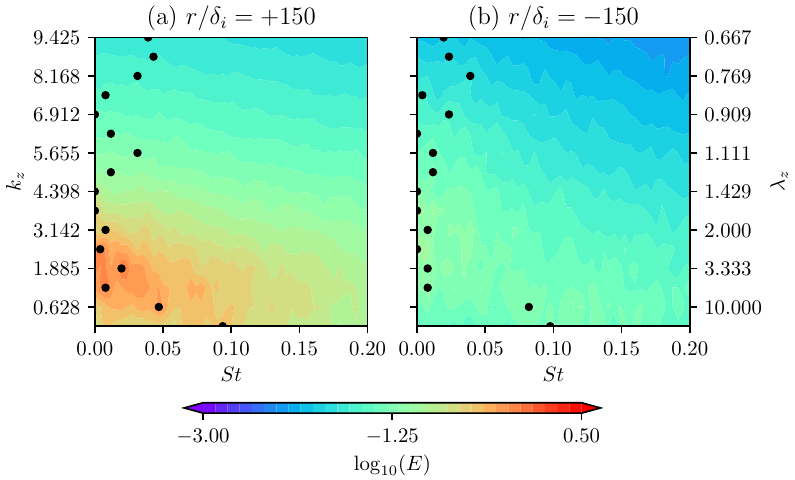}
\end{overpic}
\caption{SPOD energy spectra as a function of Strouhal number for the high-curvature configurations ($r/\delta_i = \pm 150$). Black dots denote the most energetic frequency for each spanwise wavenumber.}
\label{fig:SPOD_spectrum}
\end{figure}

In this section, the dominant coherent structures within the boundary layer are characterized using SPOD, computed over the first 15 spanwise wavenumbers. Figure~\ref{fig:SPOD_spectrum} presents the SPOD energy spectra as a function of the Strouhal number ($St = f \delta_{i} / U_\infty$) for the strongest concave and convex curvatures ($r/\delta_{i} = \pm 150$). Information about the convergence of the SPOD results is provided in the appendix \ref{app:spod_convergence}. The most energetic frequency for each wavenumber is denoted by a black dot. As expected, the concave configuration exhibits higher energy levels than its convex counterpart. It is important to note that this amplitude disparity is inherently tied to the use of the compressible Chu norm, which weights the velocity fluctuations by the local mean density. Nevertheless, the wavenumbers corresponding to the most energetic modes over the concave wall are in agreement with the preceding TKE and bispectral analyses. By leveraging the energy-optimality of the leading SPOD modes, this decomposition provides a statistically robust representation of the precise coherent structures that govern these kinetic energy budgets and triadic interactions \citep{Nekkanti_Colonius_Schmidt_2025, schiavo}.

\begin{figure}[htbp]
\centering
\begin{overpic}[trim = 0cm 0cm 0cm 0cm, clip, width=0.49\textwidth]{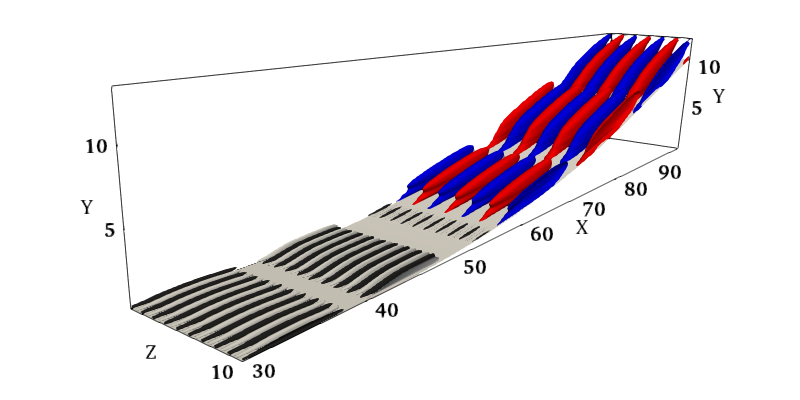}
\put(20,450){(a)}
\put(100,450){$k_z = 1.88; 5.65$}
\end{overpic}
\begin{overpic}[trim = 0cm 0cm 0cm 0cm, clip, width=0.49\textwidth]{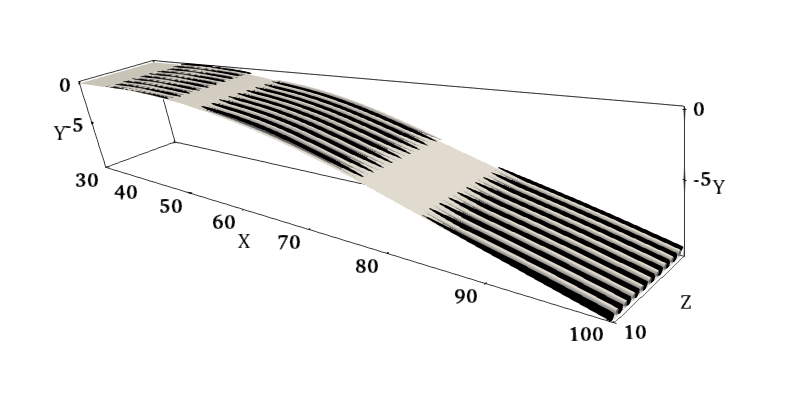}
\put(20,450){(b)}
\put(100,450){$k_z = 5.65$}
\end{overpic}
\begin{overpic}[trim = 0cm 0cm 0cm 0cm, clip, width=0.49\textwidth]{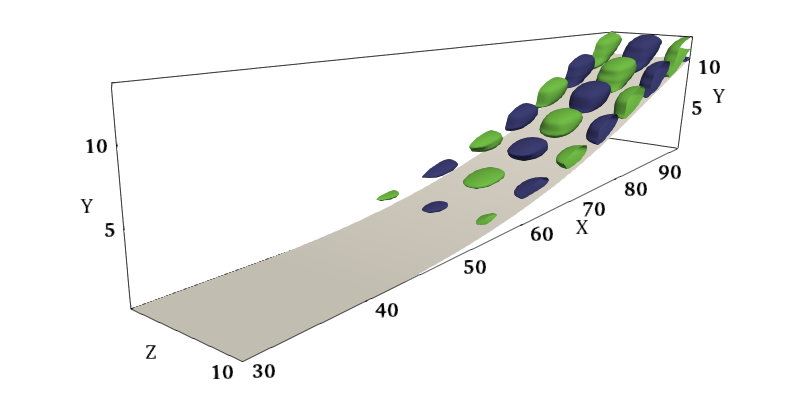}
\put(20,450){(c)}
\put(100,450){$k_z = 0.63$}
\end{overpic}
\begin{overpic}[trim = 0cm 0cm 0cm 0cm, clip, width=0.49\textwidth]{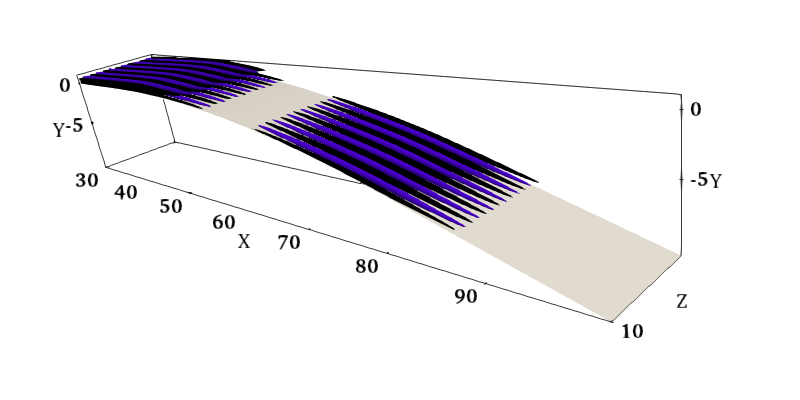}
\put(20,450){(d)}
\put(100,450){$k_z = 3.77$}
\end{overpic}
\begin{overpic}[trim = 0cm 0cm 0cm 0cm, clip, width=0.49\textwidth]{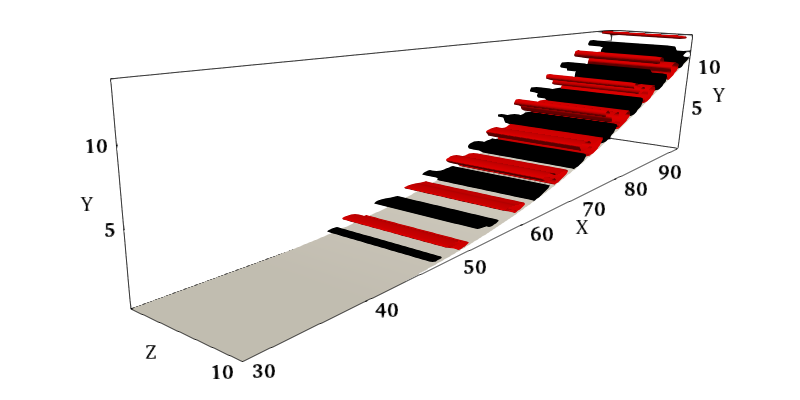}
\put(20,450){(e)}
\put(100,450){$k_z = 0.0$}
\end{overpic}
\begin{overpic}[trim = 0cm 0cm 0cm 0cm, clip, width=0.49\textwidth]{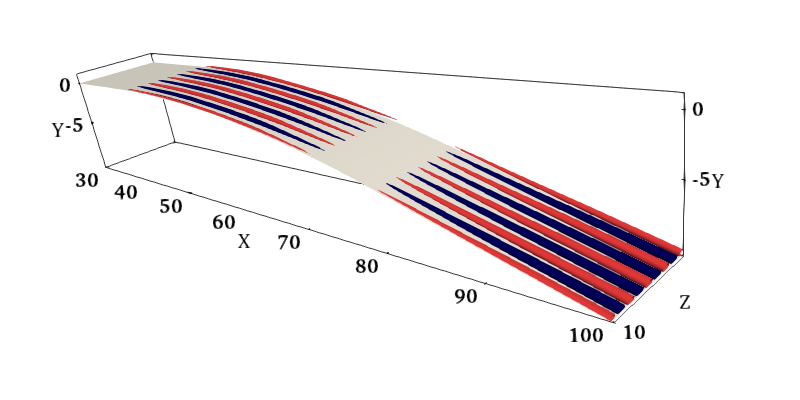}
\put(20,450){(f)}
\put(100,450){$k_z = 2.51$}
\end{overpic}
\caption{Spatial support of the most energetic SPOD modes for the high-curvature configurations ($r/\delta_i = \pm 150$). The left column (a, c, e) depicts the structural distribution over the concave surface, illustrating the streamwise vortex breakdown across decreasing wavenumbers. The right column (b, d, f) depicts the modes over the convex surface, highlighting their persistent streamwise elongation. The annotated values denote the specific spanwise wavenumber $k_z$ corresponding to each panel.}
\label{fig:SPOD_modes}
\end{figure}

To further elucidate the physical mechanisms underpinning the spectral energy transport, figure~\ref{fig:SPOD_modes} illustrates the spatial support of the dominant SPOD modes at selected wavenumbers. The three-dimensional structures visualized in these panels are isosurfaces of the tangential velocity fluctuation ($u_{\xi}''$), reconstructed from the complex SPOD modes via an inverse Fourier transform along the homogeneous spanwise direction. The left column displays the results of the concave configuration, while the right column shows the convex case. For the concave geometry, these specific scales were chosen to visually trace the energy transfer pathways identified in the TKE budget (figure~\ref{fig:tke-bl-integration}) and the bispectrum (figure~\ref{fig:tke-bispectrum}), particularly the cascade towards the largest spanwise scales ($k_z = 0.63$) and the spanwise-uniform mode ($k_z = 0$).

Panels (a) and (b) present the modes corresponding to the largest accumulation of TKE, driven by the combined effects of production and inter-scale transfer. Over the concave surface (panel a), the moderate-scale $k_z = 5.65$ mode (visualized in black and white) manifests primarily in the initial portion of the curved section. Further downstream, this structure is superseded by the larger $k_z = 1.88$ mode (red and blue). This downstream spatial progression is in excellent agreement with the spatial evolution of turbulent production and inter-scale energy transfer depicted previously in figure~\ref{fig:tke-bl-integration}.

Crucially, examining the spatial support of the largest dominant mode ($k_z = 0.63$, panel c) and the spanwise-uniform mode ($k_z = 0$, panel e) clarifies the true physical nature of this spectral energy transfer. Although the bispectrum indicates a robust transfer of energy towards larger spanwise wavelengths, the spatial modes reveal that energy transferred from moderate spanwise motions (such as $k_z = 1.88$) ultimately induces a vortex breakdown in the streamwise direction. Therefore, the spanwise upscale energy transfer does not represent a global three-dimensional inverse energy cascade. Instead, interactions between similar-sized structures generate larger spanwise scales ($k_z = 0.63$), while self-interactions of identical wavenumbers yield spanwise-uniform motions ($k_z = 0$). As visualized, both of these resulting large-scale structures undergo severe fragmentation and breakdown across the streamwise and wall-normal dimensions. This streamwise downscale is also observed by \cite{Gungor_Maciel_Gungor_2022} for flat boundary layers under APG.

The convex configuration exhibits a different structural topology. As shown in the right-hand panels (b, d, and f), the $k_z = 5.65$ mode is amplified primarily within the downstream flat recovery section, which is clearly reflected in its spatial support. Furthermore, regardless of the selected spanwise wavenumber, all dominant SPOD modes over the convex surface manifest as highly elongated structures in the streamwise direction \citep{Wang_2020}. This persistent geometric stretching accurately reflects the strong suppression of transversal fluctuations and the consequent shift towards one-component turbulence discussed previously in \S~\ref{sec:turbulence_statistics}.

\subsection{Resolvent analysis}
\label{sec:resolvent_analysis}

In this section, global resolvent analysis is employed to investigate the role of linear amplification mechanisms in governing the spatial organization and energy distribution of coherent structures over the concave and convex surfaces. Consistent with the preceding SPOD analysis, we focus on the configurations with $r/\delta_i = \pm 150$, where curvature effects are more pronounced. The analysis is conducted on a subdomain of the DNS solution spanning the streamwise interval $25 < \xi/\delta_i < 100$. 

\begin{figure}[htbp]
\centering
\begin{overpic}[trim = 0cm 0cm 0cm 0cm, clip, width=0.49\textwidth]{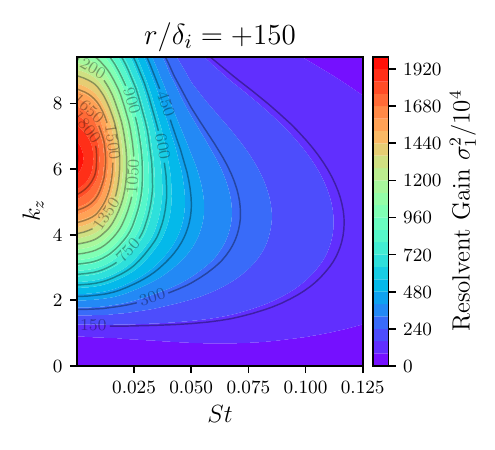}
\put(0,60){(a)}
\end{overpic}
\begin{overpic}[trim = 0cm 0cm 0cm 0cm, clip, width=0.49\textwidth]{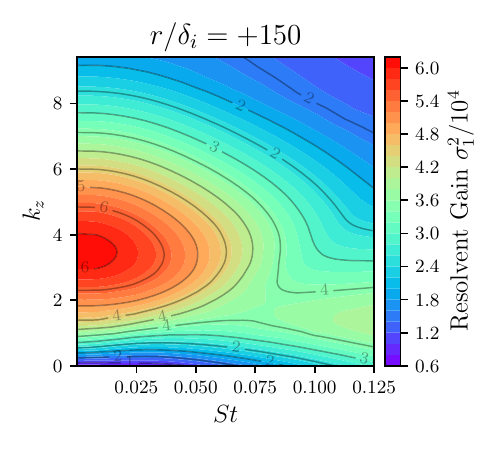}
\put(0,60){(b)}
\end{overpic}
\begin{overpic}[trim = 0cm 0cm 0cm 0cm, clip, width=0.49\textwidth]{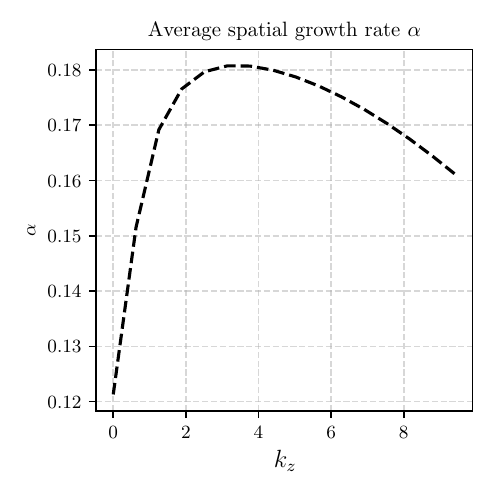}
\put(0,60){(c)}
\end{overpic}
\begin{overpic}[trim = 0cm 0cm 0cm 0cm, clip, width=0.49\textwidth]{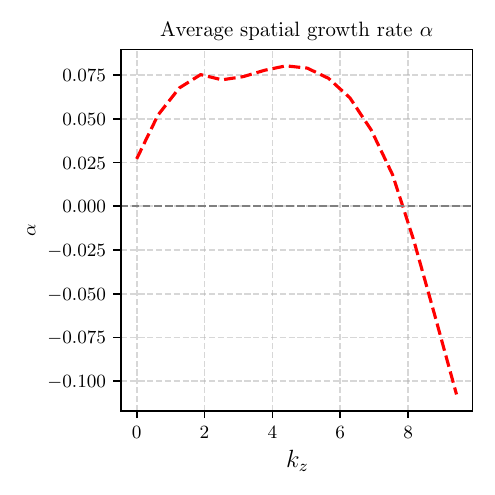}
\put(0,60){(d)}
\end{overpic}
\caption{Leading resolvent gains as a function of Strouhal number and spanwise wavenumber (a, b), alongside the corresponding streamwise spatial amplification rates $\alpha(k_z)$ (c, d), for the high-curvature concave ($r/\delta_i = +150$, left) and convex ($r/\delta_i = -150$, right) configurations.
}
\label{fig:resolvent_gains}
\end{figure}

We begin by evaluating the leading resolvent gains ($\sigma_1$), presented in figures~\ref{fig:resolvent_gains}(a) and \ref{fig:resolvent_gains}(b), which quantify the optimal harmonic amplification arising from the non-normality of the linearized Navier--Stokes operator. To complement the global gains and isolate spatial growth along the curved wall, figures~\ref{fig:resolvent_gains}(c) and \ref{fig:resolvent_gains}(d) present the empirical spatial amplification rate, $\alpha(k_z)$, computed from the leading resolvent response modes using a normal-mode ansatz across the curved section:
\begin{equation}
    \alpha(k_z) = St_{\text{avg}}\left[ \frac{1}{40\delta_i}\int_{\xi = 40\delta_i}^{\xi = 80\delta_i} \frac{1}{\kappa(\xi, k_z, St)} \frac{\partial \kappa(\xi, k_z,St)}{\partial \xi} \, \mathrm{d}\xi\right]
    \label{eq:spatial_amplification}
\end{equation}
where $\kappa(\xi, k_z, St)$ denotes the modal kinetic energy integrated across the boundary layer thickness:
\begin{equation}
    \kappa(\xi, k_z, St) =  \int_0^{\delta} \bar{\rho}(\xi, \eta) \, \widehat{u_i'}(\xi, \eta, k_z, St) \, \widehat{u_i'}^*(\xi, \eta, k_z, St) \, \mathrm{d}\eta.
    \label{eq:modal_tke}
\end{equation}

It is important to note that although the empirical amplification is spatially averaged along the curvature section, we emphasize that $(1/\kappa)\partial\kappa/\partial\xi \approx\text{constant}$. Comparing the resolvent gains in figure~\ref{fig:resolvent_gains} with the SPOD energy spectra in figure~\ref{fig:SPOD_spectrum} reveals distinct behavioral differences between the two curvature regimes. For the concave configuration, the peak resolvent gains are concentrated at moderate wavenumbers ($5 < k_z < 7$), whereas the SPOD energy peaks at lower wavenumbers ($1.8 < k_z < 3$), reflecting the nonlinear shift toward larger spanwise scales documented in \S~\ref{sec:spectral_transport}. However, when examining the spatial amplification rate $\alpha(k_z)$ in figure~\ref{fig:resolvent_gains}(c), the strongest spatial growth occurs at lower wavenumbers, aligning more closely with the large-scale spanwise motions observed in the DNS. This indicates that the linear amplification of these large-scale structures over the concave wall is primarily governed by spatially developing convective amplification rather than global temporal amplification.

In contrast, for the convex configuration, the distribution of the resolvent gains agrees closely with the SPOD energy spectra. Amplification is concentrated at low-to-moderate wavenumbers ($2.51 < k_z < 6$) and low frequencies, which is characteristic of the linear lift-up mechanism responsible for generating the elongated streamwise streaks observed in this flow. Furthermore, this close correspondence between the linear resolvent gains and the empirical SPOD spectra is a direct consequence of the quasi-relaminarization induced by the stabilizing convex curvature and favorable pressure gradient, which attenuates nonlinear triadic interactions and allows linear mechanisms to dominate the boundary layer dynamics. Spatial amplification contours further reveal that growth is confined to structures within the range $0 < k_z < 8$, whereas finer scales are spatially damped. This selective damping of fine scales directly corroborates the relaminarization tendencies observed in the DNS.

\begin{figure}[htbp]
\centering
\begin{overpic}[trim = 0cm 0cm 0cm 0cm, clip, width=0.49\textwidth]{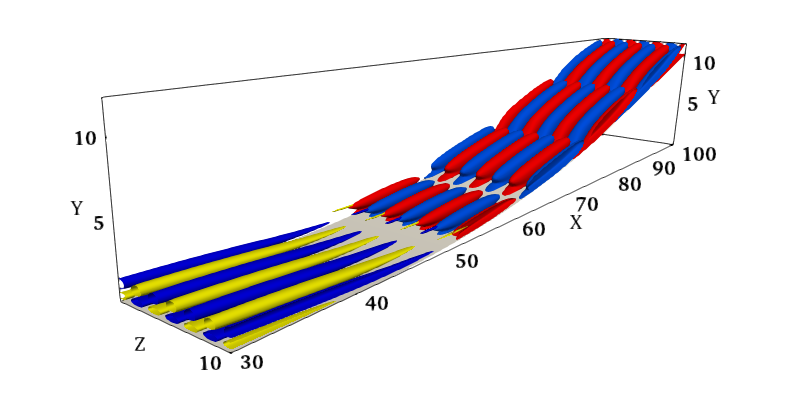}
\put(20,450){(a)}
\put(100,450){$k_z = 1.88$}
\end{overpic}
\begin{overpic}[trim = 0cm 0cm 0cm 0cm, clip, width=0.49\textwidth]{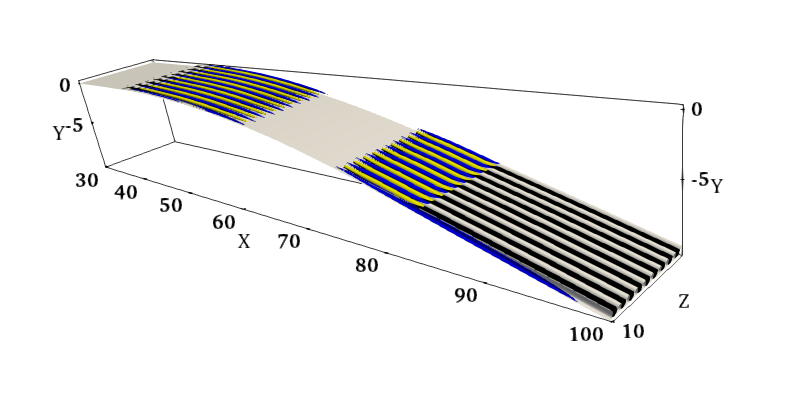}
\put(20,450){(b)}
\put(100,450){$k_z = 5.65$}
\end{overpic}
\begin{overpic}[trim = 0cm 0cm 0cm 0cm, clip, width=0.49\textwidth]{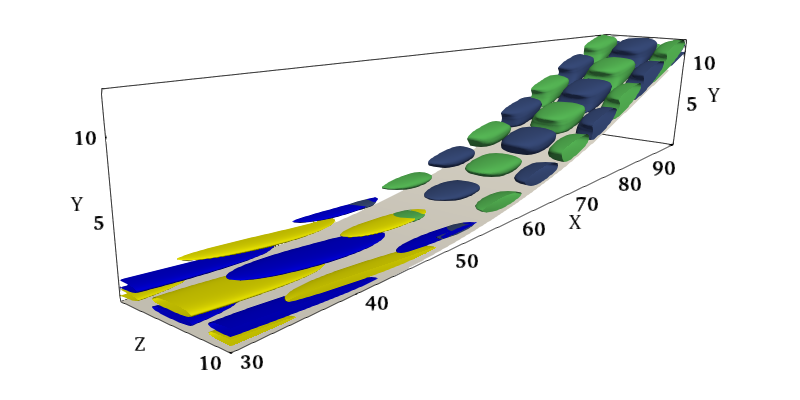}
\put(20,450){(c)}
\put(100,450){$k_z = 0.63$}
\end{overpic}
\begin{overpic}[trim = 0cm 0cm 0cm 0cm, clip, width=0.49\textwidth]{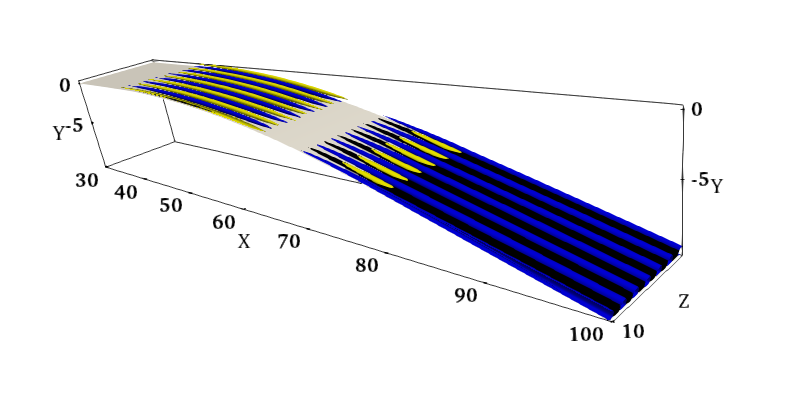}
\put(20,450){(d)}
\put(100,450){$k_z = 3.77$}
\end{overpic}
\begin{overpic}[trim = 0cm 0cm 0cm 0cm, clip, width=0.49\textwidth]{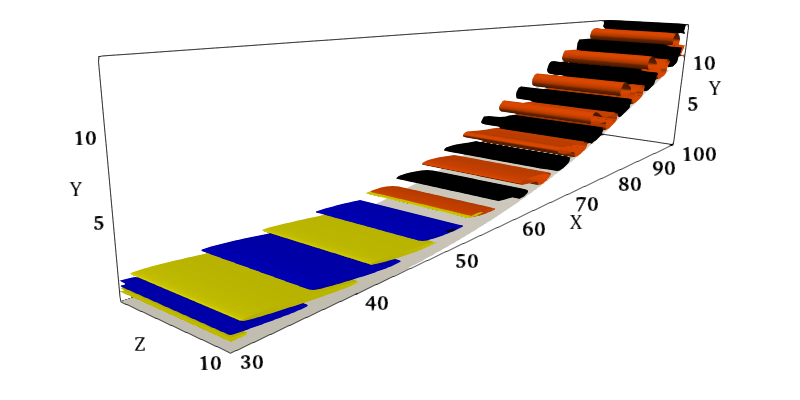}
\put(20,450){(e)}
\put(100,450){$k_z = 0.0$}
\end{overpic}
\begin{overpic}[trim = 0cm 0cm 0cm 0cm, clip, width=0.49\textwidth]{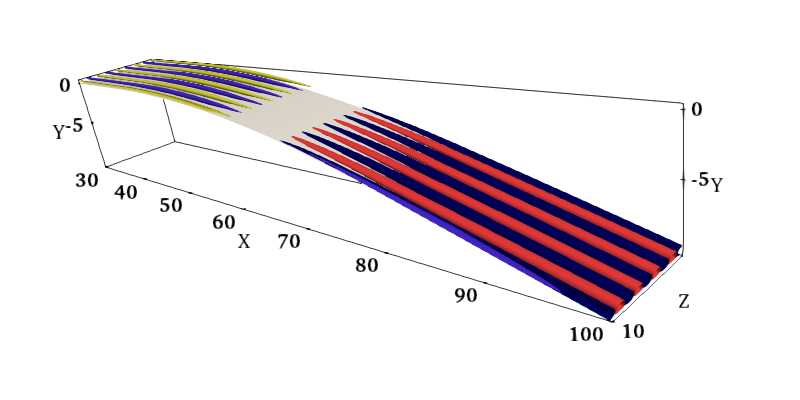}
\put(20,450){(f)}
\put(100,450){$k_z = 2.51$}
\end{overpic}
\caption{Spatial support of the optimal resolvent forcing modes (blue and yellow) and response modes (colors similar to SPOD modes in figure \ref{fig:SPOD_modes}) corresponding to the most energetic SPOD modes for the high-curvature configurations ($r/\delta_i = \pm 150$). The left column (a, c, e) depicts the modes over the concave surface, and the right column (b, d, f) depicts the modes over the convex surface. The annotated values denote the spanwise wavenumber $k_z$ for each panel.}
\label{fig:resolvent_modes}
\end{figure}

To compare the predicted spatial structures against the empirical modes, figure~\ref{fig:resolvent_modes} displays the optimal resolvent forcing modes (blue and yellow isosurfaces) and corresponding response modes at the same wavenumber-Strouhal pairs selected for the SPOD visualizations in figure~\ref{fig:SPOD_modes}. Over the concave surface (left column), the resolvent response modes exhibit a remarkable resemblance to the leading SPOD modes, even at the lowest wavenumbers ($k_z = 0.63$ and $0.0$) that are heavily energized by nonlinear inter-scale transfer in the DNS. Specifically, the linear response modes accurately capture both the spatial organization and the downstream structural breakdown observed in the empirical data. This agreement indicates that while nonlinear triadic interactions redistribute energy across scales to select the dominant amplitudes, the underlying modal shapes at a given frequency and wavelength are dictated primarily by the receptivity of the mean flow.

A similar qualitative agreement between the SPOD and resolvent response modes is observed over the convex surface (right column of figure~\ref{fig:resolvent_modes}). However, a critical distinction emerges in the relative spatial placement of the forcing and response modes. In the concave configuration, the optimal forcing modes are localized entirely upstream of the curved section, indicative of a purely convective amplification process where upstream perturbations are amplified as they advect downstream. In the convex configuration, the optimal forcing and response modes exhibit a substantial spatial overlap within the flat recovery region downstream of the curvature. Following \citet{Skene_Yeh_Schmid_Taira_2022}, this spatial overlap between forcing and response is characteristic of a resolvent wavemaker, signifying a localized, self-sustaining linear mechanism that actively energizes the downstream boundary layer.

\begin{figure}
\centering
\begin{overpic}[trim = 0cm 0cm 0cm 0cm, clip, width=0.6\textwidth]{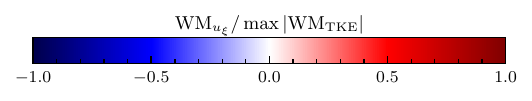}
\end{overpic}
\begin{overpic}[trim = 0cm 0cm 0cm 0cm, clip, width=\textwidth]{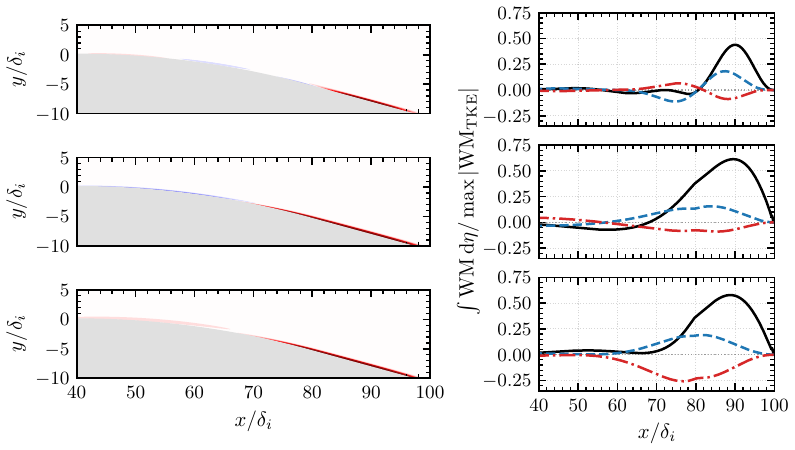}
\put(300,215){$k_z = 2.51$}
\put(300,380){$k_z = 3.77$}
\put(300,545){$k_z = 5.65$}
\end{overpic}
\begin{overpic}[trim = 0cm 0cm 0cm 0cm, clip, width=0.8\textwidth]{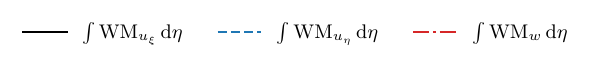}
\end{overpic}
\caption{Resolvent wavemaker modes for the high-curvature convex configuration ($r/\delta_i = -150$). Left: spatial distribution of the wavemaker for the tangential velocity component. Right: wall-normal integrated wavemaker distributions for the streamwise, wall-normal, and spanwise velocity components, normalized by the maximum value of the TKE wavemaker.}
\label{fig:resolvent_wavemaker}
\end{figure}

To substantiate this mechanism, the resolvent wavemaker corresponding to the leading modes of the convex case 
is presented in figure~\ref{fig:resolvent_wavemaker}. The contours on the left display the spatial distribution of the wavemaker for the tangential velocity component, while the panels on the right illustrate the wall-normal integrated wavemaker profiles for the streamwise, wall-normal, and spanwise velocity components. All quantities are normalized by the maximum value of the wavemaker computed for the turbulent kinetic energy. The spatial contours clearly exhibit a pronounced overlap concentrated within the flat recovery section of the boundary layer, a feature that is further substantiated by the wall-normal integrated profiles. This strong spatial coincidence confirms that the localized resurgence and amplification of the streamwise velocity fluctuations downstream of the convex bend is governed by a linear amplification mechanism, directly explaining the localized Reynolds stress amplification reported in \S~\ref{sec:turbulence_statistics}.

\section{Conclusions}
\label{sec:conclusions}

Direct numerical simulations are employed to investigate the structural and dynamical modifications of supersonic turbulent boundary layers ($M_\infty = 2.0$, $Re_{\delta_i} = 7894$) subjected to continuous concave ($r/\delta_i = +150, +200$) and convex ($r/\delta_i = -150, -200$) surface curvatures, followed by a straight recovery section. By combining mean-flow and Reynolds-stress diagnostics with scale-resolved spectral TKE budgets, triadic energy-transfer bispectra, SPOD, and global resolvent analysis, the results demonstrate how compressibility and pressure gradients, as well as linear and nonlinear mechanisms, govern curved high-speed boundary layers.

Bulk dilatation and compression play a dominant role in dictating the integral boundary-layer parameters, reversing the shape-factor trends classically expected in incompressible pressure-gradient flows. Over convex surfaces, the FPG accelerates the flow and reduces the velocity deficit, decreasing the momentum thickness ($\theta$). However, the accompanying bulk dilatation causes a severe drop in near-wall density that produces a substantial mass-flux deficit and thickens the boundary layer, driving a pronounced increase in the displacement thickness ($\delta^*$) and kinematic shape factor ($H = \delta^*/\theta$), alongside a reduction in  skin friction ($C_f$). Over concave surfaces, the APG decelerates the flow while bulk compression increases the local density. These competing effects preserve the normalized streamwise momentum profile near the flat-plate baseline, resulting in minor variations in $\delta^*$ while the increased velocity deficit substantially raises $\theta$, thereby reducing the shape factor $H$.

The turbulence structure and Reynolds-stress anisotropy diverge sharply between the two curvature regimes, governed by the inter-component pressure-strain redistribution ($\Pi$). Over convex surfaces, stabilizing curvature and bulk dilatation dampen cross-stream velocity fluctuations and diminish the mean-density-weighted and velocity-gradient source terms of the pressure Poisson equation, causing a collapse of root-mean-square pressure fluctuations ($p_{\mathrm{rms}}$). This attenuates the pressure-strain redistribution term, preventing the transfer of energy from the streamwise component into the wall-normal and spanwise directions. Consequently, the boundary layer shifts toward a highly anisotropic one-component state, disrupting the autonomous near-wall turbulent cycle and sequestering kinetic energy within the streamwise fluctuations ($\widetilde{u_\xi'' u_\xi''}$). Over concave surfaces, destabilizing curvature and bulk compression amplify $p_{\mathrm{rms}}$ and enhance near-wall pressure diffusion and pressure-strain redistribution, efficiently transferring energy into cross-stream fluctuations and shifting the near-wall turbulence toward a two-component state.

Scale-resolved spectral TKE budgets demonstrate that across all configurations, turbulent production is predominantly governed by the mean shear component ($P_{\xi\eta}$). While bulk density variations thermodynamically modulate the overall magnitude of the density-weighted budget terms, concave curvature induces a distinct secondary production peak in the outer layer concentrated at low spanwise wavenumbers ($k_z \in [1.88, 2.51]$), accompanied by enhanced outward spatial turbulent diffusion ($DT$) into the wake region. 
To resolve the phase-coupled inter-scale pathways over the concave wall, the TKE bispectrum isolates a structured, multi-step inverse spanwise transfer from intermediate scales ($k_z \in [1.26, 3.14]$, harmonic modes 2--5) into the fundamental large-scale mode ($k_z = 0.63$, mode 1), reaching peak intensity near the end of the curved section via the $3 \to 1$ triad, while self-interactions of individual harmonic modes transfer energy into unsteady, spanwise-uniform oscillations ($k_z = 0$). Reconstructing the three-dimensional coherent structures via SPOD clarifies that this upscale spanwise transfer does not correspond to a global three-dimensional inverse cascade. Instead, as energy pools into the largest spanwise scales ($k_z = 0.63$ and $0.0$), the structures undergo severe streamwise fragmentation and vortex breakdown. In contrast, SPOD modes over the convex surface remain highly elongated in the streamwise direction across all wavenumbers.

Finally, global resolvent analysis elucidates the linear amplification mechanisms underlying these coherent structures. Over the concave surface, whereas global temporal resolvent gains peak at moderate wavenumbers ($5 < k_z < 7$), streamwise spatial amplification rates peak at low wavenumbers with optimal forcing localized upstream of the curve, demonstrating that large-scale spanwise structures are potentially linearly amplified by spatially developing convective instabilities. Over the convex surface, quasi-relaminarization suppresses nonlinear interactions and spatially damps fine scales ($k_z > 8$), yielding a close correspondence between the linear resolvent gains and empirical SPOD spectra at low frequencies and low-to-moderate wavenumbers characteristic of the linear lift-up mechanism. Across both geometries, the optimal resolvent response modes closely reproduce the spatial organization of the leading SPOD modes, even at wavenumbers predominantly energized by nonlinear triadic transfer in the concave surface ($k_z = 0.63$ and $0.0$). This confirms that while nonlinear triadic interactions redistribute energy across scales to select the dominant mode amplitudes, the spatial topology of the coherent structures at a given scale is dictated by the linear receptivity of the mean flow. Moreover, in the convex configuration, a pronounced spatial overlap between the optimal forcing and response modes within the straight recovery section reveals a resolvent wavemaker, demonstrating that a localized, self-sustaining linear mechanism drives the downstream amplification of streamwise velocity fluctuations.

\section*{Acknowledgments}
The authors acknowledge the support of the São Paulo Research Foundation (FAPESP) under grants No.\ 2013/08293-7, 2019/17874-0, 2021/06448-0, 2024/20547-9, 2025/13174-4 and 2025/25614-9. The authors thank the Center for Computing in Engineering and Sciences (CCES), the National Center for High Performance Computing (CENAPAD-SP), and the National Laboratory for Scientific Computing (LNCC) for providing computational resources through the Coaraci Supercomputer, cluster Ada Lovelace (project 551), and cluster SDumont (project SIMTURB), respectively.

\begin{appen}


\section{Governing Equations}
\label{app:governing-equations}

The present framework solves the three-dimensional compressible Navier-Stokes equations formulated in generalized curvilinear coordinates. The governing equations are integrated in time using the conservative variables $\mathbf{Q} = [\rho, \rho u, \rho v, \rho w, E_t]^{\mathrm{T}}$, which comprise density, momentum, and total energy per unit volume. By mapping the physical domain in Cartesian coordinates $(x,y,z)$ onto a computational domain defined by the curvilinear coordinates $(\xi, \eta, \zeta)$, the governing equations take the form
\begin{equation}
\frac{\partial \mathbf{Q}}{\partial t} = \frac{1}{\mathcal{J}} \left( - \frac{\partial \hat{\mathbf{F}}}{\partial \xi} - \frac{\partial \hat{\mathbf{G}}}{\partial \eta} - \frac{\partial \hat{\mathbf{H}}}{\partial \zeta} + \frac{\partial \hat{\mathbf{F}}_v}{\partial \xi} + \frac{\partial \hat{\mathbf{G}}_v}{\partial \eta} + \frac{\partial \hat{\mathbf{H}}_v}{\partial \zeta} \right) \mbox{ ,}
\end{equation}

\noindent
where $\mathcal{J} = \det(\partial(x,y,z)/\partial(\xi,\eta,\zeta))$ is the determinant of the Jacobian matrix. The metric terms of the coordinate transformation are denoted using subscripts, such that for an arbitrary curvilinear coordinate $\psi \in \{\xi, \eta, \zeta\}$, we have $\psi_x = \partial \psi / \partial x$ (or more generally $\psi_{x_j} = \partial \psi / \partial x_j$ for $x_j \in \{x_1, x_2, x_3\} = \{x, y, z\}$).
To reflect the computational implementation, these metric terms are scaled by the Jacobian, denoted as $\hat{\psi}_{x_j} = \mathcal{J} \psi_{x_j}$.
Throughout the appendix sections, the Einstein summation convention is employed for repeated indices, and the velocity components are written interchangeably as $(u_1, u_2, u_3) = (u, v, w)$.

The transformed convective fluxes ($\hat{\mathbf{F}}$, $\hat{\mathbf{G}}$, $\hat{\mathbf{H}}$) and viscous fluxes ($\hat{\mathbf{F}}_v$, $\hat{\mathbf{G}}_v$, $\hat{\mathbf{H}}_v$) are then explicitly defined by introducing the scaled contravariant velocity components $\hat{U}, \hat{V}, \hat{W}$
\begin{equation}
    \begin{aligned}
        \hat{U} &= \hat{\xi}_x u + \hat{\xi}_y v + \hat{\xi}_z w \mbox{ ,}\\
        \hat{V} &= \hat{\eta}_x u + \hat{\eta}_y v + \hat{\eta}_z w \mbox{ ,}\\
        \hat{W} &= \hat{\zeta}_x u + \hat{\zeta}_y v + \hat{\zeta}_z w \mbox{ .}
    \end{aligned}
\end{equation}

The convective flux vectors are given by
\begin{equation}
    \hat{\mathbf{F}} = \begin{bmatrix}
        \rho \hat{U} \\
        \rho u \hat{U} + p \hat{\xi}_x \\
        \rho v \hat{U} + p \hat{\xi}_y \\
        \rho w \hat{U} + p \hat{\xi}_z \\
        (E_t + p)\hat{U}
    \end{bmatrix}, \quad
    \hat{\mathbf{G}} = \begin{bmatrix}
        \rho \hat{V} \\
        \rho u \hat{V} + p \hat{\eta}_x \\
        \rho v \hat{V} + p \hat{\eta}_y \\
        \rho w \hat{V} + p \hat{\eta}_z \\
        (E_t + p)\hat{V}
    \end{bmatrix}, \quad
    \hat{\mathbf{H}} = \begin{bmatrix}
        \rho \hat{W} \\
        \rho u \hat{W} + p \hat{\zeta}_x \\
        \rho v \hat{W} + p \hat{\zeta}_y \\
        \rho w \hat{W} + p \hat{\zeta}_z \\
        (E_t + p)\hat{W} 
    \end{bmatrix},
\end{equation}
and, similarly, the transformed viscous flux vectors are expressed as
\begin{equation}
    \hat{\mathbf{F}}_v = \begin{bmatrix}
        0 \\
        \hat{\xi}_{x_j} \tau_{1j} \\
        \hat{\xi}_{x_j} \tau_{2j} \\
        \hat{\xi}_{x_j} \tau_{3j} \\
        \hat{\xi}_{x_j} \beta_j
    \end{bmatrix}, \quad
    \hat{\mathbf{G}}_v = \begin{bmatrix}
        0 \\
        \hat{\eta}_{x_j} \tau_{1j} \\
        \hat{\eta}_{x_j} \tau_{2j} \\
        \hat{\eta}_{x_j} \tau_{3j} \\
        \hat{\eta}_{x_j} \beta_j
    \end{bmatrix}, \quad
    \hat{\mathbf{H}}_v = \begin{bmatrix}
        0 \\
        \hat{\zeta}_{x_j} \tau_{1j} \\
        \hat{\zeta}_{x_j} \tau_{2j} \\
        \hat{\zeta}_{x_j} \tau_{3j} \\
        \hat{\zeta}_{x_j} \beta_j
    \end{bmatrix},
\end{equation}
where the terms $\beta_j$ account for the viscous work and heat conduction
\begin{equation}
    \beta_j = u_i \tau_{ij} - q_j \mbox{ .}
\end{equation}
Here, $\tau_{ij}$ is the Newtonian viscous stress tensor, defined as
\begin{equation}
    \tau_{ij} = \frac{\mu M_{\infty}}{Re} \left( \frac{\partial u_i}{\partial x_j} + \frac{\partial u_j}{\partial x_i} - \frac{2}{3} \delta_{ij} \frac{\partial u_k}{\partial x_k} \right) \mbox{ ,}
\end{equation}
where $\mu$ is the non-dimensional dynamic viscosity and $\delta_{ij}$ is the Kronecker delta. Furthermore, $p$ is the static pressure, obtained from the ideal gas equation of state $p = (\gamma - 1)(E_t - \frac{1}{2}\rho(u^2 + v^2 + w^2))$, and $q_j$ is the heat flux vector obeying Fourier's law, given in non-dimensional form by:

\begin{equation}
    q_j = - \frac{\mu M_{\infty}}{Re Pr} \frac{\partial T}{\partial x_j}  \mbox{ .}
\end{equation}

For an arbitrary scalar field $\phi$ (here, $u_i$ or $T$), the physical derivatives appearing in $\tau_{ij}$ and $q_j$ are evaluated in the computational space through the chain rule,
\begin{equation}
    \frac{\partial \phi}{\partial x_j}
    = \xi_{x_j}\frac{\partial \phi}{\partial \xi}
    + \eta_{x_j}\frac{\partial \phi}{\partial \eta}
    + \zeta_{x_j}\frac{\partial \phi}{\partial \zeta}
    \mbox{ .}
\end{equation}

\section{Correction to the spectral mass-diffusion term}
\label{app:mass_diffusion}

In the study of shock wave-turbulent boundary layer interactions by \citet{Yu_Zhao_Tang_Yuan_Xu_2022}, a spectral transport equation for turbulent kinetic energy (TKE) was derived based on the two-point correlation framework of \cite{Arun_Sameen_Srinivasan_Girimaji_2021}. While the overarching methodology is robust, an algebraic oversight is present in their derivation of the mass-flux term ($B_{ij}$, Eq. 3.15 in their manuscript). This appendix outlines the origin of this discrepancy and provides the mathematically rigorous, corrected spectral formulation.

The oversight stems from the algebraic treatment of the spatial gradients in the velocity equations. The instantaneous velocity equation is governed by:
\begin{equation}
    \frac{\partial u_i}{\partial t} = -u_k \frac{\partial u_i}{\partial x_k} + \frac{1}{\rho} \left( -\frac{\partial p}{\partial x_i} + \frac{\partial \tau_{ik}}{\partial x_k} \right) \mbox{ ,}
\end{equation}
whereas the mean Favre-averaged velocity, $\tilde{u}_i$, is given by:
\begin{equation}
    \frac{\partial \tilde{u}_i}{\partial t} = -\tilde{u}_k \frac{\partial \tilde{u}_i}{\partial x_k} + \frac{1}{\bar{\rho}} \left( -\frac{\partial \bar{p}}{\partial x_i} + \frac{\partial \bar{\tau}_{ik}}{\partial x_k} - \frac{\partial R_{ik}}{\partial x_k} \right) \mbox{ .}
\end{equation}
Crucially, while the pressure and viscous stress gradients appear in both the instantaneous and mean equations (scaled by $1/\rho$ and $1/\bar{\rho}$, respectively), the Reynolds stress gradient, $\partial R_{ik}/\partial x_k$ (where $R_{ik} = \overline{\rho u''_i u''_k}$), emerges solely in the mean momentum equation. Consequently, it is strictly scaled by the mean density, $1/\bar{\rho}$.

Subtracting the mean momentum equation from the instantaneous one yields the transport equation for the velocity fluctuations ($u''_i = u_i - \tilde{u}_i$):
\begin{equation}
    \frac{\partial u''_i}{\partial t} = \dots + \left( \frac{1}{\rho} - \frac{1}{\bar{\rho}} \right) \left( -\frac{\partial \bar{p}}{\partial x_i} + \frac{\partial \bar{\tau}_{ik}}{\partial x_k} \right) + \frac{1}{\rho} \left( -\frac{\partial p'}{\partial x_i} + \frac{\partial \tau'_{ik}}{\partial x_k} \right) + \frac{1}{\bar{\rho}} \frac{\partial R_{ik}}{\partial x_k} \mbox{ .}
\end{equation}
As demonstrated by \cite{Arun_Sameen_Srinivasan_Girimaji_2021}, the Reynolds stress gradient is never scaled by $1/\rho$. To construct the two-point correlation equation, this fluctuation equation at location $x$ is multiplied by the displaced state $\rho(x+r)u''_j(x+r)$ and ensemble averaged. Applying this operation to the Reynolds stress gradient yields the correct spatial correlation:
\begin{equation}
    \overline{ \rho(x+r) u''_j(x+r) \left[ \frac{1}{\bar{\rho}(x)} \frac{\partial R_{ik}}{\partial x_k} \right] } = \frac{\overline{\rho(x+r)u''_j(x+r)}}{\bar{\rho}(x)} \frac{\partial  R_{ik}}{\partial x_k} \mbox{ .}
\end{equation}
In Equation (3.15) of \citet{Yu_Zhao_Tang_Yuan_Xu_2022}, however, the authors apply the $1/\rho(x)$ operator to the Reynolds stress gradient, assuming it shared the same algebraic structure as the pressure and viscous gradients. Furthermore,  this term involves the local correlation $\overline{\rho(x+r) u''_i(x+r)}$, which evaluates identically to zero by the fundamental definition of Favre averaging ($\overline{\rho u''} = 0$).

Because of this algebraic oversight in physical space, the Fourier transform presented in their Equation (3.27) yields fictitious spectral components:
\begin{equation}
\begin{split}
    \hat{B}_{ij} =\ & \frac{1}{2}\Re \left[ \widehat{u''_j}^* - \frac{\widehat{\rho} \widehat{u''_j}^*}{\bar{\rho}} + \widehat{\left(\frac{1}{\rho}\right)} \widehat{\rho u''_j}^* \right] \left( - \frac{\partial \bar{p}}{\partial x_i} + \frac{\partial \bar{\tau}_{ik}}{\partial x_k} \right) \\
    &+ \frac{1}{2} \Re \left[ \frac{\widehat{\rho}^* \widehat{u''_j}}{\bar{\rho}} + \widehat{\left(\frac{1}{\rho}\right)} \widehat{\rho u''_j} \right] \frac{\partial (\bar{\rho} R_{ik})}{\partial x_k} + \dots
\end{split}
\end{equation}
This expression contains two inconsistencies. First, the term $\widehat{(1/\rho)} \widehat{\rho u''_j}$ multiplying the gradient of the Reynolds stress tensor is a mathematical artifact of the incorrect $1/\rho$ scaling. When integrated over all wavenumbers, the true physical term must recover the spatial mean budget. The presence of this term prevents the spectral budget from correctly recovering the mean TKE budget. The term $\widehat{\rho}^* \widehat{u''_j}/\bar{\rho}$ is correct since it comes from the product of the instantaneous fluctuating momentum equation, where the Reynolds stress is scaled by the local density $\rho$ and the fluctuating velocity, therefore yielding the product $\widehat{\rho}^* \widehat{u''_j}$ in the spectral domain.

Second, the gradient term $\partial (\bar{\rho} R_{ik}) / \partial x_k$ introduces a dimensional error. Because the Reynolds stress tensor is defined as a density-weighted covariance ($R_{ik} = \overline{\rho u''_i u''_k}$ equation 2.4b in their manuscript), multiplying it by the mean density $\bar{\rho}$ yields incorrect physical dimensions (an extra factor of density). We respectfully note that this dimensional error is also reproduced in Equation (4.1) of \citet{Yu_Zhao_Tang_Yuan_Xu_2022}, where the spatial turbulent kinetic energy budget is defined based on the formulation of \cite{Pirozzoli_bernardini_2011b}.

Applying the rigorous spatial derivation and correcting the spectral transforms, the true and mathematically consistent formulation for the trace 
of the mass-flux term is:
\begin{equation}
    B = \frac{B_{ii}}{2} =
    \ \frac{1}{2} \Re \left[ \widehat{u''_i} - \frac{\widehat{\rho} \widehat{u''_i}^*}{\bar{\rho}} + \widehat{\left(\frac{1}{\rho}\right)} \widehat{\rho u''_i}^* \right] \left( - \frac{\partial \bar{p}}{\partial x_i} + \frac{\partial \bar{\tau}_{ik}}{\partial x_k} \right) + \frac{1}{2} \Re \left[ \frac{\widehat{\rho}^* \widehat{u''_i}}{\bar{\rho}} \right] \frac{\partial R_{ik}}{\partial x_k} \mbox{ .}
\label{eq:appendix_Bii}
\end{equation}

\section{Coordinate transformation and tensor rotation}
\label{app:tensor-rotation}

To analyze the turbulent energy cycles relative to the curved surface, the Cartesian budget terms must be projected into a local frame of reference. By defining the local wall-tangential unit vector $\mathbf{t} = [t_x, t_y]^\mathrm{T}$ and wall-normal unit vector $\mathbf{n} = [n_x, n_y]^\mathrm{T}$, any generic second-rank Cartesian tensor $\mathbf{A}$ is rotated into its wall-tangential ($\xi$) and wall-normal ($\eta$) components as follows
\begin{equation}
\begin{split}
    A_{\xi\xi} &= A_{xx}t_x^2 + A_{xy}t_x t_y + A_{yx}t_y t_x + A_{yy}t_y^2 \mbox{ ,} \\
    A_{\eta\eta} &= A_{xx}n_x^2 + A_{xy}n_x n_y + A_{yx}n_y n_x + A_{yy}n_y^2 \mbox{ ,} \\
    A_{\xi\eta} &= A_{xx}t_x n_x + A_{xy}t_x n_y + A_{yx}t_y n_x + A_{yy}t_y n_y \mbox{ ,} \\
    A_{\eta\xi} &= A_{xx}n_x t_x + A_{xy}n_x t_y + A_{yx}n_y t_x + A_{yy}n_y t_y \mbox{ .}
\end{split}
\label{eq:tensor_rotation}
\end{equation}
Applying this coordinate transformation to the uncontracted forms of the spectral budget terms yields the formulations for each individual rotated component. For instance, the uncontracted Cartesian turbulence production tensor is defined as $P_{ij} = - \hat{Q}_{ik}\partial \tilde{u}_j/\partial x_k - \hat{Q}_{jk}\partial \tilde{u}_i/\partial x_k$. Evaluating equation \eqref{eq:tensor_rotation} for this tensor yields the local tangential, normal, and shear components:
\begin{equation}
\begin{split}
    P_{\xi\xi} &= P_{xx}t_x^2 + P_{xy}t_x t_y + P_{yx}t_y t_x + P_{yy}t_y^2 \mbox{ ,} \\
    P_{\eta\eta} &= P_{xx}n_x^2 + P_{xy}n_x n_y + P_{yx}n_y n_x + P_{yy}n_y^2 \mbox{ ,} \\
    P_{\xi\eta} &= P_{xx}t_x n_x + P_{xy}t_x n_y + P_{yx}t_y n_x + P_{yy}t_y n_y \mbox{ ,} \\
    P_{\eta\xi} &= P_{xx}n_x t_x + P_{xy}n_x t_y + P_{yx}n_y t_x + P_{yy}n_y t_y \mbox{ .}
\end{split}
\end{equation}
Similarly, defining the uncontracted pressure-strain tensor as $\Pi_{ij} = \frac{1}{2} \Re [ \widehat{p'} (\partial \widehat{u_i'}/\partial x_j)^* + \dots ]$, the individual rotated formulations are given by:
\begin{equation}
\begin{split}
    \Pi_{\xi\xi} &= \Pi_{xx}t_x^2 + \Pi_{xy}t_x t_y + \Pi_{yx}t_y t_x + \Pi_{yy}t_y^2 \mbox{ ,} \\
    \Pi_{\eta\eta} &= \Pi_{xx}n_x^2 + \Pi_{xy}n_x n_y + \Pi_{yx}n_y n_x + \Pi_{yy}n_y^2 \mbox{ .}
\end{split}
\end{equation}

\section{Statistical convergence of the SPOD}
\label{app:spod_convergence}

To rigorously assess the statistical convergence of the SPOD modes, a comparative sensitivity analysis is performed. The time-series datasets are sampled at a constant non-dimensional time interval of $\Delta t = 0.5$ (non-dimensionalized by the inflow boundary layer thickness $\delta_{i}$ and freestream velocity $U_\infty$). Accordingly, the Strouhal number is defined as $St = f \delta_{\mathrm{i}} / U_\infty$. The spatial structures are compared across varying Welch block sizes ($N_{\mathrm{FFT}} = 256$ and $512$) and data ensemble fractions ($75\%$ and $100\%$ of the available time-series). Employing the augmented snapshot method with a $66\%$ block overlap, the resulting number of spectral blocks for both the concave and convex configurations are summarized in table~\ref{tab:spod_parameters}. The configuration employing the highest spectral resolution and maximum data span ($N_{\mathrm{FFT}}=512$ using $100\%$ of the data) is established as the baseline reference.

The structural convergence is evaluated by comparing the leading SPOD mode of the streamwise velocity, $u$, at the most energetic frequencies. Because the modes are decoupled along the homogeneous spanwise direction, this convergence analysis is performed independently for each spanwise wavenumber ($k_z$). For a given $k_z$, the reference SPOD energy spectrum is examined to identify the Strouhal number corresponding to the peak energy density, denoted $St_{\mathrm{max}}$. Because variations in the block size $N_{\mathrm{FFT}}$ inherently alter the discrete frequency resolution ($\Delta f$) of the spectral bins, this target frequency $St_{\mathrm{max}}$ is mapped to the nearest available discrete frequency bin in the respective comparison cases.

Once the frequencies are matched, the spatial structure of the leading complex SPOD mode is extracted. The degree of structural convergence is quantified using the magnitude of the complex cosine similarity, defined as
\begin{equation}
    S_u = \frac{\left| \langle u_{\mathrm{ref}}, u_{\mathrm{comp}} \rangle \right|}{\| u_{\mathrm{ref}} \|_2 \| u_{\mathrm{comp}} \|_2} = \frac{\left| \sum u_{\mathrm{ref}} \cdot u_{\mathrm{comp}}^* \right|}{\sqrt{ \left(\sum |u_{\mathrm{ref}}|^2\right) \left(\sum |u_{\mathrm{comp}}|^2\right) }},
    \label{eq:spod_similarity}
\end{equation}
where $u_{\mathrm{ref}}$ and $u_{\mathrm{comp}}$ are the complex spatial modes of the reference and comparison cases, respectively, and summation is performed over the discrete spatial domain. The asterisk ($^*$) denotes the complex conjugate. A similarity value approaching $S_u \to 1.0$ indicates agreement of the spatial structures.
\begin{figure}[htbp]
\centering
\begin{overpic}[trim = 0cm 0cm 0cm 0cm, clip, width=0.99\textwidth]{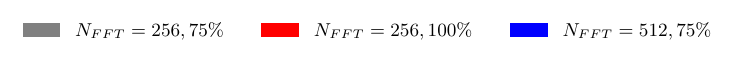}
\end{overpic}
\begin{overpic}[trim = 0cm 0cm 0cm 0cm, clip, width=0.89\textwidth]{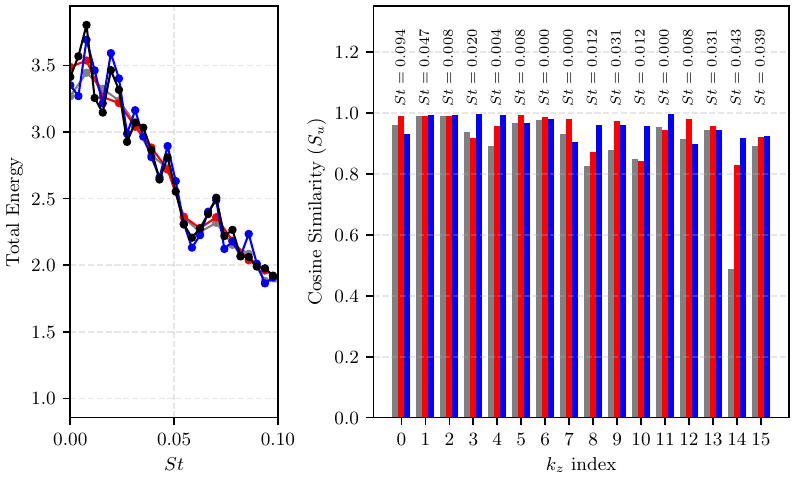}
\end{overpic}
\begin{overpic}[trim = 0cm 0cm 0cm 0cm, clip, width=0.89\textwidth]{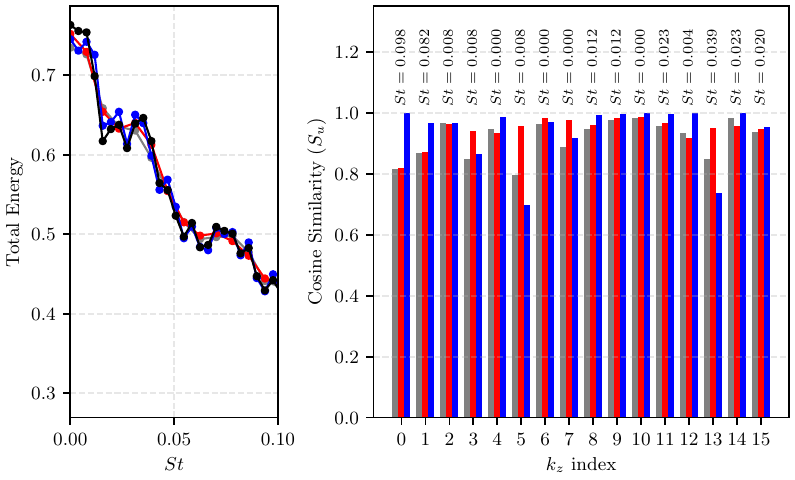}
\end{overpic}
\begin{overpic}[trim = 0cm 0cm 0cm 0cm, clip, width=0.99\textwidth]{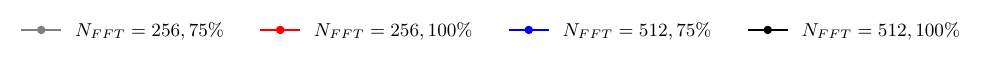}
\end{overpic}
\caption{Statistical convergence of the SPOD analysis. Left: Total energy spectrum obtained by integrating the SPOD spectra across all 16 spanwise wavenumbers (Fourier indices 0--15) for the reference and comparison cases. Right: Cosine similarity ($S_u$) for the leading spatial mode at each spanwise wavenumber evaluated at the dominant frequency $St_{\mathrm{max}}$ (annotated above each bar).}
\label{fig:SPODconvergence}
\end{figure}

The results of this convergence analysis are summarised in figure~\ref{fig:SPODconvergence}. The left panel displays the total energy spectrum, obtained by summing the individual SPOD spectra over all 16 computed spanwise wavenumbers. The spectral estimations demonstrate excellent agreement across all configurations, confirming that the overall energy distribution is statistically converged. To evaluate spatial convergence, the right panel presents the cosine similarity $S_u$ for the dominant mode at each wavenumber. The spatial modes exhibit exceptional structural consistency, with almost all cases yielding a similarity approaching unity. While marginally larger deviations are observed for the shorter $N_{\mathrm{FFT}} = 256$ blocks, the similarity remains above $80\%$ for most wavenumbers. The only notable discrepancy occurs at $k_z$ index 14 for the concave geometry case utilizing $N_{\mathrm{FFT}} = 256$ and $75\%$ of the data, where the similarity drops below $60\%$. These results confirm that the baseline reference configuration ($N_{\mathrm{FFT}} = 512$, $100\%$ data) is robust and converged.

\begin{table}
  \centering
  \vspace{0.2cm}
  \begin{tabular}{lccc}
    \toprule
    & & \multicolumn{2}{c}{Number of blocks ($N_{\mathrm{bl}}$)} \\
    \cmidrule(lr){3-4}
    $N_{\mathrm{FFT}}$ & Data fraction (\%) & Concave ($N_t = 1953$) & Convex ($N_t = 1855$) \\
    \midrule
    256 & 75  & 30 & 28 \\
    256 & 100 & 40 & 38 \\
    \addlinespace
    512 & 75  & 12 & 12 \\
    512 & 100 & 18 & 16 \\
    \bottomrule
  \end{tabular}
  \caption{Parameters used for the statistical convergence assessment of the SPOD modes. The number of blocks ($N_{\mathrm{bl}}$) is computed using the augmented snapshot method with a $66\%$ overlap, applied to the total available snapshots ($N_t$) for each configuration.}
  \label{tab:spod_parameters}
\end{table}

\end{appen}

\bibliographystyle{jfm}
\bibliography{jfm}


\end{document}